\documentclass[
aps,
prr,
twocolumn,
superscriptaddress,
floatfix,
longbibliography,
nofootinbib
]{revtex4-2}
\usepackage[dvipdfmx]{graphicx}
\usepackage{dcolumn}
\usepackage{bm}
\usepackage{ulem} 
\usepackage{amsmath}
\usepackage{amssymb}
\usepackage{txfonts}
\usepackage{hyperref}
\usepackage{color} 
\usepackage{xcolor}
\usepackage{listings}
\usepackage{cprotect}

\usepackage{tikz}
\usetikzlibrary{quantikz2}

\newcommand{\Tr}{{\rm Tr}}
\newcommand{\e}{{\rm e}}
\newcommand{\imag}{{\rm i}}

\newcommand{\im}{\mathrm{i}}
\newcommand{\diff}{\mathrm{d}}
\newcommand{\calD}{\mathcal{D}}
\newcommand{\calE}{\mathcal{E}}
\newcommand{\calH}{\mathcal{H}}
\newcommand{\calJ}{\mathcal{J}}
\newcommand{\calL}{\mathcal{L}}
\newcommand{\calO}{\mathcal{O}}
\newcommand{\calU}{\mathcal{U}}
\newcommand{\bE}{\mathbb{E}}
\newcommand{\p}{\partial}
\newcommand{\Var}{\mathrm{Var}}
\newcommand{\mF}{\mathrm{F}}
\newcommand{\tagH}{\calH}
\newcommand{\tagJ}{\calJ}
\newcommand{\Esample}{\mathbb{E}^{\prime}}
\newcommand{\Vsample}{{\rm Var}^{\prime}}

\hypersetup{
  colorlinks=true,
  linkcolor=[rgb]{0.60,0.00,0.00},
  citecolor=[rgb]{0.00,0.00,0.60},
  urlcolor=[rgb]{0.00,0.00,0.60},
  setpagesize=false
}

\definecolor{codegreen}{rgb}{0,0.6,0}
\definecolor{codegray}{rgb}{0.5,0.5,0.5}
\definecolor{codepurple}{rgb}{0.58,0,0.82}
\definecolor{backcolour}{rgb}{0.95,0.95,0.92}

\lstdefinestyle{mystyle}{
    backgroundcolor=\color{backcolour},   
    commentstyle=\color{codegreen},
    keywordstyle=\color{magenta},
    numberstyle=\tiny\color{codegray},
    stringstyle=\color{codepurple},
    basicstyle=\ttfamily\footnotesize,
    breakatwhitespace=false,         
    breaklines=true,                 
    captionpos=b,                    
    keepspaces=true,                 
    numbers=left,                    
    numbersep=5pt,                  
    showspaces=false,                
    showstringspaces=false,
    showtabs=false,                  
    tabsize=2
}

\begin{document}

\preprint{RIKEN-iTHEMS-Report-24}

\title{
  Simulating Floquet scrambling circuits on trapped-ion quantum computers
}

\author{Kazuhiro~Seki}
\email{kazuhiro.seki@riken.jp}
\affiliation{Quantum Computational Science Research Team, RIKEN Center for Quantum Computing (RQC), Saitama 351-0198, Japan}

\author{Yuta Kikuchi}
\email{yuta.kikuchi@quantinuum.com}
\affiliation{Quantinuum K.K., Otemachi Financial City Grand Cube 3F, 1-9-2 Otemachi, Chiyoda-ku, Tokyo, Japan}
\affiliation{Interdisciplinary Theoretical and Mathematical Sciences Program (iTHEMS), RIKEN, Wako, Saitama 351-0198, Japan}

\author{Tomoya Hayata}
\email{hayata@keio.jp}
\affiliation{Departments of Physics, Keio University School of Medicine, 4-1-1 Hiyoshi, Kanagawa 223-8521, Japan}
\affiliation{Interdisciplinary Theoretical and Mathematical Sciences Program (iTHEMS), RIKEN, Wako, Saitama 351-0198, Japan}

\author{Seiji Yunoki}
\email{yunoki@riken.jp}
\affiliation{Quantum Computational Science Research Team, RIKEN Center for Quantum Computing (RQC), Saitama 351-0198, Japan}
\affiliation{Computational Quantum Matter Research Team, RIKEN Center for Emergent Matter Science (CEMS), Wako, Saitama 351-0198, Japan}
\affiliation{Computational Materials Science Research Team,
RIKEN Center for Computational Science (R-CCS), Kobe, Hyogo 650-0047, Japan}
\affiliation{Computational Condensed Matter Physics Laboratory,
RIKEN Cluster for Pioneering Research (CPR), Saitama 351-0198, Japan}

\begin{abstract}
    Complex quantum many-body dynamics spread initially localized quantum information across the entire system. Information scrambling refers to such a process, whose simulation is one of the promising applications of quantum computing.
    We demonstrate the Hayden-Preskill recovery protocol and the interferometric protocol for calculating out-of-time-ordered correlators to study the scrambling property of a one-dimensional kicked-Ising model on 20-qubit trapped-ion quantum processors.
    The simulated quantum circuits have a geometrically local structure that exhibits the ballistic growth of entanglement, resulting in the circuit depth being linear in the number of qubits for the entire state to be scrambled.
    We experimentally confirm the growth of signals in the Hayden-Preskill recovery protocol and the decay of out-of-time-ordered correlators at late times.
    As an application of the created scrambling circuits, we also experimentally demonstrate the calculation of the microcanonical expectation values of local operators adopting the idea of thermal pure quantum states.
    Our experiments are made possible by extensively utilizing one of the highest-fidelity quantum processors currently available and, thus, should be considered as a benchmark for the current status of the most advanced quantum computers.
\end{abstract}

\date{\today}

\maketitle


\section{Introduction}
\label{sec:introduction}

The spreading or scrambling of quantum information is a characteristic feature of complex quantum many-body dynamics.
Understanding the properties of information scrambling is crucial for addressing fundamental problems in quantum mechanics, including thermalization and chaos.
As a quantum state evolves under unitary dynamics, it gradually loses microscopic details, ultimately retaining only macroscopic information, 
a process known as thermalization.  
As the system further evolves, quantum information is scrambled in the sense that initially localized information spreads, making it increasingly harder 
to recover from local measurements of the evolved state.
The difficulty of information retrieval is closely tied to the complexity of quantum dynamics. 
For instance, it is intuitive to understand that 
information thrown into a fast scrambler, such as a black hole, would be highly challenging to recover.
Indeed, the information scrambling was initially investigated to elucidate the complex black hole dynamics~\cite{Hayden2007, Sekino2008, Shenker2013, Shenker2014, Maldacena2015}, and it is nowadays used to diagnose nonequilibrium properties of quantum many-body dynamics~\cite{Hosur2015, Roberts2017, Swingle2018, Xu2022}.
Extensive studies have been devoted to quantifying the information scrambling and understanding its quantum information-theoretic aspects by computing various indicators, including spectral form factors~\cite{Berry1985, Sieber2001, Sieber2002, Muller2004}, Loschmidt echo~\cite{Peres1984, Goussev2012, Gorin2006}, and out-of-time-ordered correlators (OTOCs)~\cite{Larkin1969, Shenker2013, Roberts2014, Hosur2015, Swingle2018otoc, Xu2022}.

Along with advances in analytical and numerical techniques for studying scrambling quantum dynamics, quantum computing has emerged as a promising tool for efficiently simulating such systems. 
Despite such anticipation and rapid developments of quantum processors, the full potential of quantum computing remains unrealized largely because of errors caused by interaction with environments and imperfect control of quantum operations.
Extracting signals from currently available quantum hardware requires precise error characterization for each hardware. Furthermore, the way that errors affect the final outcome depends not only on the hardware but also heavily on the type of algorithm or quantum circuits employed, even when the same device is used.
To realize large-scale quantum simulation of complex scrambling dynamics, which is one of the promising applications of quantum computing, it is crucial to assess the feasibility of proposed quantum algorithms on current hardware. Such assessments are essential for accurately estimating the computational resources required to scale up the system size.

In this study, 
we explore information scrambling of periodically driven (Floquet) systems by taking the kicked-Ising model as an example~\cite{Prosen2002, Prosen2007}. 
In the context of non-equilibrium physics, the model is known to exhibit Floquet heating, a phenomenon where a quantum system is heated to the infinite temperature after evolution under periodic driving~\cite{Lazarides2014, DAlessio2014, Abanin2015, Mori2016, Kuwahara2016, Mori2018}. 
To assess the protocols characterizing and utilizing this heating phenomenon, we carry out three experiments fully exploiting the highest-fidelity trapped-ion quantum processors~\cite{Eckstein2023}. 
The present studies demonstrate the capability of currently available quantum devices for studying quantum dynamics, and pave the way to scale up experiments and potentially reach the realm of classically intractable system sizes and evolution times.

In the first experiment, we employ the Hayden-Preskill recovery (HPR) protocol~\cite{Hayden2007, Yoshida2017, Yoshida2019}. 
So far, HPR protocols~\cite{Landsman2018, Blok2021} and teleportation protocols inspired by gravitational theory~\cite{Shapoval2022, Jafferis2022} have been demonstrated on other programmable quantum devices. 
Here, to fully explore the information scrambling of the Floquet circuits, we conduct the protocol across all the time steps until late-time scrambling behavior is confirmed.
The raw experimental results clearly show the growth of HPR signals at late times, indicating the scrambling of information as theoretically anticipated.
Moreover, assuming a simple noise model~\cite{Yoshida2019, Bao2020, Hayata2021}, we mitigate the noise effect and demonstrate that the error-mitigated results better agree with the exact results obtained by classical simulations. 

In the second experiment, we perform the interferometric protocol for calculating OTOCs~\cite{Swingle2016, Swingle2018}.
We note that several experimental efforts have been devoted to calculating OTOCs~\cite{Li2017, Garttner2016, Joshi2020, Mi2021, Asaduzzaman2023}.
We address spatial and temporal dependence of the OTOCs executing the circuits involving a maximum of 367 entangling gates causally connected to the final measurements. 
The results indicate that information scrambling grows ballistically during the Floquet dynamics at a chosen set of kicked-Ising parameters, which is clearly seen after applying the error mitigation~\cite{Swingle2018, Mi2021}. 
While the OTOCs and the HPR signals are closely related and carry similar information, there are some pragmatic differences in experiments: the HPR protocol uses two copies of the original system by using almost twice as many qubits as the interferometric protocol, although it is robust against decoherence~\cite{Yoshida2019, Bao2020, Hayata2021}.

In the third experiment, we calculate the thermal expectation values of a local operator as an application of the scrambling circuits.
In statistical mechanics, ensembles are represented by probabilistic mixtures of a macroscopically large number of eigenstates of a Hamiltonian. 
Randomized protocols have been developed to recover such thermal properties using randomly sampled pure states, including thermal pure quantum (TPQ) states for microcanonical and canonical ensembles~\cite{Bocchieri1959, Jaklic1994, Jaklic2000, Hams2000, Iitaka2004, Popescu2006, Sugiura2012, Sugiura2013, Sugiura2017}. 
To make use of the TPQ states in quantum algorithms, unitary or state designs play key roles in preparing relevant random states efficiently on quantum computers~\cite{Nakata2017math, Jin2021, Richter2021, Seki2022, Powers2021, Coopmans2023, Coopmans2023}.

Giving the implications of the first two experimental observations, 
which suggest the states evolved under chaotic dynamics 
approximate a state 2-design~\cite{Kaneko2020, Choi2021, Cotler2023, Fava2023},\footnote{
    We consider an ensemble of states evolved under a Floquet dynamics, 
    instead of a Hamiltonian dynamics~\cite{Kaneko2020}. 
    }
we proceed to calculate the expectation value of local operators 
with respect to the microcanonical ensemble by employing the scrambling circuits as a subroutine.
However, it should be noted that, from the viewpoint of gate 
complexity, use of our Floquet scrambling circuits instead of, e.g., 
the Clifford scrambler does not offer significant advantages on 
devices featuring native long-range gate operations such as trapped-ion 
devices. Nevertheless, our protocol should be favored on hardware 
with limited connectivity because the evolution operator of 
the kicked-Ising model does not require long-range entangling gates.

The rest of this paper is organized as follows.
In Sec.~\ref{sec:model}, we provide a brief review of 
the kicked-Ising model as a periodically driven quantum system.
In Sec.~\ref{sec:method}, we discuss the quantum algorithms, 
which we will demonstrate using the trapped-ion quantum computers. 
We introduce the HPR protocol and interferometric protocol for 
OTOCs, both of which serve as measures of information scrambling.
Additionally, we describe a method for computing thermal expectation 
values based on the concept of TPQ states using the scrambling 
circuits.
The experimental setups and results obtained are presented in Sec.~\ref{sec:experiments}.
Finally, we provide conclusions and outlooks in Sec.~\ref{sec:conclusion}. 
In Appendix~\ref{app:mitigate}, the noise model we adopted to mitigate the hardware noise in the experiments is explained.
Appendix~\ref{app:numerical_results} provides classical numerical results of OTOCs to supplement the experimental results presented in Sec.~\ref{sec:experiments}.
The rest of the appendices are devoted to elaborating the computation of thermal expectation values using scrambling circuits.
We numerically justify the use of scrambling circuits to estimate the normalized trace in Appendix~\ref{app:Loschmidt}. 
In Appendix~\ref{app:TPQ}, we carefully study the error in microcanonical TPQ expectation values.
The asymptotic resource analysis for the imaginary time evolution algorithm used in the present experiments is found in Appendix~\ref{app:imaginary}.
Finally, we provide additional experimental results of thermal expectation values in Appendix~\ref{app:LZZ}.

\section{Kicked-Ising model as a Floquet circuit}
\label{sec:model}

Floquet evolution is described by a time-dependent Hamiltonian periodic in time, $H(t)=H(t+T)$, with the period~$T$.
Due to the periodic driving, the system does not conserve energy. 
Furthermore, such periodically driven nonintegrable systems are expected to eventually heat up to infinite temperature by acquiring energy from the driving force, known as the Floquet heating, 
occurring after an exponentially long time in their driving frequency~\cite{Lazarides2014, DAlessio2014, Mori2016, Kuwahara2016, Mori2018}. 
A notable example is a Hamiltonian dynamics approximated by 
the first-order product (Lie-Trotter) formula~\cite{Lloyd1996},
\begin{align}
    \left(\prod_{i=1}^{K}\e^{-\im H_i \frac{T}{K}} \right)^{mK},
\end{align}
for a static Hamiltonian $H=\sum_{i=1}^{K} H_i$ ($H_i$ are non-commutative with each other) and total evolution time~$mT$ with an integer $m$. This can be considered as the periodically driven dynamics with the driving period $T$ over $m$ cycles.
It has been observed that, in the Trotter time evolution of an Ising spin chain, there exists a transition or crossover between many-body localized and quantum chaotic regimes as a function of the driving period $T$, or equivalently the inverse driving frequency, characterized by the inverse participation ratio and OTOCs~\cite{Heyl2019, Sieberer2019, Vernier2023}.

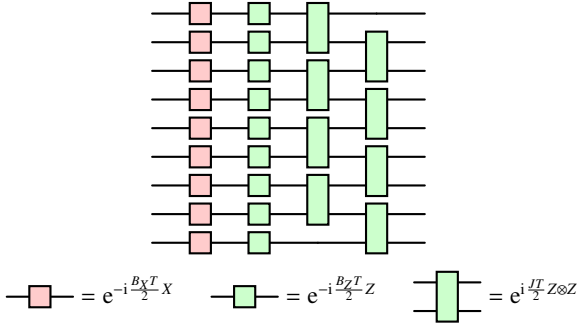
\begin{figure}
\begin{quantikz}[column sep=.5cm, row sep=.1cm]
     & \gate[style={fill=red!20}]{} 
     & \gate[style={fill=green!20}]{} 
     & \gate[2, style={fill=green!20}]{}
     & 
     & 
     \\
     & \gate[style={fill=red!20}]{} 
     & \gate[style={fill=green!20}]{}
     & 
     & \gate[2, style={fill=green!20}]{}
     & 
     \\
     & \gate[style={fill=red!20}]{}  
     & \gate[style={fill=green!20}]{}
     & \gate[2, style={fill=green!20}]{}
     & 
     & 
     \\
     & \gate[style={fill=red!20}]{} 
     & \gate[style={fill=green!20}]{} 
     & 
     & \gate[2, style={fill=green!20}]{}
     & 
     \\
     & \gate[style={fill=red!20}]{} 
     & \gate[style={fill=green!20}]{}
     & \gate[2, style={fill=green!20}]{}
     & 
     & 
     \\
     & \gate[style={fill=red!20}]{} 
     & \gate[style={fill=green!20}]{}
     & 
     & \gate[2, style={fill=green!20}]{}
     & 
     \\
     & \gate[style={fill=red!20}]{} 
     & \gate[style={fill=green!20}]{}
     & \gate[2, style={fill=green!20}]{}
     & 
     & 
     \\
     & \gate[style={fill=red!20}]{}  
     & \gate[style={fill=green!20}]{}
     & 
     & \gate[2, style={fill=green!20}]{}
     & 
     \\
     & \gate[style={fill=red!20}]{} 
     & \gate[style={fill=green!20}]{} 
     & 
     & 
     & 
\end{quantikz}
\\
\begin{quantikz}[column sep=.3cm, row sep=.1cm]
&\gate[style={fill=red!20}]{}&
\end{quantikz}$ 
= \e^{-\imag \frac{B_XT}{2} X}$
\quad
\begin{quantikz}[column sep=.3cm, row sep=.1cm]
&\gate[style={fill=green!20}]{}&
\end{quantikz}$ 
= \e^{-\imag \frac{B_Z T}{2} Z}$
\quad
\begin{quantikz}[column sep=.3cm, row sep=.1cm]
&\gate[2, style={fill=green!20}]{}&\\ &&
\end{quantikz}$ 
= \e^{\imag \frac{JT}{2} Z\otimes Z}$

\caption{The circuit representation of a single Floquet circle 
described by the Floquet operator $U_{\rm F}$ 
in Eq.~\eqref{eq:U_F} for 
a one-dimensional system consisting of $N=9$ qubits 
under open boundary conditions. 
\label{fig:UFcirc}
}
\end{figure}

In this study, we consider the kicked-Ising model on a one-dimensional chain, which is a periodically driven quantum system described by the following time-dependent Hamiltonian:
\begin{align}
\label{eq:periodic_H}
  &H(t) =
  \nonumber\\
  &\left\{
    \begin{array}{lll}
      \displaystyle{H_X = B_X\sum_{i=1}^{N} X_i,}
      & \displaystyle{t\in [0, T/2)},
      \\
      \displaystyle{H_Z = -J\sum_{i=1}^{N\text{ or }N-1}Z_iZ_{i+1} + B_Z\sum_{i=1}^{N} Z_i,}
      & \displaystyle{t\in[T/2,T)},
    \end{array}
  \right.
\end{align}
where $X_i$ ($Z_i$) is the $x$ ($z$) component of the Pauli operators at 
site $i$, and $N$ is the total number of sites. 
We will consider both open and periodic boundary conditions and 
accordingly the sum of Ising-coupling terms in $H_Z$ is properly 
chosen. 
When the periodic boundary conditions are imposed, the $N+1$th site 
is identified with the $1$st site. 
The corresponding time evolution over a single period is described by the Floquet operator, 
\begin{equation}
\label{eq:U_F}
    U_\mathrm{F}
    =
    \e^{-\im H_Z T/2} \e^{-\im H_X T/2},
\end{equation}
which consists of $\calO(N)$ single-qubit gates and two-qubit gates.
An example of the circuit representation of $U_{\rm F}$ is shown in Fig.~\ref{fig:UFcirc}. 
Depending on the model parameters, connectivity of the Ising-coupling terms, and initial states, this model may transiently exhibit discrete-time quasicrystalline orders before thermalization~\cite{Shinjo2024}.
In our experiments, however, we focus mainly on simulating the model with the parameters at and around the self-dual point$|JT|=|B_XT|=\pi/2$, where the dynamics is maximally chaotic~\cite{Bertini2018SFF, Bertini2019, Bertini2019entanglement, Poroli2020dynamics}.

\section{Methods}
\label{sec:method}

We utilize the HPR fidelity and OTOCs as indicators of information scrambling and introduce two protocols for calculating 
these quantities.
Furthermore, we discuss the calculation of thermal expectation 
values as an application of scrambling circuits.
Those three protocols are implemented on quantum hardware, 
as detailed in Section~\ref{sec:experiments}.

\subsection{Hayden-Preskill recovery (HPR) protocol}

The HPR protocol provides a means to characterize the scrambling properties of unitary dynamics~\cite{Hayden2007,Yoshida2017,Yoshida2019}.
This protocol can be understood as the teleportation of an 
input quantum state from a local subsystem $A$ to a distant 
subsystem $R'$ (see Fig.~\ref{fig:HPRcirc}). 
Remarkably, the degree of scrambling in the dynamics is directly 
correlated with the teleportation fidelity. Specifically, 
more scrambling dynamics lead to higher teleportation fidelity, 
as elucidated further below.

For illustrative purposes, we first explain the top circuit in Fig.~\ref{fig:HPRcirc}. 
Let us denote the number of qubits in subsystem $A$ as $N_A$ 
with the dimension of the corresponding Hilbert space $\calH_A$ 
being $d_A=2^{N_A}$, and similarly for other subsystems.
The sizes of the subsystem satisfy $N_A=N_{A'}=N_R=N_{R'}$, $N_B=N_{B'}$, $N_C=N_{C'}$, $N_D=N_{D'}$, and $N_A+N_B=N_C+N_D=N$.
The circuit starts with the state 
$\ket{\psi}_A\otimes \ket{\mathrm{EPR}}_{BB'}\otimes \ket{\mathrm{EPR}}_{R'A'}$, where $\ket{\psi}$ is the state 
to be teleported and the EPR states are defined as $\ket{\mathrm{EPR}}_{BB'}:=d_B^{-1/2}\sum_{i=1}^{d_B}\ket{i}_B\otimes\ket{i}_{B'}$ 
and $\ket{\mathrm{EPR}}_{R'A'}:=d_A^{-1/2}\sum_{i=1}^{d_A}\ket{i}_{R'}\otimes\ket{i}_{A'}$.
After applying $U_{AB}\otimes U^\ast_{A'B'}$ with $U_{AB}$ ($U^\ast_{A'B'}$) being a unitary evolution: $\calH_A\otimes\calH_B\rightarrow\calH_C\otimes\calH_D$ (the corresponding backward evolution: $\calH_{A'}\otimes\calH_{B'}\rightarrow\calH_{C'}\otimes\calH_{D'}$), we measure the state on $DD'$ and post-select only when it is projected onto $\ket{\mathrm{EPR}}_{DD'}$.
Upon the post-selection, we find the state $\ket{\psi'}$ 
as an output on $R'$, which is close to the input state 
$\ket{\psi}$ at late times, i.e. $\braket{\psi'}{\psi}\sim1$, 
if the unitary $U$ has scrambled the input quantum state.

\begin{figure}
\begin{quantikz}
\lstick{$\ket{\psi}_A$} 
  & \gate[2][3em]{U}\gateinput{$A$}\gateoutput{$C$}
  &
  \\
  \makeebit{$\ket{\mathrm{EPR}}_{BB'}$}
  & \gateinput{$B$}\gateoutput{$D$}
  & \meter[2]{\mathrm{EPR}}
  \\
  & \gate[2][3em]{U^*}\gateinput{$B'$}\gateoutput{$D'$}
  &
  \\
  \makeebit{$\ket{\mathrm{EPR}}_{R'A'}$}
  & \gateinput{$A'$}\gateoutput{$C'$}
  & 
  \\
  &
  & \rstick{$\ket{\psi'}_{R'}$}
\end{quantikz}
\\[2em]
\begin{quantikz}
  \makeebit{$\ket{\mathrm{EPR}}_{RA}$}
  &
  &
  & \meter[6]{\mathrm{EPR}}
  \\
  & \gate[2][3em]{U}\gateinput{$A$}\gateoutput{$C$}
  &
  \\
  \makeebit{$\ket{\mathrm{EPR}}_{BB'}$}
  & \gateinput{$B$}\gateoutput{$D$}
  & \meter[2]{\mathrm{EPR}}
  \\
  & \gate[2][3em]{U^*}\gateinput{$B'$}\gateoutput{$D'$}
  &
  \\
  \makeebit{$\ket{\mathrm{EPR}}_{R'A'}$}
  & \gateinput{$A'$}\gateoutput{$C'$}
  & 
  \\
  &
  &
  &
\end{quantikz}
\caption{\label{fig:HPRcirc}
    The quantum circuits for the HPR protocol. 
    (Top) The state $\ket{\psi}$ is approximately teleported to the bottom register on $R'$.
    (Bottom) The input state $\ket{\psi}$ is replaced by an EPR state entangled with a reference register on $R$ to eliminate the state dependence.
}
\end{figure}

To extract the scrambling property of the unitary $U$, 
independent of the input state $\ket{\psi}$, we employ another 
protocol using the circuit at the bottom of Fig.~\ref{fig:HPRcirc}, 
where $\ket{\psi}$ is replaced with another EPR state entangled 
with the reference state on $R$.
The whole state before the measurement is given by 
\begin{align}
  \ket{\Phi} 
  :=
  \big(U_{AB} \otimes U^\ast_{A'B'}\big)
  \ket{\mathrm{EPR}}_{RA}\otimes \ket{\mathrm{EPR}}_{BB'}\otimes \ket{\mathrm{EPR}}_{R'A'}. 
\end{align}
We apply the projector $\Pi_{DD'}:=\ket{\mathrm{EPR}}\bra{\mathrm{EPR}}_{DD'}$ by measuring the state on $DD'$ and post-selecting, yielding 
\begin{align}
  \ket{\Psi} := \frac{\Pi_{DD'}\ket{\Phi}}{\sqrt{P_\mathrm{EPR}}},
  \quad
  P_\mathrm{EPR} := \Tr[\Pi_{DD'}\ket{\Phi}\bra{\Phi}],
\end{align}
where $P_\mathrm{EPR}$ is referred to as the post-selection 
probability. 
We then read off the probability that the state on $RR'$ is in 
$\ket{\mathrm{EPR}}_{RR'}$ as 
\begin{align}
  F_\mathrm{EPR} := \Tr[\Pi_{RR'}\ket{\Psi}\bra{\Psi}],
\end{align}
which quantifies how close the resultant state on $R'$ is to the 
input state in terms of fidelity.
The quantity $F_\mathrm{EPR}$ is known as the recovery fidelity.

\subsection{Out-of-time-ordered correlators (OTOCs)}

An OTOC quantifies the sensitivity of a given quantum dynamics $U$ to the insertion of a butterfly operator $O_D$, measuring how $O_D$ evolves under the time evolution $U$ via a measurement operator $O_A$.
The OTOC with respect to a density operator $\rho$ is defined as 
\begin{equation}
\label{eq:def_OTOC}
    \mathrm{OTOC}_\rho
    := 
    \Tr\big[\rho U^\dag O_D^\dag U O_A^\dag U^\dag O_D U O_A \big],
\end{equation}
where $O_A$ and $O_D$ are operators supported on subsystems $A$ and $D$, respectively.
Provided that the subsystems $A$ and $D$ are disjoint, the growth of operator $U^\dag O_D U$ leads to the decay of the OTOC as it starts to overlap with $O_A$.
For scrambling circuits, the OTOC decays exponentially in $N$ at late times~\cite{Xu2022}.

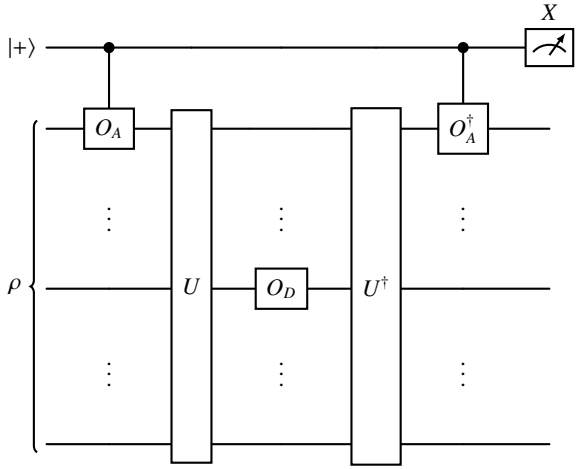
\begin{figure}
\begin{quantikz}
    \lstick{$\ket{+}$}
    & \ctrl{1}
    & 
    &
    &
    & \ctrl{1}
    & \meter{X}
    \\
    \lstick[5]{$\rho$}
    & \gate{O_A}
    & \gate[5]{U}
    & \linethrough
    & \gate[5]{U^\dag}
    & \gate{O_A^\dag}
    &
    \\
    & \vdots \wireoverride{n}
    & \wireoverride{n}
    & \vdots \wireoverride{n}
    & \wireoverride{n}
    & \vdots \wireoverride{n}
    & \wireoverride{n}
    \\
    &
    &
    & \gate{O_D}
    &
    &
    &
    \\
    & \vdots \wireoverride{n}
    & \wireoverride{n}
    & \vdots \wireoverride{n}
    & \wireoverride{n}
    & \vdots \wireoverride{n}
    & \wireoverride{n}
    \\
    &&&&&&
\end{quantikz}
\caption{\label{circ:OTOC}
The interferometric circuit employed to compute the OTOC defined in 
Eq.~\eqref{eq:def_OTOC}. The qubit at the top is the single-qubit ancillary register initialized to $\ket{+}$, while the remainder constitutes the system register with the input state $\rho$. 
At the end of the circuit, the ancillary qubit is measured in the 
$X$ basis.
}
\end{figure}

We employ the interferometric circuit shown in Fig.~\ref{circ:OTOC} to compute OTOCs~\cite{Swingle2016, Swingle2018}. 
The circuit comprises the system register and 
a single-qubit ancillary register shown at the top of the figure, 
with the composite system initialized to 
$\ket{+}\bra{+}\otimes\rho$.
At the end of the circuit, the ancillary qubit is measured in the 
$X$ basis to extract the values of the OTOCs.
For OTOCs with complex values, the measurement in the $X$ basis 
($Y$ basis) retrieves its real (imaginary) part.

We provide several remarks to make connections to the HPR protocol:
Firstly, the post-selection probability $P_\mathrm{EPR}$ in the HPR 
protocol relates to a scrambling property through the 
operator-averaged OTOC as follows~\cite{Yoshida2017}:
\begin{align}
\begin{split}
    P_\mathrm{EPR}
    &= 
    \mathbb{E}_{O_A\sim U(d_A),O_D\sim U(d_D)}\big[
      {\rm OTOC}_{I/d}
    \big]
    \\
    &=
    \frac{1}{d_A^2 d_D^2}
    \sum_{O_A\in\mathrm{Pauli}_A}
    \sum_{O_D\in\mathrm{Pauli}_D}
    {\rm OTOC}_{I/d},
\end{split}
\end{align}
where the OTOCs are evaluated with respect to the maximally mixed 
state $\rho=I^{\otimes N}/d$ with $d=2^N$ being the dimension of 
the Hilbert space.
The Haar averages of $O_A$ and $O_D$ are taken over the 
unitary operators supported on $A$ and $D$, respectively, implying 
that the operator-averaged OTOC is independent of the choice of 
operators but still depends on the choice of the supports $A$ and $D$. 
On the right-hand side, Pauli$_A$ and Pauli$_D$ denote the sets of 
all Pauli operators supported on $A$ and $D$, respectively.
In the second equality, we used the fact that the Haar average of 
the first moment of a unitary operator and its Hermitian conjugate 
can be replaced by the average over equally weighted Pauli operators.

Secondly, the recovery fidelity $F_\mathrm{EPR}$ is expressed 
as~\cite{Yoshida2017},
\begin{align}\label{eq:relation}
  F_\mathrm{EPR} = \frac{1}{d_A^2 P_\mathrm{EPR}}.
\end{align}
In particular, when the operator $U$ is a Haar random unitary operator, $P_\mathrm{EPR}$ takes the value $d_A^{-2} + d_D^{-2} - (d_A d_D)^{-2}$, resulting in the recovery fidelity being 
$4^{-\delta}$-close to unity for $N_D>N_A+\delta$.
More generally, the recovery fidelity $F_\mathrm{EPR}$ 
increases significantly at late times when $U$ scrambles the input 
state, thereby serving as an indicator of scrambling similar to 
OTOCs.

Thirdly, the recovery fidelity $F_\mathrm{EPR}$ carries equivalent information to the operator-averaged OTOC in the absence of errors.
However, the HPR protocol offers an advantage in the presence of hardware noise. In the HPR protocol, scrambling induces the growth 
of $F_\mathrm{EPR}$, while decoherence causes the fidelity to decay~\cite{Yoshida2019}. This contrasts with OTOCs, which decay 
due to both scrambling and decoherence. Consequently, distinguishing between them becomes challenging in noisy simulations. 
In extreme cases, OTOCs may decay even in the absence of scrambling 
under strong decoherence.

Finally, the advantage of the interferometric protocol for OTOCs 
over the HPR protocol lies in its capability to handle larger 
system sizes. This becomes particularly crucial when the number 
of available qubits is limited.

\subsection{Thermal expectation value from scrambling circuits}
\label{sec:TPQ}

As an application of the scrambling circuits, we calculate the expectation value of a local operator $O$ with respect to a microcanonical ensemble: 
\begin{align}
\label{eq:micro_ave}
    \langle O\rangle = \Tr[O \rho_\sigma(E)],
    \quad
    \rho_\sigma(E) = \frac{\e^{-(E-\tagH)^2/2\sigma^2}}{\Tr[\e^{-(E-\tagH)^2/2\sigma^2}]},
\end{align}
where $\calH$ denotes the Hamiltonian of a thermal system, distinct 
from the Floquet Hamiltonian $H$ in Eq.~\eqref{eq:periodic_H}.
Selecting the width parameter $\sigma\propto\sqrt{N}$ ensures 
that the expectation value converges to the correct microcanonical 
expectation value in the thermodynamic limit.
Motivated by the idea of microcanonical TPQ states~\cite{Sugiura2012}, 
we first consider a state $k$-design, defined as an ensemble $\calE$ 
of states reproducing the statistics of $k$-th moments of the Haar 
random distribution, i.e.,  
\begin{align}
    \bE_{\ket{\psi}\sim\calE}[(\ket{\psi}\bra{\psi})^{\otimes k}O]
    =
    \bE_{\ket{\psi}\sim\mathrm{Haar}}[(\ket{\psi}\bra{\psi})^{\otimes k}O],
\end{align}
for any operator $O$. 
By sampling a pure state $\ket{\psi}$ randomly from a state 2-design, 
the estimated expectation value $\langle O\rangle$ in Eq.~\eqref{eq:micro_ave} is given as 
\begin{align}
\label{eq:estimator}
    \langle O\rangle^{\mathrm{(est)}}
    =
    \frac{\bra{\psi}O\e^{-(E-\tagH)^2/2\sigma^2} \ket{\psi}}
    {\bra{\psi}\e^{-(E-\tagH)^2/2\sigma^2} \ket{\psi}}.
\end{align}

For a pure state $\ket{\psi}$ sampled from a state $k$-design with $k\geqslant 2$,\footnote{
    Similarly, a unitary $k$-design is an ensemble ${\cal F}$ of 
    unitary operators that reproduces the statistics of $k$-th moments of the Haar random distribution, i.e., $\bE_{U\sim{\cal F}}[U^{\dag\otimes k}OU^{\otimes k}]=\bE_{U\sim\mathrm{Haar}}[U^{\dag\otimes k}OU^{\otimes k}]$ holds for any operator $O$.
} 
the target expectation value is estimated with exponentially small 
variance in system size $N$. 
More precisely, the following inequality
\begin{align}
\label{eq:success_prob}
\begin{split}
    &\mathrm{Pr}_{\psi\sim\text{2-design}}\big[
        |
            \langle O\rangle^{\mathrm{(est)}}
            -
            \langle O\rangle 
        | 
        \le \epsilon
    \big]
    \\
    &\ge
    1-\calO\big(\sqrt{\Tr[\rho_\sigma(E)^2]}/\epsilon^2\big)
    =
    1-\calO(\e^{-c N}/\epsilon^2),
\end{split}
\end{align}
follows from the Chebyshev inequality for some positive constant~$c$~\cite{Sugiura2012, Sugiura2013}.
Here, the left-hand side represents the probability that the 
estimator $\langle O\rangle^\mathrm{(est)}$ approximates 
$\langle O\rangle$ within an error of at most $\epsilon$, 
considering a pure state $\ket{\psi}$ randomly sampled from 
a state 2-design.\footnote{
    A pure state $\ket{\phi}$ sampled from an ensemble $\Phi$ is called a microcanonical TPQ state if it satisfies the condition: 
    $\mathrm{Pr}_{\phi\sim\Phi}\big[
    |\bra{\phi} O_i\ket{\phi} - \langle O\rangle | \ge \epsilon 
    \big]
    \le \calO(\e^{-c N})$
    for a predefined set of operators $\{O_i\}_i$.
    Therefore, $\e^{-(E-\tagH)^2/4\sigma^2}\ket{\psi}/\|\e^{-(E-\tagH)^2/4\sigma^2}\ket{\psi}\|$ with $\ket{\psi}$ sampled 
    from a state 2-design is considered a microcanonical TPQ state. }
In the last equality, we used that the purity $\Tr[\rho_\sigma(E)^2]$ decays exponentially, that is, the thermal entropy of $\rho_\sigma(E)$ follows a volume law, which holds for a typical state~\cite{Sugiura2012} (see Appendix~\ref{app:TPQ} for details).
Therefore, the inequality states that 
$\langle O\rangle^\mathrm{(est)}$ provides an $\epsilon$-precise 
estimator with a probability exponentially close to 1.
We prove this statement in Appendix~\ref{app:TPQ}.
Consequently, a single instance of a random state suffices when the associated purity of a targeted microcanonical state rapidly decays.

On the other hand, the quantum states generated after the scrambling time evolution approximate a state $k$-design~\cite{Kaneko2020, Choi2021, Cotler2023, Fava2023}. 
Considering that the Floquet operator $U_\mathrm{F}$ exhibits 
scrambling properties, we expect that $(U_\mF)^m\ket{\psi_0}$ 
behaves as a state sampled from a state 2-design. 
In other words, the ensemble
\begin{align}
\label{eq:Floquet_ensemble}
    \Psi = \{(U_\mF)^m\ket{\psi_0}\mid m\in\mathbb{N}\}
\end{align}
approximates a state 2-design if the evolution time $mT$ 
exceeds the scrambling time. Supplementary numerical analyses 
supporting this assertion can be found in Appendix~\ref{app:Loschmidt}. 

Combining these two observations, we attempt to compute the microcanonical expectation value of $O$ with the single state $\ket{\psi} = (U_\mathrm{F})^{m}\ket{\psi_0}$~\cite{Wall1995,Chowdhury2016,Lu2021,Yang2022,Seki2022,Schuckert2023,Ghanem2023}.
To implement $\e^{-(E-\tagH)^2/2\sigma^2}$, we rewrite the operator 
as follows: 
\begin{align}
\label{eq:amplitude}
\begin{split}
    \e^{-(E-\tagH)^2/2\sigma^2}
    &= 
    \frac{\sigma}{\sqrt{2\pi}}\int_{-\infty}^{\infty}\diff t\,
    \e^{-\sigma^2 t^2/2}\e^{\im E t}\e^{-\im \tagH t}
    \\
    &\approx
    \frac{\sigma}{\sqrt{2\pi}}\sum_{s=-S}^{S}\Delta t\,
    \e^{-\sigma^2(s\Delta t)^2/2}\e^{\im Es\Delta t}\e^{-\im \tagH s\Delta t},
\end{split}
\end{align}
where the integral is discretized and the cutoff $S\in\mathbb{N}$ 
is introduced to truncate the infinite sum.
Therefore, the task on the quantum processor boils down to 
evaluating the following amplitudes: 
\begin{align}
    \calL_O(s\Delta t)
    :=
    \bra{\psi_0}(U_\mF)^{\dag m}O\e^{-\im \tagH s\Delta t}(U_\mF)^m\ket{\psi_0},
\end{align}
and    
\begin{align}
\label{eq:Loschmidt_method}
    \calL(s\Delta t)
    :=
    \bra{\psi_0}(U_\mF)^{\dag m}\e^{-\im \tagH s\Delta t}(U_\mathrm{F})^m\ket{\psi_0},
\end{align}
for $s=-S, -S+1, \dots, S$.
The quantity in Eq.~\eqref{eq:Loschmidt_method} is known as 
Loschmidt amplitude~\cite{Peres1984, Goussev2012, Gorin2006}.
We then perform classical post-processing of the outcomes of 
quantum computations using Eq.~\eqref{eq:amplitude}, i.e., 
\begin{align}
\label{eq:D_O(E)}
    D_O(E)
    :=
    \frac{\sigma}{\sqrt{2\pi}}\sum_{s=-S}^{S}\Delta t\,
    \e^{-\sigma^2(s\Delta t)^2/2}\e^{\im Es\Delta t}\calL_O(s\Delta t),
\end{align}
and    
\begin{align}
\label{eq:D(E)}
    D(E)
    :=
    \frac{\sigma}{\sqrt{2\pi}}\sum_{s=-S}^{S}\Delta t\,
    \e^{-\sigma^2(s\Delta t)^2/2}\e^{\im Es\Delta t}\calL(s\Delta t).
\end{align}
Notice that $D(E)$ is the density of states of $\tagH$ with respect 
to the energy $E$.
Using these quantities, the microcanonical expectation value of $O$ 
is estimated as
\begin{align}
\label{eq:estimator_circuit}
    \langle O\rangle^{\mathrm{(est)}}
    =
    \frac{D_O(E)}{D(E)}.
\end{align}

Three remarks are in order. 
First, evaluating $\calL_O(s\Delta t)$ and $\calL(s\Delta t)$ to estimate the normalized traces $\Tr[O\e^{-\im\tagH s\Delta t}]/2^N$ and $\Tr[\e^{-\im\tagH s\Delta t}]/2^N$, respectively, with an additive error of $1/\mathrm{poly}(N)$ can be hard tasks for classical computation.
The potential hardness is based on the fact that the normalized trace estimation is a DQC1 complete problem, for which no polynomial-time classical algorithm is likely to exist~\cite{Knill1998, Datta2005, Shepherd2006, Shor2008}. See also Refs.~\cite{Brandao2008,Chowdhury2021} for the normalized partition function problem with one-clean qubit.
Second, the protocol presented here, nevertheless, requires a non-scalable shot overhead to evaluate the expectation value in Eq.~\eqref{eq:estimator} within a given uncertainty.
This is because a scrambled state typically has a small overlap with the state $\rho_\sigma(E)$ in Eq.~\eqref{eq:micro_ave}, particularly for small $E$ (see Appendix~\ref{app:imaginary} for details).
We note that efforts have been made to reduce the overhead by classically sampling the input states to address the low-energy microcanonical states as well as the low-temperature canonical states~\cite{Lu2021,Yang2022,Schuckert2023,Ghanem2023}.
Third, quadratic improvement of asymptotic runtime for calculating the expectation value in Eq.~\eqref{eq:estimator} can be achieved, for instance, by adopting the algorithm proposed in Ref.~\cite{Chowdhury2016}.
However, we employ the present protocol to experimentally demonstrate the property of scrambling circuits as an approximate state 2-design on a currently available noisy quantum device (see Appendix~\ref{app:Loschmidt} for supplementary analyses).

\section{Results}
\label{sec:experiments}

We experimentally demonstrate the aforementioned protocols on Quantinuum H1-1 and H1-2 trapped-ion quantum computers.
The specifications of H1 quantum computers at the time of the experiments are as follows: both H1 quantum computers consist of 20 qubits, each qubit comprised of the $S_{1/2}$ hyperfine clock states 
of a trapped $^{171}\mathrm{Yb}^+$ ion.
The H1 systems natively implement single-qubit rotation gates and 
two-qubit gates $\e^{-\im\frac{\theta}{2}Z\otimes Z}$ parametrized 
by a real angle~$\theta$.
All gate operations are performed in one of five gate zones, 
each operating in parallel. 
The two-qubit native gates can be applied to an arbitrary pair of 
qubits by shuttling ions to gate zones, thus realizing all-to-all 
connectivity.
Average single- and two-qubit gate infidelities are about $0.004\%$ 
and $0.2\%$, respectively. 
Specifically, the average infidelity of the native two-qubit gate, $\e^{-\im\frac{\theta}{2}Z\otimes Z}$, is currently characterized as 
\begin{align}
\label{eq:pt}
    p_\mathrm{2Q}(\theta)
    = 
    p \left(p_a \frac{|\theta|}{\pi} + p_b\right),
\end{align}
with $p_a=1.651$, $p_b=0.175$, and $p=1.38\times 10^{-3}$ for 
the H1-1 system, and $p_a=1.651$, $p_b=0.175$, and $p=2.97\times 10^{-3}$ for the H1-2 system. 
State preparation and measurement errors are $0.3\%$ on average.
See Ref.~\cite{H1datasheet} for more details.
All the quantum circuits used in the experiments are compiled with TKET 1.24 for the native gate set of the Quantinuum H1 quantum computers at optimization level~$2$~\cite{sivarajah2020}.

\subsection{Hayden-Preskill recovery protocol}

\begin{figure*}[t]
  \includegraphics[width=.45\textwidth]
  {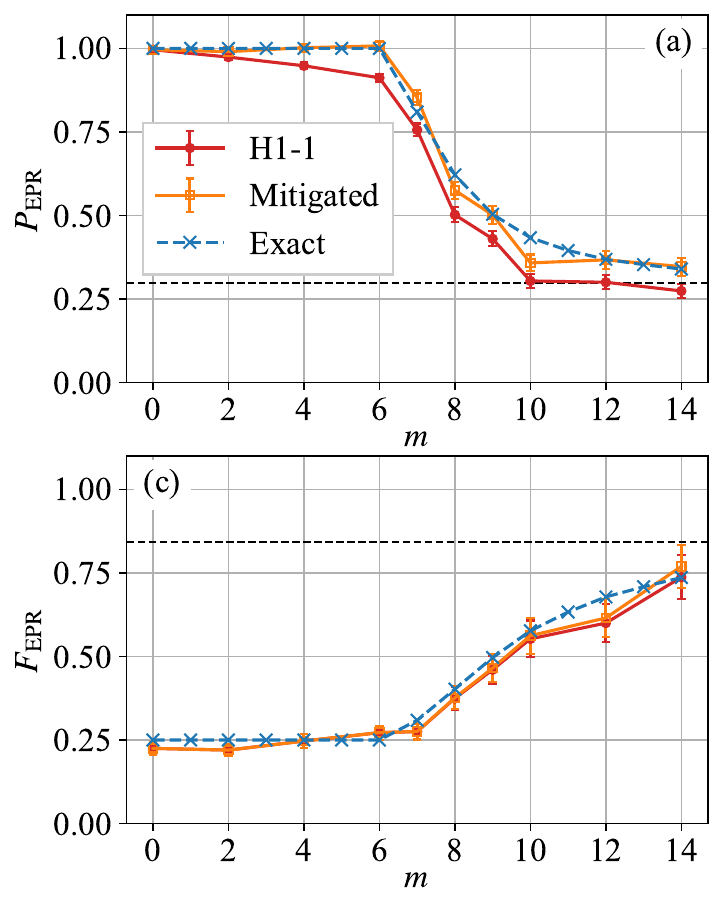}
  \includegraphics[width=.45\textwidth]
  {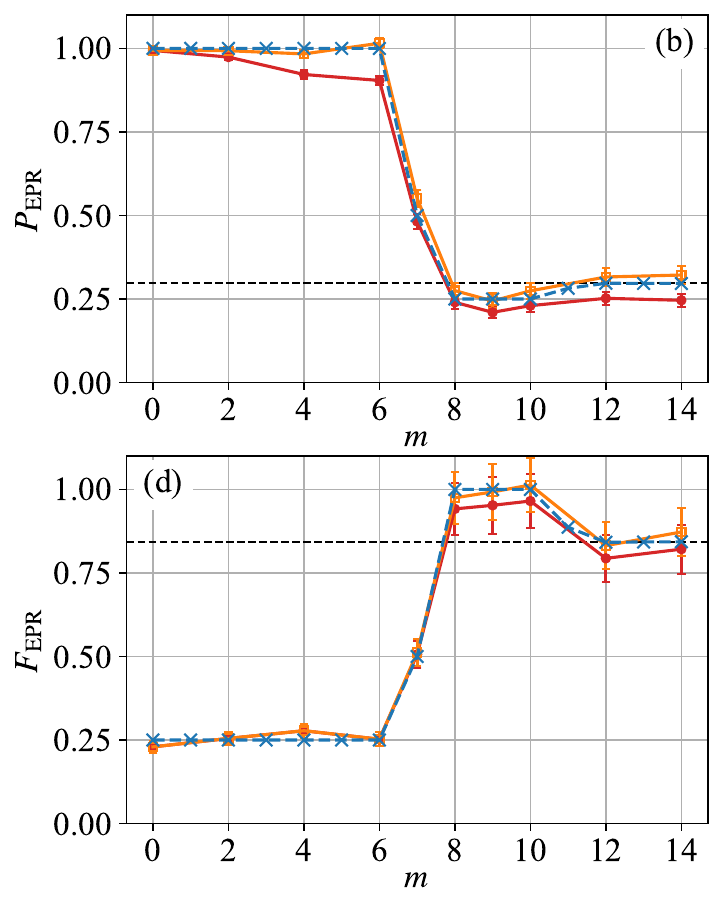}
  \caption{
    \label{fig:HPR_H11}
        (a,b) The post-selection probability $P_{\rm EPR}$ 
        and (c,d) the recovery fidelity $F_{\rm EPR}$ 
        of the HPR protocol for the Floquet circuit of (a,c) $JT=1.3$ 
        and (b,d) $JT=\pi/2$.
        The red circles show the raw data, $P_\mathrm{EPR}^\mathrm{(noisy)}$ and  $F_\mathrm{EPR}^\mathrm{(noisy)}$, 
        of the experiments on the Quantinuum H1-1 system, while 
        the error-mitigated results, 
        $P_\mathrm{EPR}^\mathrm{(mit)}$ and  $F_\mathrm{EPR}^\mathrm{(mit)}$, are shown 
        by the orange squares.  
        For comparison, the exact results obtained from classical 
        calculations and also plotted by the blue crosses.  
        The dashed horizontal lines indicate the values obtained 
        with the Haar random unitary circuit used in place of 
        $(U_\mF)^m$. 
  }
\end{figure*}

Figure~\ref{fig:HPR_H11} shows the experimentally obtained 
post-selection probability $P_{\rm EPR}^{\rm (noisy)}$ and 
recovery fidelity $F_{\rm EPR}^{\rm (noisy)}$ as a function 
of the number of Floquet cycles $m$. 
The experiments are performed by executing the circuit shown in 
the bottom of Fig.~\ref{fig:HPRcirc} 
on the Quantinuum H1-1 system. 
We use two copies of the one-dimensional spin chain composed of 
$N=9$ sites under open boundary conditions, with subsystems of 
$N_A=1$ and $N_D=2$ sites each located on the opposite ends 
of the chain, thus utilizing a total of 
$2(N_R + N)=20$ qubits. Exact results are obtained 
from classical simulations. 
The parameter $T$ is set to two different values, 
$JT=1.3$ and $\pi/2$, while the other 
parameters are fixed at $(B_X/J,B_Z/J)=(1,1.3)$.
Note that the circuit at $JT = 1.3$ is expected to be less scrambling than at the self-dual point $JT = \pi/2$. 
Comparison between the two allows us to study whether and how the error mitigation based on the depolarizing noise model works for scrambling and non-scrambling cases.

First of all, the qualitative agreement between the experimental raw data and the exact results firmly demonstrates the ability of the quantum computer to simulate quantum many-body dynamics involving unitary circuits and measurements.
As shown in Figs.~\ref{fig:HPR_H11}(c) and \ref{fig:HPR_H11}(d), 
the raw data of $F_{\rm EPR}^{\rm (noisy)}$  
agrees with the exact results within the statistical uncertainty, 
confirming the robustness of $F_{\rm EPR}^{\rm (noisy)}$ 
against incoherent noise~\cite{Yoshida2019}. 
The discrepancy between the experimental and exact results 
for $P_{\rm EPR}^{\rm (noisy)}$ in Figs.~\ref{fig:HPR_H11}(a) 
and \ref{fig:HPR_H11}(b)
shall be accounted for by considering the depolarizing noise below.

\begin{figure*}[ht]
  \includegraphics[width=.4\textwidth]{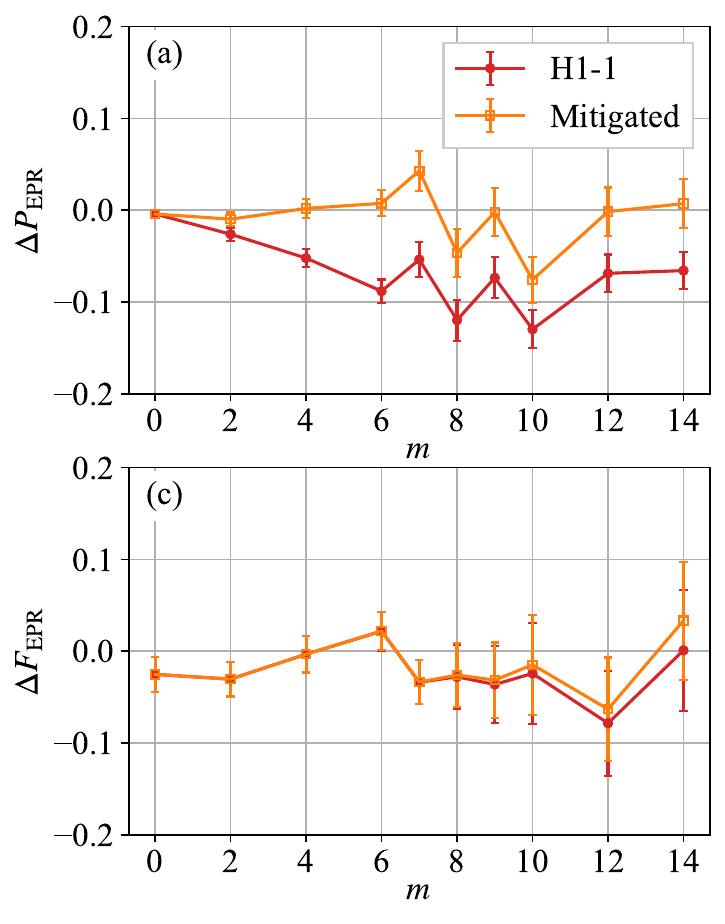}
  \includegraphics[width=.4\textwidth]{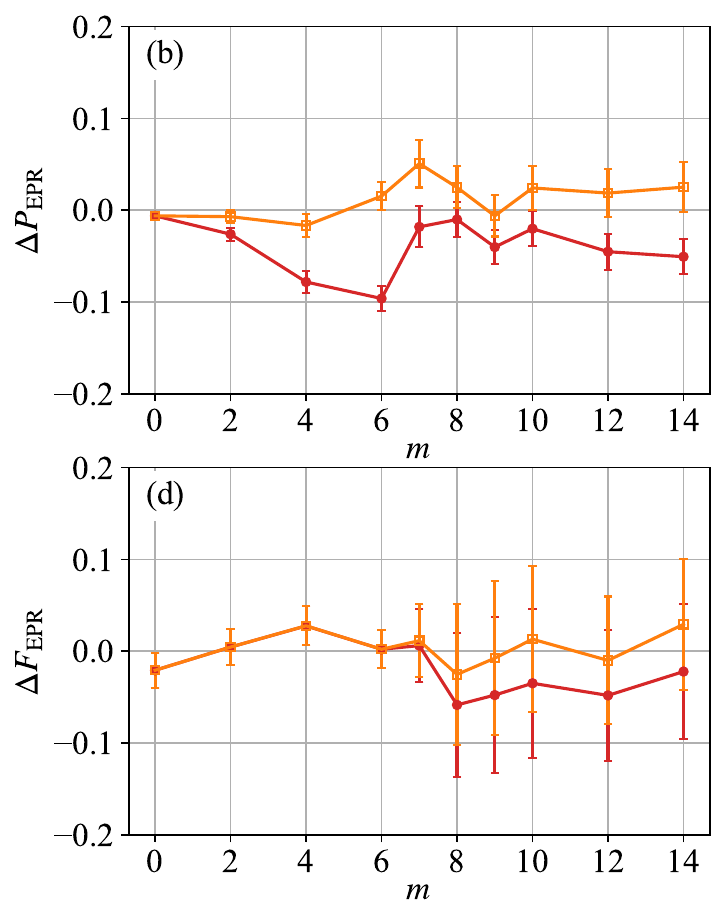}
  \caption{
    (a,b) The post-selection probability subtracted by the exact results, $\Delta P_{\rm EPR}$, and (c,d) the recovery fidelity subtracted by the exact results, $\Delta F_{\rm EPR}$, as a function of the Floquet cycles $m$ 
    for (a,c) $JT=1.3$ and (b,d) $JT=\pi/2$.
    The red circles and orange squares show the raw and the error-mitigated data, respectively, subtracted by the exact results. 
    \label{fig:hp_diff}
  }
\end{figure*}

We perform error mitigation on $P_{\rm EPR}^{\rm (noisy)}$ and $F_{\rm EPR}^{\rm (noisy)}$ assuming the depolarizing noise model of the form in Refs.~\cite{Yoshida2019, Hayata2021}. 
As described in more details in Appendix~\ref{app:mitigateHPR}, 
the error-mitigated post-selection probability 
$P^\mathrm{(mit)}_\mathrm{EPR}$ 
and recovery fidelity $F^\mathrm{(mit)}_\mathrm{EPR}$ are 
respectively given by 
\begin{align}
    P^\mathrm{(mit)}_\mathrm{EPR}
    &=
    \frac{P^\mathrm{(noisy)}_\mathrm{EPR}}{f^2} - \frac{1-f^2}{d_D^2 f^2}
    \label{eq:Pmit}
\end{align}
and
\begin{align}
    F^\mathrm{(mit)}_\mathrm{EPR}
    &=
    \left[\frac{1}{F^\mathrm{(noisy)}_\mathrm{EPR}}\left(1 + \frac{1-f^2}{d_D^2 f^2}\right) - d_A^2 \frac{1-f^2}{d_D^2 f^2}\right]^{-1}.
    \label{eq:Fmit}
\end{align}
Here, $f\in[0,1]$ is a parameter characterizing the depolarizing noise in the application of $U$ or $U^*$ (see Appendix~\ref{app:mitigateHPR} for details). 
A value of $f$ closer to 1 indicates more accurate execution of 
the corresponding circuit. We choose the depolarizing parameter 
as $f = (1-p^\mathrm{(ent)}_\mathrm{2Q}(\theta))^{N_\mathrm{2Q}}$, 
where $p^\mathrm{(ent)}_\mathrm{2Q}(\theta)=\frac{5}{4}p_\mathrm{2Q}(\theta)$ is the entanglement infidelity and $N_\mathrm{2Q}$ is 
the number of the two-qubit gates involved in $(U_\mF)^m$~\cite{Proctor2020, Baldwin2022}.
This choice of $f$ is based on the observation that the butterfly velocity is nearly equal to the light-cone velocity in our Floquet circuits (as can be confirmed from OTOC results in Appendix~\ref{app:numerical_results}), implying that the effective number of the gates affecting the measurements should be nearly equal to the number of the gates within the light cone.
The rotation angle of two-qubit gates in $(U_{\rm F})^m$ is $|\theta|=JT$, and the number of the corresponding two-qubit gates 
in $(U_{\rm F})^m$ is $N_\mathrm{2Q} = \{0, 6, 20, 36, 44, 52, 60, 68, 84, 100\}$ for $m=\{0, 2, 4, 6, 7, 8, 9, 10, 12, 14\}$. 
This also applies to those in $(U_{\rm F}^*)^m$. 
Note that $N_\mathrm{2Q}$ counts the number of two-qubit gates 
causally connected to the measurements and thus it is smaller 
than the total gate count $(N-1)m=8m$ in the entire unitary operator 
for the $N$-qubit one-dimensional system under open boundary 
conditions. 
To highlight the effect of error mitigation, we show $P_{\rm EPR}$ and $F_{\rm EPR}$ subtracted by the exact results, denoted respectively as $\Delta P_{\rm EPR}$ and $\Delta F_{\rm EPR}$, in Fig~\ref{fig:hp_diff}.

In Figs.~\ref{fig:HPR_H11}(b) and \ref{fig:HPR_H11}(d), 
both $P_{\rm EPR}^{\rm (noisy)}$ and $F_{\rm EPR}^{\rm (noisy)}$ saturate to values consistent with those from the Haar random unitary, implying the scrambling at late times for $JT=\pi/2$.
Conversely, the results for $JT=1.3$ in Figs.~\ref{fig:HPR_H11}(a) and \ref{fig:HPR_H11}(c) do not exhibit saturation to the Haar 
values up to $m=14$, indicating slower information scrambling 
in the Floquet circuit with $JT=1.3$ compraed to $JT=\pi/2$. 
This observation is consistent with the fact that the parameters 
at $JT=B_X T=\pi/2$ correspond to a maximally chaotic parametrization~\cite{Bertini2018SFF, Bertini2019, Bertini2019entanglement, Poroli2020dynamics}, which we 
further validate with classical simulations in  Appendix~\ref{app:numerical_results}. 
In the remaining experiments in Sec.~\ref{sec:otoc} and 
Sec.~\ref{sec:exp_thermal}, 
we will focus on this maximally chaotic parametrization.

The improved agreements between the experimental results and the 
exact results upon applying the error mitigation [Fig.~\ref{fig:hp_diff}] imply that the 
noise effects on $P_\mathrm{EPR}$ and $F_\mathrm{EPR}$ are primarily 
captured by the depolarizing noise channel.
The less significant improvement of $F_{\mathrm{EPR}}$ by error mitigation [Fig.~\ref{fig:hp_diff}(c,d)] is again consistent with the robustness of $F_{\mathrm{EPR}}$ against incoherent noise as compared with $P_{\mathrm{EPR}}$.
In Fig.~\ref{fig:HPR_H11}, we observe that the mitigation works better for the data with $JT=\pi/2$ than for the data with $JT=1.3$. 
This can be attributed to the property of the circuit, where the scrambling unitary operation converts local coherent errors into 
global incoherent errors~\cite{Arute2019, Dalzell2021}. Thus, the 
global depolarizing channel would more accurately capture the noise 
effects of such circuits, which aligns with the fact that 
the circuit with $JT=\pi/2$ has the particularly strong scrambling 
property.

\subsection{Out-of-time-ordered correlators}
\label{sec:otoc}

To explore information scrambling in a larger system, we evaluate
the evolution of an OTOC with respect to the input state $\rho=|0^N\rangle\langle0^N|$ and the unitary $U=(U_\mF)^m$, i.e., 
\begin{equation}
\label{eq:experiment_OTOC}
    \langle 0^N| 
    (U_\mF)^{\dag m} X_n (U_\mF)^{m} Z_1 (U_\mF)^{\dag m} X_n (U_\mF)^{m}Z_1
    |0^N\rangle.
\end{equation}
It should be noted that, when we construct the corresponding quantum circuit, the right-most $Z_1$ is omitted based on the fact that $Z_1|0^N\rangle=|0^N\rangle$.
We consider a one-dimensional system consisting of $N=19$ qubits 
under open boundary conditions. Adding a single ancillary qubit 
for the interferometry (see Fig.~\ref{circ:OTOC}), 
we make use of all the 20 qubits of the Quantiuum H1-2 system. 
We run 500 shots of the circuit for each data point with a different 
time step $m$. 
The sizes of measurement and butterfly operators are set to $N_A=1$ and $N_D=1$, respectively. 
The measurement operator $O_A$ is fixed to $Z_1$ and the site $n$ of the butterfly operator $O_D=X_n$ is varied in the range of $1 \le n \le 12$ to explore how the information propagates. 
Although it is desirable to vary $n$ up to $N$ to fully study the information propagation in the entire system,  we restrict the range of $n$ as mentioned above due to the limitations of quantum computing resources available to us.
We set the driving period $JT=\pi/2$ and the Hamiltonian parameters $(B_X/J, B_Z/J)=(1, 1.3)$. 
The circuits with $n=10$ and $11$ contain the maximum number of 
two-qubit gates, 367, inside the causal cone when $m=15$.

\begin{figure*}
  \includegraphics[width=.9\textwidth]
  {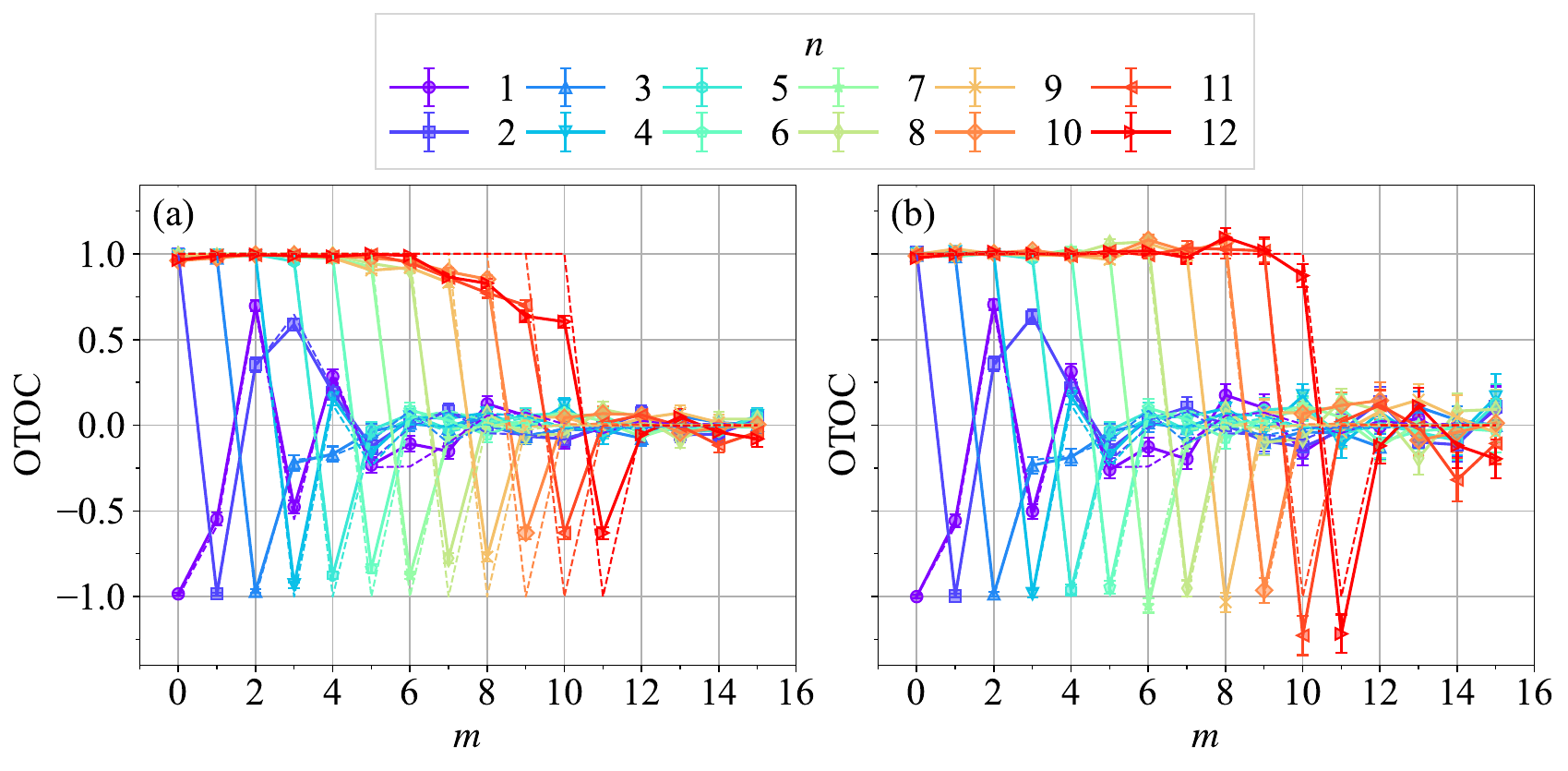}
  \caption{
    \label{fig:OTOC_H11}
        OTOCs defined in Eq.~\eqref{eq:experiment_OTOC} plotted 
        as a function of the number $m$ of Floquet cycles 
        (horizontal axis) 
        for various positions $n$ of the butterfly operator.
        (a) The raw data obtained through experiments performed 
        on the Quantinuum H1-2 system and 
        (b) the error-mitigated results. 
        Dashed lines represent the exact results obtained from classical calculations. 
        The one-dimensional chain consisting of $N=19$ spins 
        under open boundary conditions is considered.
  }
\end{figure*}

In the raw data shown in Fig.~\ref{fig:OTOC_H11}(a), a sharp drop 
of the OTOCs around $m=n$ is observed, followed by remaining around 
zero. This behaviour is consistent with the ballistic operator growth 
in a geometrically local circuit and scrambling at late times.
To mitigate the noise effect, we additionally compute the 
normalization factor by running the circuit where the butterfly 
operator $X_n$ is removed while the rest of the circuit remains 
the same. The normalization factor becomes 1 in the absence of 
noise~\cite{Swingle2018, Mi2021} (also see 
Appendix~\ref{app:mitigateOTOC}).
Figure~\ref{fig:OTOC_H11}~(b) shows the normalized OTOCs, 
the raw data divided by the normalization value.
The agreement with the exact results is improved at early times 
by the mitigation. However, the normalization induces larger 
variances of the mitigated estimators, which particularly amplifies 
the statistical fluctuation at late times.
%

\begin{figure*}[t]
  \includegraphics[height=6cm]{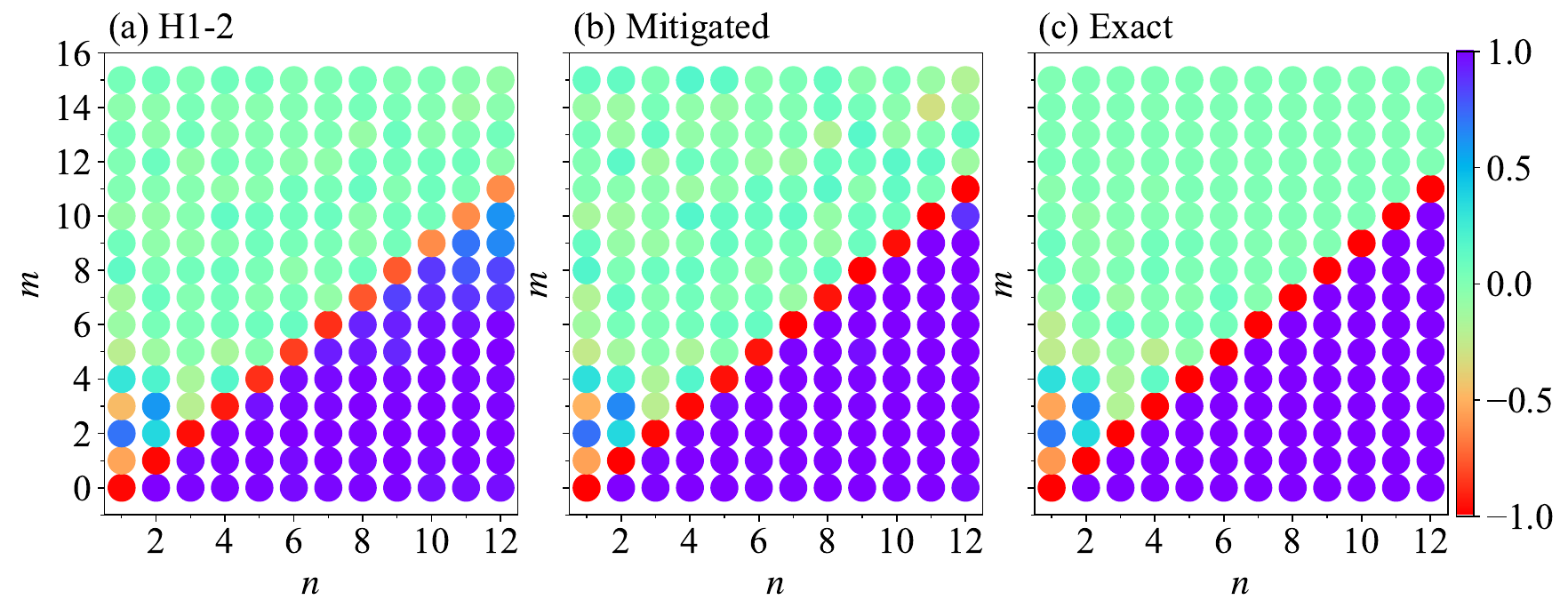}\\
  \includegraphics[height=6cm]{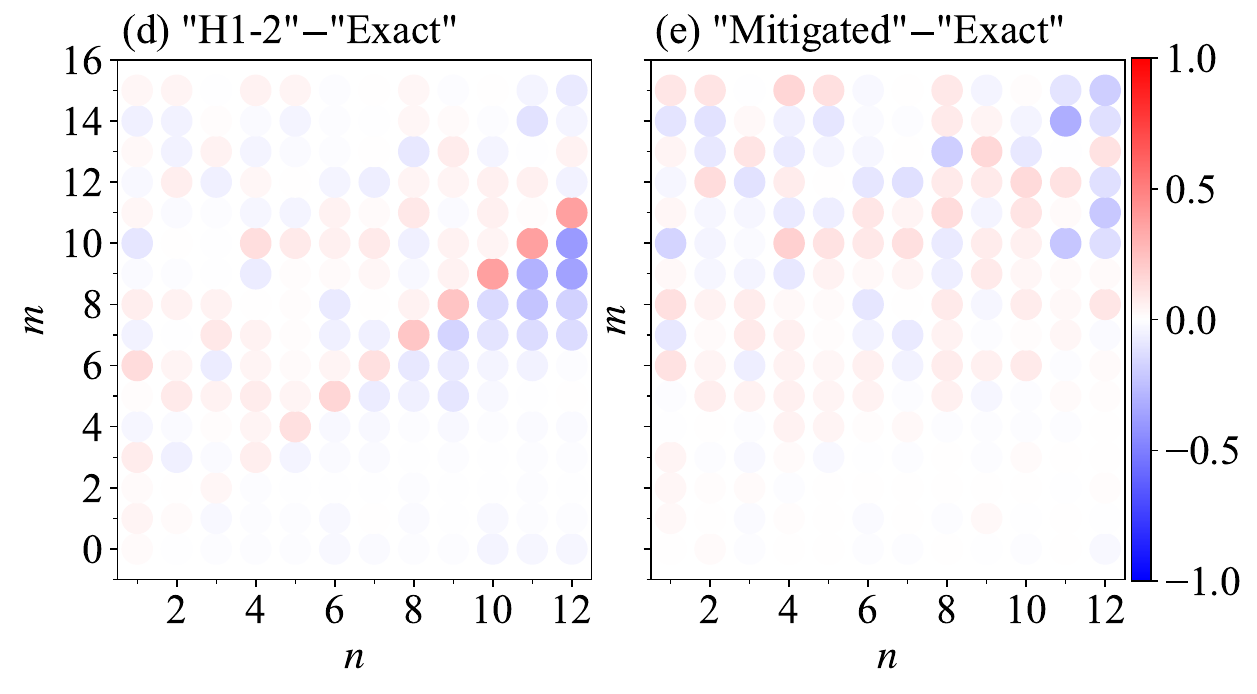}
  \caption{
    \label{fig:OTOC_H1-2_cmap}
    OTOCs defined in Eq.~\eqref{eq:experiment_OTOC} plotted 
    as functions of 
    the position $n$ of the butterfly operator (horizontal axis) 
    and the number $m$ of Floquet cycles (vertical axis).  
    (a) The raw data obtained through experiments performed on 
    the Quantinuum H1-2 system, 
    (b) the error-mitigated data, and
    (c) the exact results obtained from classical calculations. 
    The date presented here are the same as those in Fig.~\ref{fig:OTOC_H11}.
    The raw and the mitigated data subtracted by the exact data are shown in (d) and (e), respectively.
  }
\end{figure*}

To visualize the ballistic growth of the butterfly operator more 
effectively, Fig.~\ref{fig:OTOC_H1-2_cmap} shows the OTOCs as 
functions of the position $n$ of the butterfly operator 
(on the horizontal axis) and the number $m$ of Floquet cycles 
(on the vertical axis). 
The results shown in Fig.~\ref{fig:OTOC_H1-2_cmap} are identical 
to those in Fig.~\ref{fig:OTOC_H11}, but the statistical errors 
are disregarded to enhance clarity. 
The causal-cone structure of the one-dimensional system 
with adjacent interactions, characterized by the abrupt drop of 
the OTOCs at $m=n-1$, is already observed in the raw data 
[see Fig.~\ref{fig:OTOC_H1-2_cmap}(a)]. 
In the normalized OTOCs, the gradual decrease in the absolute value 
of the OTOC observed in the raw data for $n \gtrsim 8$ and 
$m \leq n-1$ is corrected, 
as shown in Fig.~\ref{fig:OTOC_H1-2_cmap}(b), 
resulting in improved agreement with 
the exact results [see Fig.~\ref{fig:OTOC_H1-2_cmap}(c)]. 
The effect of the error mitigation is highlighted in Figs.~\ref{fig:OTOC_H1-2_cmap}(d,e). 
We observe that the mitigation improves the agreement with the exact results around the light front for $n\gtrsim 8$. 
On the other hand, we also observe a slight increase in error away from the light front after applying the mitigation. 
This is attributed to the increased statistical error due to the mitigation, as discussed for the OTOC results at late times shown in Fig.~\ref{fig:OTOC_H11}.
A systematic analysis of OTOCs for other parameters using classical 
calculations is provided in Appendix~\ref{app:numerical_results}.

%

\subsection{Application: thermal expectation value}
\label{sec:exp_thermal}

\begin{figure}
\begin{quantikz}[row sep =0.2cm]
    & \wireoverride{n}
    & \wireoverride{n}
    & \lstick{$\ket{+}$} \wireoverride{n}
    & \ctrl{1}
    & \ctrl{1}
    & \meter{X}
    \\
    \lstick[4]{$\ket{\psi_0}$}
    &
    & \gate[4]{(U_{\mathrm F})^m}
    &
    & \gate[4]{\calU(t)} 
    & \gate[2]{Z_1Z_2}
    &
    \\
    &
    &
    &
    &
    &
    &
    \\
    & \vdots \wireoverride{n}
    & \wireoverride{n}
    & \wireoverride{n}
    & \wireoverride{n}
    & \vdots \wireoverride{n}
    & \wireoverride{n}
    \\
    &
    &
    &
    &
    &
    &
\end{quantikz}
\caption{\label{circ:TPQ}
The quantum circuit used to calculate the real part of 
$\calL_{Z_1Z_2}(s\Delta t)$ in Eq.~\eqref{eq:Loschmidt_ZZ}, 
where $\calU(t):=\e^{-\im \tagH_\mathrm{even}s\Delta t}\e^{-\im \tagH_\mathrm{odd}s\Delta t}$. The top qubit is the single-qubit ancillary register initialized to $\ket{+}$, while the remaining 
qubits constitute the system register with the input state 
$\ket{\psi_0}$. At the end of the circuit, the ancillary qubit is 
measured in the $X$ basis. The imaginary part is calculated 
by employing the same circuit with the measurement basis $X$ 
replaced by the $Y$ basis. 
}
\end{figure}

As an application of the Floquet scrambling circuit, we evaluate 
the expectation value of $O=Z_{1}Z_{2}$ within the microcanonical 
ensemble of the one-dimensional spin-1/2 isotropic Heisenberg model, 
employing the protocol described in Sec.~\ref{sec:TPQ}. 
The Hamiltonian is given by 
\begin{equation}\label{eq:H_xxx}
\tagH_{XXX} = \frac{\tagJ}{2}\sum_{i=1}^{N}(X_{i}X_{i+1}+Y_{i}Y_{i+1}+Z_{i}Z_{i+1}+I_{i}I_{i+1})
\end{equation}
under periodic boundary conditions, where the $N+1$th and 
1st sites are identified. 
We evaluate the Loschmidt amplitude  
\begin{equation}
\label{eq:Loschmidt}
    {\cal L}(s\Delta t) 
    = 
    \bra{\psi(m)} \e^{-\im \tagH_\mathrm{even}s\Delta t}\e^{-\im \tagH_\mathrm{odd}s\Delta t} \ket{\psi(m)}
\end{equation}
and the corresponding amplitude
\begin{equation}
\label{eq:Loschmidt_ZZ}
    {\cal L}_{Z_1 Z_2}(s\Delta t) 
    = 
    \bra{\psi(m)} Z_1 Z_2 \e^{-\im \tagH_\mathrm{even}s\Delta t}\e^{-\im \tagH_\mathrm{odd}s\Delta t} \ket{\psi(m)}
\end{equation}
using the state evolved with the Floquet circuit,  
where $\ket{\psi(m)} = (U_\mF)^m\ket{\psi_0}$ and $\tagH_\mathrm{even/odd}=(\tagJ/2)\sum_{i:\mathrm{even/odd}}(X_{i}X_{i+1}+Y_{i}Y_{i+1}+Z_{i}Z_{i+1}+I_{i}I_{i+1})$. The initial product state $\ket{\psi_0}$ is prepared as $\ket{\psi_0}=\prod_{i=1}^{N}{\rm e}^{-\imag\phi_i Y_i/2}|0^N\rangle$, where $\{\phi_i\}_{i=1}^{N}$ is a set of independent random angles, each of which is drawn uniformly from $[0,4\pi)$. 
We consider a system of $N=16$ qubits and set the number of Floquet cycles to $m=8$ and $m=16$ so that $|\psi(m)\rangle$ approximately forms a state 2-design under periodic boundary conditions (see Appendix~\ref{app:Loschmidt}).

As already indicated in Eqs.~(\ref{eq:Loschmidt}) and 
(\ref{eq:Loschmidt_ZZ}), to evaluate the two amplitudes 
${\cal L}(t)$ and ${\cal L}_{Z_1 Z_2}(t)$ with $t=s\Delta t$, 
we approximate the time-evolution operator 
$\e^{-\im\tagH_{XXX} s\Delta t}$ by a single-step first-order 
product (Trotter) formula, $\e^{-\im \tagH_{XXX} s\Delta t}\simeq\e^{-\im \tagH_\mathrm{even}s\Delta t}\e^{-\im \tagH_\mathrm{odd}s\Delta t}$. 
The computed amplitudes are then classically post-processed as in  Eqs.~\eqref{eq:D_O(E)} and \eqref{eq:D(E)} with $S=40$ and $\Delta t=0.05/\tagJ$. The real and imaginary parts of ${\cal L}(t)$ 
and ${\cal L}_{Z_1 Z_2}(t)$ are estimated separately with 
the Hadamard-test circuits (see Fig.~\ref{circ:TPQ}) on 
the Quantinuum H1-2 system. 
For the trace estimation, we generate eight instances of the Floquet 
scrambled states $|\psi(m)\rangle$, consisting of four independent 
realizations of $|\psi_0\rangle$ and two Floquet cycles $m=8$ and $m=16$.
We take 100 shots to evaluate $\calL(t)$ and $\calL_{Z_1 Z_2}(t)$ 
for each instance of $|\psi(m)\rangle$, thus taking 800 shots
in total at each data point.
Furthermore, we symmetrize (antisymmetrize) the real (imaginary) 
parts of $\calL(t)$ and $\calL_{Z_1 Z_2}(t)$ with respect to $t$ 
by using the data obtained for $t > 0$ and $t<0$.
The symmetrizations (antisymmetrizations) are made so that $\calL(t)$ and $\calL_{Z_1 Z_2}(t)$ share with $\Tr[\e^{-\imag \tagH_{XXX}t}]/d$ and $\Tr[Z_1 Z_2 \e^{-\imag \tagH_{XXX}t}]/d$ the common property that their real (imaginary) parts are even (odd) functions of $t$. Here, $d=2^N$ is the dimension of the Hilbert space for the system $\tagH_{XXX}$. 
For this reason, we will only show the results for $t\geq0$.

We note a previous study~\cite{Summer2023} using a similar 
Hadamard-test circuit on entangled random states to assess 
the density of states in spin chains comprising 12 and 18 spins, 
performed using the Quantinuum H1-1 system. 
Additionally, a more recent study evaluated the Loschmidt 
amplitude of the Fermi-Hubbard model with 32 orbitals in reference 
to product states using 
the Quantinuum H2-1 system, as reported in Ref.~\cite{Hemery2023}. 
They estimated the microcanonical expectation value of the double 
occupancy using the importance sampling techniques on initial product 
states~\cite{Lu2021,Yang2022,Schuckert2023,Ghanem2023}. 
Here, we further proceed in this direction by evaluating 
the microcanonical expectation value of the spin correlation function 
$\langle Z_1 Z_2\rangle^{(\rm est)}$ for the Heisenberg spin chain.

\begin{figure*}
  \includegraphics[width=1.0\textwidth]
  {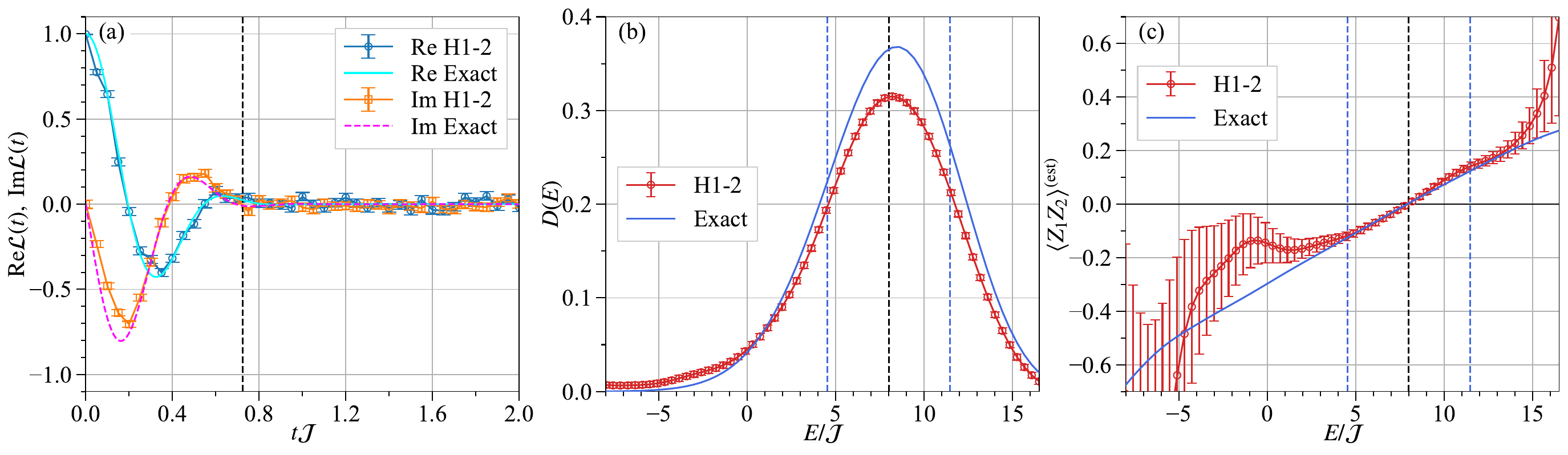}
  \caption{
    \label{fig:H12_Z1Z2}
    (a) Real and imaginary parts of the Loschmidt amplitude ${\cal L}(t)$ for the one-dimensional Heisenberg model $\tagH_{XXX}$ with $N=16$ sites under periodic boundary conditions. These 
    experimental results are obtained for the Floquet scrambled 
    states $|\psi(m)\rangle$ with $m=8$ and $16$
    using the Quantinuum H1-2 system. 
    The vertical line indicates $t=\sigma^{-1}$, where $\sigma$ is 
    defined in Eq.~(\ref{eq:sigma}). For comparison, the ideal 
    values obtained from classical calculations are also shown 
    by solid and dashed curves for the real and imaginary parts, 
    respectively. 
    (b) Density of states $D(E)$ obtained via Eq.~(\ref{eq:D(E)})
    as a Fourier transform of the Gaussian-weighted Loschmidt 
    amplitude ${\cal L}(t)$ shown in (a). 
    (c) Microcanonical expectation value of the nearest-neighbor 
    spin correlation function $\langle Z_1 Z_2 \rangle^{(\rm est)}$ 
    for the one-dimensional Heisenberg model $\tagH_{XXX}$ with 
    $N=16$ sites under periodic boundary conditions 
    (also see Appendix~\ref{app:LZZ}).
    In (b) and (c), the black dashed vertical line indicates 
    $E=E_{\infty}=8\tagJ$, and the blue dashed vertical lines 
    indicate $E=E_{\infty}\pm \sigma_\tagH = 8\tagJ \pm 2\sqrt{3}\tagJ$, 
    where $\sigma_\tagH$ is defined in Eq.~(\ref{eq:sigma_H}). 
    In (b) and (c), the ideal values obtained from classical 
    calculations are also shown by blue solid curves.
  }
\end{figure*}

Figures~\ref{fig:H12_Z1Z2}(a) and \ref{fig:H12_Z1Z2}(b) show 
the Loschmidt amplitude $\calL(t)$ obtained from the experiments 
using the Quantinuum H1-2 system 
and the associated density of states $D(E)$ evaluated via 
Eq.~(\ref{eq:D(E)}), respectively. 
For comparison, we carry out classical calculations of 
the Loschmidt amplitude, labeled as "Exact" in the figures, 
where $\ket{\psi(m)}$ is substituted by a Haar-random state 
and the results are averaged over eight instances of 
Haar-random states. 
Qualitative agreement with the classical results is observed 
without error mitigation. 
%
The parameter $\sigma$ in Eq.~(\ref{eq:D(E)}) 
[also see Eq.~(\ref{eq:micro_ave})] is selected as 
\begin{equation}\label{eq:sigma}
\sigma = \frac{\sigma_\tagH}{\sqrt{2\pi}}, 
\end{equation}
where 
\begin{equation}\label{eq:sigma_H}
\sigma_\tagH^2 :=\frac{1}{d}\Tr[\tagH_{XXX}^2]-\left(\frac{1}{d}\Tr[\tagH_{XXX}]\right)^2 
\end{equation}
is the energy variance with respect to the maximally mixed state, 
and the factor $1/\sqrt{2\pi}$ is introduced accordingly to the 
definition of the width of the energy window described 
in Ref.~\cite{Seki2022}. 
The selection of $\sigma$ is motivated by the notion that 
$\sigma_\tagH$ roughly signifies the breadth of the density of states 
in energy, suggesting that the Loschmidt amplitude decays 
around the time $t \sim \sigma_\tagH^{-1}$ 
(for details, see Appendix~\ref{app:Loschmidt}). 
Moreover, the variance $\sigma_\tagH^2$ scales linearly with the 
system size $N$, ensuring the convergence of $\langle O\rangle$ 
in Eq.~(\ref{eq:micro_ave}) towards the correct value in the 
thermodynamic limit.
In the present case with $N=16$, the energy variance is determined as 
$\sigma_\tagH^2=12\tagJ^2$, resulting in 
$\sigma = \sqrt{6/\pi}\tagJ\approx 1.38\tagJ$.

Figure~\ref{fig:H12_Z1Z2}(c) shows the microcanonical expectation 
value of the spin correlation function 
$\langle Z_1 Z_2 \rangle^{(\rm est)}$ 
[also see Eqs.~(\ref{eq:estimator}) and (\ref{eq:estimator_circuit})]. 
Here, we have evaluated ${\cal L}_{Z_1 Z_2}(t)$ and $D_{Z_1 Z_2}(E)$ 
in a manner similar to ${\cal L}(t)$ and $D(E)$. 
The experimental results of ${\cal L}_{Z_1 Z_2}(t)$ 
using the Quantinuum H1-2 system and the associated 
$D_{Z_1 Z_2}(E)$ are provided in Appendix~\ref{app:LZZ}. 
We observe a good agreement with the exact results within 
the statistical uncertainty for the energy range $E_{\infty} - \sigma_\tagH \lesssim E \lesssim E_{\infty} + \sigma_\tagH$, 
which lies between the two vertical dashed lines in 
Figs.~\ref{fig:H12_Z1Z2}(b) and \ref{fig:H12_Z1Z2}(c). 
Here, $E_{\infty} :=\Tr[\tagH_{XXX}]/d = N\tagJ/2$ is the energy 
corresponding to the maximally mixed state, or equivalently the 
energy at infinite temperature, at which 
$\langle {Z_1 Z_2} \rangle^{(\rm est)}$ changes its sign.
This sign change can be understood by noting that the correlation 
vanishes at infinite temperature. Furthermore, 
the nearest-neighbor spin correlation is antiferromagnetic at 
positive temperatures, resulting in 
$\langle {Z_1 Z_2} \rangle^{(\rm est)} < 0$ for 
$E\lesssim E_{\infty}$, and ferromagnetic at negative temperatures, 
resulting in $\langle {Z_1 Z_2} \rangle^{(\rm est)} > 0$ 
for $E\gtrsim E_{\infty}$.\footnote{
    In general, negative-temperature states, corresponding to $E\ge E_\infty$, are not realized in thermal equilibrium but can be realized as nonequilibrium states. 
}
The sign structure of $\langle {Z_1 Z_2} \rangle^{(\rm est)}$ around 
$E_{\infty}$ is thus correctly captured by the experimental results.

The agreement for $E_{\infty} - \sigma_\tagH \lesssim E \lesssim E_{\infty} + \sigma_\tagH$ is attributed to the concentration of
the density of states in this energy span, 
as shown in  Fig.~\ref{fig:H12_Z1Z2}(b), where $D(E)$ peaks 
around $E_{\infty}$ with a width $\sigma_\tagH$. 
Indeed, the density of states $D(E)$ decays rapidly beyond 
this energy range. 
The decay of $D(E)$ implies  an increase in the purity 
$\Tr[\rho_\sigma(E)^2]$ of the associated microcanonical density 
matrix. This, in turn, leads to an increase in the failure 
probability in Eq.~\eqref{eq:success_prob} 
(also see Appendix~\ref{app:TPQ}) as well as the shot noise 
in $\langle Z_1 Z_2 \rangle^{(\rm est)}$ 
(for details, see Appendix~\ref{app:imaginary}). This explains 
the deviation from the exact results with larger statistical 
errors beyond this energy range. 
Finally, we observe that the density of states shown in Fig.~\ref{fig:H12_Z1Z2}(b) are less accurate than the thermal expectation values in Fig.~\ref{fig:H12_Z1Z2}(c) within the range $E_{\infty} - \sigma_\tagH \lesssim E \lesssim E_{\infty} + \sigma_\tagH$. The better results in Fig.~\ref{fig:H12_Z1Z2}(c) are attributed to the cancellation of depolarizing noise effects between the numerator $D_{Z_1 Z_2}(E)$ and denominator $D(E)$ in the thermal expectation value $\langle Z_1 Z_2 \rangle^{(\rm est)}=D_{Z_1 Z_2}(E)/D(E)$.

\section{Discussion}
\label{sec:conclusion}

We performed three experiments on trapped-ion quantum computers: the 
HPR protocol, the interferometric protocol, and the calculation of 
the thermal expectation value of a local operator.
In the first two experiments, we characterized the information 
scrambling of the kicked-Ising model at and around 
the self-dual point.
The HPR protocol involved manipulating two copies of the 9-qubit 
spin chains augmented by a two-qubit ancillary register, 
resulting in a total of 20 qubits. We verified the high-fidelity 
teleportation of quantum information, as expected from the chaotic 
unitary dynamics. Additionally, We tested the incoherent error 
mitigation assuming the depolarizing noise model. The improved 
performance with the mitigation suggests that the depolarizing noise 
model effectively captures the influence of the noise on the HPR 
protocol.

In the second experiment, we scaled up the system size to 19 qubits 
and calculated the OTOCs using the interferometric protocol. 
Successful execution of the circuits confirmed the sharp decay of 
OTOCs at late times, consistent with theoretical predictions. 
Despite the maximum two-qubit gate count being as large as 367 
inside the causal cone of the measured qubits, we observed 
agreements between the experimental data and the exact numerical 
values. 
Lastly, having confirmed the scrambling property of the unitary 
circuits, we employed them to calculate the thermal expectation 
values of local operators, borrowing the idea of microcanonical 
TPQ states.

While the present studies nicely demonstrate the capability of 
currently available quantum devices and the proposed quantum 
algorithms, there is still the question of how far these protocols 
can be scaled up to potentially reach the realm of classically 
intractable system sizes and evolution times. 
Foreseeing the interferometric experiments on larger systems, 
let us make a rough estimate of the required resources. 
Recall that our 20-qubit interferometric circuit for the kicked Ising 
model under open boundary conditions contains at most 367 two-qubit 
gates inside the causal cone.
Assuming the dominant source of error is two-qubit gate operations,  
with their average gate infidelity of $0.1\%$, we now discuss the 
fidelity of a 60-qubit interferometric circuit. 
For the same one-dimensional geometrically local model under 
periodic boundary conditions to exhibit scrambling, 
the Floquet system 
needs to evolve for $3/2$ times more cycles than the 20-qubit 
system with open boundary conditions. This leads to a two-qubit 
gate counts of $367 \times 3 \times 3/2 \approx1652$. 
Consequently, the circuit fidelity is roughly estimated as 
$f\sim 0.999^{1652}\approx 0.2$. Evaluating the normalized OTOC 
with a statistical uncertainty $\epsilon$ requires a shot overhead of 
$\calO((f \epsilon)^{-2})$. 
Alternatively, one can introduce long-range interactions to reduce 
the scrambling time, which is suitable for the hardware equipped 
with long-range connectivity. For instance, fast scrambling systems 
like the Sachdev–Ye–Kitaev 
model~\cite{Sachdev1993,Kitaev2015,Maldacena2016} have a scrambling 
time $~\calO(\log N)$, contrasting with $N$ in geometrically local 
systems.
We leave more comprehensive studies for larger-scale experiments 
in the future. 

\section*{Data Availability}
The data that support the findings of this study are available at Zenodo~\cite{Zenodo}.

\section*{Code Availability}
The code used to create the figures in this paper is available from the authors upon reasonable request.

\begin{acknowledgments}
Y.K. is grateful to Luuk Coopmans, Matthew DeCross, Henrik Dreyer, Michael Foss-Feig, Etienne Granet, Enrico Rinaldi, Matthias Rosenkranz, Kentaro Yamamoto, and Yasuyoshi Yonezawa for fruitful discussions.
A portion of this work is based on results obtained from project JPNP20017, subsidized by the New Energy and Industrial Technology Development Organization (NEDO). 
This study is also supported by JSPS KAKENHI 
Grants No. JP19K23433,
No. JP21H01007, 
No. JP21H01084,
No. JP21H04446, 
No. JP22K03520, and
No. JP24K00630.
We are also grateful for the funding received from JST COI-NEXT (Grant No. JPMJPF2221) and the Program for Promoting Research of the Supercomputer Fugaku (Grant No. MXP1020230411) from MEXT, Japan.  
Additionally, we acknowledge the support from the UTokyo Quantum Initiative, 
the RIKEN TRIP project, and
the COE research grant in computational science from Hyogo Prefecture and Kobe City through the Foundation for Computational Science.
A part of the numerical simulations has been performed using the HOKUSAI supercomputer at RIKEN, cluster computers at iTHEMS in RIKEN, and Yukawa-21 at YITP in Kyoto University.
\end{acknowledgments} 

\section*{Author contributions}
K.S., Y.K., and T.H. conceived and designed the study. K.S., Y.K., and T.H. performed analytic calculations, and K.S. carried out numerical studies and hardware experiments. All authors analysed the data, created the figures, interpreted the results and wrote the manuscript.

\section*{Competing interests}
The authors declare no competing financial or non-financial interests.

\bibliography{scrambling}
\newpage
\clearpage
\onecolumngrid
\appendix

\section{Error mitigation}\label{app:mitigate}

\subsection{Hayden-Preskill recovery protocol}
\label{app:mitigateHPR}

To model the noise effect in the HPR protocol, we consider 
a depolarizing channel $\calD_U$ acting on the unitary operator $U$, 
which acts on an $N$-qubit input state $\rho$ as follows: 
\begin{align}
\label{eq:depolarizing}
    \calD_U[\rho]
    =
    f U\rho U^\dag + (1-f) \frac{I^{\otimes N}}{d},
\end{align}
where $d=2^N$ and $1-f$ represents the depolarizing error rate 
associated with the noisy application of $U$.
By substituting the unitary $U$ with the corresponding channel 
$\calD_U$, we can express the post-selection probability 
$P^\calD_\mathrm{EPR}$ and the recovery fidelity 
$F^\calD_\mathrm{EPR}$ of $\calD_U$ 
using the ideal values $P_\mathrm{EPR}$ and $F_\mathrm{EPR}$ 
as follows: 
\begin{align}
    P^\calD_\mathrm{EPR}
    &=
    f^2 P_\mathrm{EPR} + \frac{1-f^2}{d_D^2}
\end{align}
and
\begin{align}
    F^\calD_\mathrm{EPR}
    &=
    \frac{1}{d_A^2 P^\calD_\mathrm{EPR}}\left[f^2 + \frac{1-f^2}{d_D^{2}}\right]
    =
    \frac{f^2 + d_D^{-2}(1-f^2)}{f^2 /F_\mathrm{EPR} + d_A^2 d_D^{-2}(1-f^2)}, 
\end{align}
respectively~\cite{Yoshida2019}. 
Assuming that the results $P^\mathrm{(noisy)}_\mathrm{EPR}$ and 
$F^\mathrm{(noisy)}_\mathrm{EPR}$ obtained from noisy experiments 
are approximated by $P^\calD_\mathrm{EPR}$ and $F^\calD_\mathrm{EPR}$, 
we calculate the mitigated quantities $P^\mathrm{(mit)}_\mathrm{EPR}$ 
and $F^\mathrm{(mit)}_\mathrm{EPR}$ as given in Eqs.~(\ref{eq:Pmit}) 
and (\ref{eq:Fmit}), respectively. 
Those mitigated quantities are shown along with the raw data in Fig.~\ref{fig:HPR_H11}.

\begin{figure*}
\includegraphics[width=0.45\textwidth]{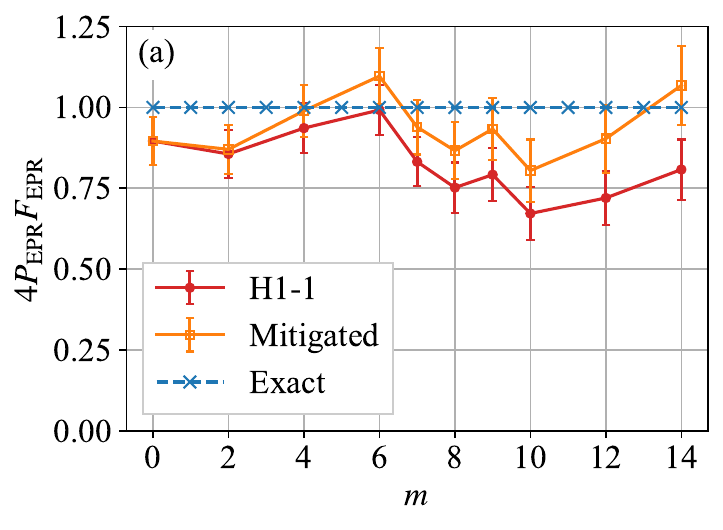}
\includegraphics[width=0.45\textwidth]{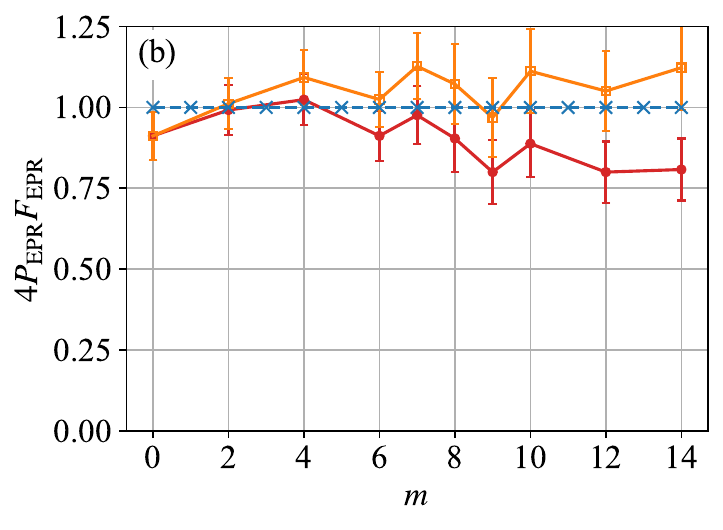}
\caption{
    \label{fig:4PF}
    The product of the post-selection probability $P_{\mathrm{EPR}}$ and the recovery fidelity $F_{\mathrm{EPR}}$, multiplied by 
    $d_A^2=4$, for (a) $JT=1.3$ and (b) $JT=\pi/2$. These results 
    correspond to those shown in Fig.~\ref{fig:HPR_H11}. 
}
\end{figure*}

Noting that
\begin{align}
    d_A^2 P^\calD_\mathrm{EPR}F^\calD_\mathrm{EPR}
    =
    f^2 + \frac{1-f^2}{d_D^2},
\end{align}
which is 1 in the absence of depolarizing noise as in 
Eq.~(\ref{eq:relation}), we can extract the strength of the noise by
\begin{align}
    d_A^2 P^\mathrm{(noisy)}_\mathrm{EPR}F^\mathrm{(noisy)}_\mathrm{EPR}.
\end{align}
Figure~\ref{fig:4PF} shows $4 P_{\mathrm{EPR}} F_{\mathrm{EPR}}$ 
obtained from the HPR protocol experiment with $d_A=2$ using the 
Quantinuum H1-1 system, corresponding to the results shown in 
Fig.~\ref{fig:HPR_H11}. 
Despite the presence of large statistical errors, we can find 
that $4 P^\mathrm{(noisy)}_{\mathrm{EPR}} F^\mathrm{(noisy)}_{\mathrm{EPR}}$ tends to deviate from 1 as 
$m$ increases.

\subsection{Interferometric protocol for OTOCs}
\label{app:mitigateOTOC}

To mitigate the noise effect, following 
Refs.~\cite{Swingle2018,Mi2021}, we calculate the normalized OTOC 
as 
\begin{equation}
\label{eq:normalized_OTOC}
    \frac{
        \langle 0^N| U X_n U^\dag Z_1 U X_n U^\dag |0^N\rangle
    }{
        \langle 0^N| U I_n U^\dag Z_1 U I_n U^\dag |0^N\rangle
    }.
\end{equation}
where $U=(U_\mF)^m$ describes the unitary dynamics of the system 
[see Eq.~(\ref{eq:experiment_OTOC})], and 
the denominator takes the value 1 in the absence of errors.

To analyze the behavior of the normalized OTOC in the presence of 
depolarizing noise, modeled by Eq.~(\ref{eq:depolarizing}), 
we compute the numerator and denominator with $U$ replaced by 
$\calD_U$ in the circuit shown in Fig.~\ref{circ:OTOC}, 
yielding respectively 
\begin{align}
\begin{split}
    \langle 0^N| U X_n U^\dag Z_1 U X_n U^\dag |0^N\rangle^\calD
    &=
    f^2\langle 0^N| U X_n U^\dag Z_1 U X_n U^\dag |0^N\rangle,
    \\
    \langle 0^N| U I_n U^\dag Z_1 U I_n U^\dag |0^N\rangle^\calD
    &=
    f^2\langle 0^N| U I_n U^\dag Z_1 U I_n U^\dag |0^N\rangle
    = 
    f^2.
\end{split}
\end{align}
Thus, their ratio provides the OTOC without the depolarizing 
noise, $\langle 0^N| U X_n U^\dag Z_1 U X_n U^\dag |0^N\rangle$.

\section{Classical numerical results of OTOCs}
\label{app:numerical_results}

\begin{figure*}
  \includegraphics[width=1\columnwidth]{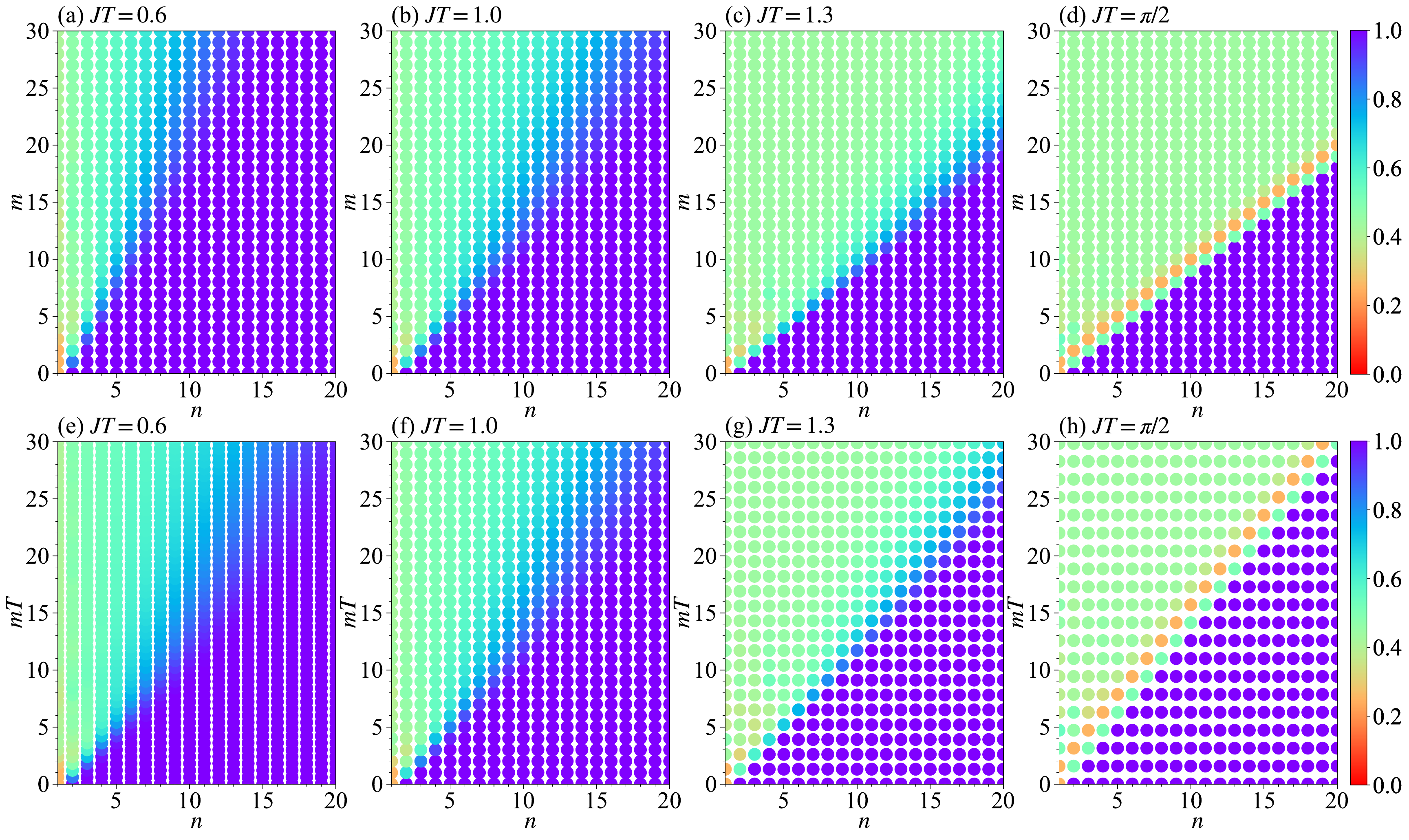}
  \caption{
    \label{fig:OTOC_butterfly_obc}
        (a-d) Operator-averaged OTOCs, defined in 
        Eq.~(\ref{eq:oa_OTOC}), 
        as a function of the position $n$ of the butterfly operator 
        and the number $m$ of Floquet cycles, calculated classically 
        for different $JT$ values. 
        The system consists of 20 spins under open boundary 
        conditions.
        (e-h) Same as (a-d) but as a function of the evolution 
        time $mT$, instead of the number of cycles $m$, in the 
        vertical axis. 
  }
\end{figure*}

Figure~\ref{fig:OTOC_butterfly_obc} shows the results of the 
operator-averaged OTOCs given by 
\begin{equation}
\label{eq:oa_OTOC}
    \frac{1}{4^{N_A + N_D}}\sum_{O=\{I,X,Y,Z\}}\sum_{\bar O=\{I,X,Y,Z\}}\langle 0^N| 
    (U_\mF)^{\dag m} O_n (U_\mF)^{m} {\bar O}_1 (U_\mF)^{\dag m} O_n (U_\mF)^{m} {\bar O}_1
    |0^N\rangle, 
\end{equation}
for the kicked Ising model with $N=20$ spins under open 
boundary conditions, calculated classically for various $JT$ 
values. 
Similar to Figs.~\ref{fig:OTOC_H11} and \ref{fig:OTOC_H1-2_cmap}, 
the other parameters are fixed to $B_X/J=\pi/2$ and $B_Z/J=1.3$, 
with the subsystem sizes set as $N_A=1$ and $N_D=1$. Here, 
subsystem $A$ on which operator $\bar O$ acts remains fixed at one 
of the edges of the one-dimensional system, while  
the location $n$ of subsystem $D$ on which operator $O$ acts 
is varied. 
We confirm that at $JT=\pi/2$, corresponding to the dual unitary 
point, the Floquet evolution induces the most rapid and pronounced 
decay of OTOCs among all cases shown in 
Fig.~\ref{fig:OTOC_butterfly_obc}. 

\begin{figure*}
  \includegraphics[width=1\columnwidth]{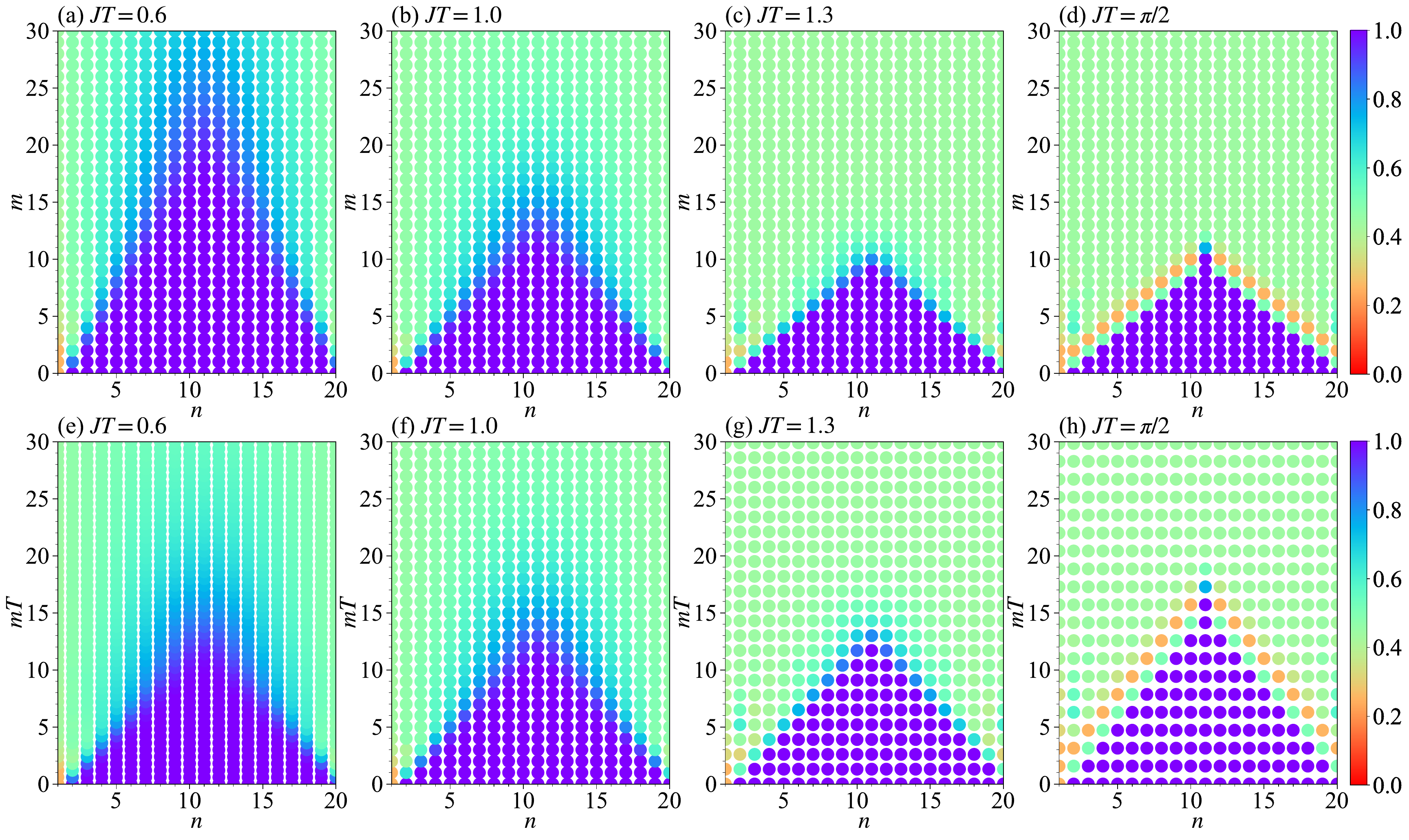}
  \caption{
    \label{fig:OTOC_butterfly_pbc}
    Same as Fig.~\ref{fig:OTOC_butterfly_obc} but for periodic boundary conditions.
  }
\end{figure*}

Figure~\ref{fig:OTOC_butterfly_pbc} shows the same results 
but under periodic boundary conditions. The boundary effect 
approximately reduces the necessary number of Floquet cycles for 
complete system scrambling by half 
compared to the cases of open boundary conditions. 
Other than this point, we observe essentially the same results.

\section{Statistical errors of normalized trace estimation}
\label{app:Loschmidt}

In this Appendix, we analyze how well the scrambled state 
approximates a state 2-design.
To this end, we first show that the ensemble
\begin{align}
\label{eq:floquet_ensemble}
    \Psi = \{(U_\mF)^m\ket{\psi_0}\mid m\in\mathbb{N}, m_\mathrm{min}\le m< m_\mathrm{max}\}
\end{align}
behaves like a state 2-design by numerically calculating 
the variance of Loschmidt amplitude $\calL(t)$ in Eq.~\eqref{eq:Loschmidt_method}.
We then provide justification for employing a single state sampled 
from the ensemble in Eq.~\eqref{eq:floquet_ensemble} to accurately 
estimate the normalized trace $\Tr[\calU(t)]/d$, where $\calU(t)$ is 
a time-evolution operator of the system and $d$ is the dimension of 
the corresponding Hilbert space.

Throughout this appendix, we denote the expectation value and sample 
mean by $\bE[\cdot]$ and $\Esample[\cdot]$, respectively, while 
$\Var[\cdot]$ and $\Vsample[\cdot]$ represent the corresponding 
variances. 
To avoid shot noise, we employ classical simulations in this 
appendix and concentrate on the statistical properties of Floquet 
scrambled states. 
The numerical results presented here are obtained 
using the Floquet circuit, with a single Floquet cycle described 
by $U_\mF$ as in Eq.~(\ref{eq:U_F}), where the parameters are 
set to $B_z/J=1$, $B_z/J=1.3$, and $JT=\pi/2$. 
The time-evolution operator $\calU(t)$ in the Loschmidt amplitude 
$\calL(t)$ corresponds to the one-dimensional Heisenberg model 
$\tagH_{XXX}$ under periodic boundary conditions, as defined in 
Eq.~(\ref{eq:H_xxx}). 
In this section, we treat the time-evolution operator 
$\calU(t)$ numerically exactly, instead of  
employing the single Trotter step approximation as described in 
Eq.~(\ref{eq:Loschmidt})

\subsection{Scrambling circuits form an approximate state 2-design}

\begin{figure}
  \includegraphics[width=.48\textwidth]{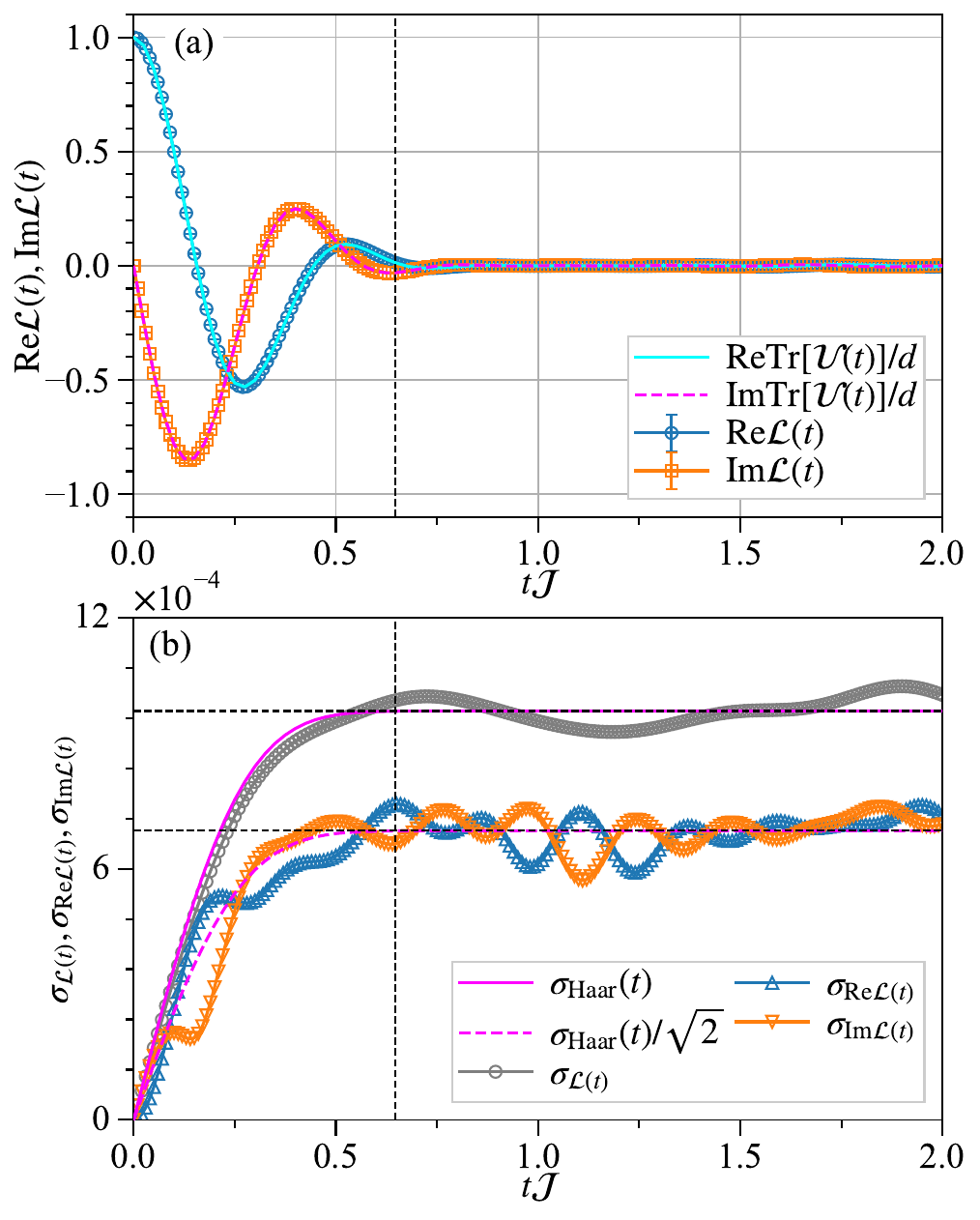}
  \caption{
    \label{fig:sample_m}
    (a) Averaged real and imaginary parts of the Loschmidt 
    amplitude, $\bE_{\psi\sim\Psi}[\mathrm{Re}\calL(t)]$ and 
    $\bE_{\psi\sim\Psi}[\mathrm{Im}\calL(t)]$, denoted by 
    circles and squares, respectively, 
    for the one-dimensional Heisenberg model $\tagH_{XXX}$ 
    with $N=20$ sites under periodic boundary conditions. 
    The average is taken over the ensemble $\Psi$ in 
    Eq.~(\ref{eq:floquet_ensemble}). For comparison, 
    the exact results of 
    ${\rm Re Tr}[{\cal U}(t)]/d$ and 
    ${\rm Im Tr}[{\cal U}(t)]/d$ are also shown by solid 
    and dashed curves, respectively. 
    (b) Standard deviations of the Loschmidt amplitude, 
    $\sigma_{{\cal L}(t)}$, and its real and imaginary parts, 
    $\sigma_{{\rm Re} {\cal L}(t)}$ and 
    $\sigma_{{\rm Im} {\cal L}(t)}$, respectively, corresponding 
    to the results shown in (a). 
    The solid and dashed curves indicate the standard deviation 
    $\sigma_{\rm Haar}(t)$ and $\sigma_{\rm Haar}(t)/\sqrt{2}$, 
    respectively, calculated for Haar random states. 
    The two dashed horizontal lines indicate $1/\sqrt{d+1}$ and 
    $1/\sqrt{2(d+1)}$.
    The dashed vertical lines in (a) and (b) indicate $t=\sigma^{-1}$, where $\sigma$ is defined in 
    Eq.~(\ref{eq:sigma}).
  }
\end{figure}

Figure~\ref{fig:sample_m}(a) shows the Loschmidt amplitude $\calL_m(t)= \langle \psi_0|(U_\mF^\dag)^m \calU(t) (U_\mF)^m \ket{\psi_0}$ averaged over states in the ensemble $\Psi$ in Eq.~\eqref{eq:floquet_ensemble}, i.e., 
\begin{align}\label{eq:Epsi}
    \bE_{\psi\sim\Psi}[\calL(t)] 
    :=
    \frac{1}{M} \sum_{m=m_{\rm min}}^{m_{\rm max}-1} {\cal L}_m(t),
\end{align}
for $N=20$.
Here, $M=m_{\rm max}-m_{\rm min}$ is the number of states in 
ensemble $\Psi$ and we set $m_{\rm min}=10$ and $m_{\rm max}=200$. 
We choose $m_{\rm min}=N/2=10$ to ensure that  
the Floquet states are sufficiently scrambled,  
while $m_{\rm max}=200$ is selected to obtain nearly 
converged results for the standard deviation. 
For ensemble $\Psi$ to approximate a state 2-design, $M$ must be 
sufficiently large so that it contains samples whose time intervals are longer than the scrambling time. 
However, it should be noted that obtaining converged results 
for the mean of Loschmidt amplitude requires only a small number 
$M$ of samples or even a single sample.
The initial state $|\psi_0\rangle$ is fixed to a product state 
of the form 
$\ket{\psi_0}=\prod_{i=1}^{N}{\rm e}^{-\imag\phi_{i} Y_i/2}|0^N\rangle$, where site-dependent angles 
$\{\phi_i\}_{i=1}^{N}$ are chosen randomly. 
As shown in Fig.~\ref{fig:sample_m}(a), 
the ensemble average in 
Eq.~(\ref{eq:Epsi}) coincides with the normalized trace of the time 
evolution operator, $\Tr[{\cal U}(t)]/d$, within statistical errors, 
where $d=2^N$ is the dimension of the Hilbert space.

Figure~\ref{fig:sample_m}(b) shows the standard deviations of the 
real and imaginary parts of the Loschmidt amplitude, denoted as 
$\sigma_{{\rm Re} {\cal L}(t)}:=\sqrt{\Var_\psi[{\rm Re} {\cal L}(t)]}$ and $\sigma_{{\rm Im} {\cal L}(t)}:=\sqrt{\Var_\psi[{\rm Im} {\cal L}(t)]}$, respectively. They are associated with the variances
\begin{align}
    \Var_{\psi\sim\Psi}[\mathrm{Re}\calL(t)]
    &= 
    \bE_{\psi\sim\Psi}[\mathrm{Re}\calL(t)^2]
    - 
    (\bE_{\psi\sim\Psi}[\mathrm{Re}\calL(t)])^2
\end{align}
and
\begin{align}
    \Var_{\psi\sim\Psi}[\mathrm{Im}\calL(t)]
    &= 
    \bE_{\psi\sim\Psi}[\mathrm{Im}\calL(t)^2]
    - 
    (\bE_{\psi\sim\Psi}[\mathrm{Im}\calL(t)])^2.  
\end{align}
Additionally, Fig.~\ref{fig:sample_m}(b) shows the standard 
deviation of the Loschmidt amplitude itself, denoted as 
$\sigma_{{\cal L}(t)}:=\sqrt{\Var_\psi[{\cal L}(t)]}$, with 
the variance $\Var_\psi[{\cal L}(t)]$ calculated as the sum of the 
variances of its real and imaginary parts: 
\begin{equation}
    \Var_{\psi\sim\Psi}[\calL(t)]
    = 
    \Var_{\psi\sim\Psi}[\mathrm{Re}\calL(t)] 
    + 
    \Var_{\psi\sim\Psi}[\mathrm{Im}\calL(t)].
\end{equation}
Although mostly invisible because they are smaller than the size of 
symbols, the error bars in Fig.~\ref{fig:sample_m}(a) represent 
are the standard deviations of the mean, calculated as 
$\sqrt{\Var_\psi[{\rm Re} {\cal L}(t)]/M}$ and $\sqrt{\Var_\psi[{\rm Im} {\cal L}(t)]/M}$ for 
the real and imaginary parts of the Loschmidt amplitude, 
respectively.

To gain insights into the numerical results of the standard 
deviations, let us consider the case where Haar random states 
are used instead of $\ket{\psi(m)}=\{(U_\mF)^m\ket{\psi_0}$. 
In this case, the variance 
$\Var_{\psi\sim\mathrm{Haar}}[\calL(t)]$ is given by
\begin{equation}
\label{eq:var_Haar}
    \Var_{\psi\sim\mathrm{Haar}}[\calL(t)]
    = 
    \bE_{\psi\sim\mathrm{Haar}}[|\calL(t)|^2]
    - 
    \big(\bE_{\psi\sim\mathrm{Haar}}[\calL(t)]\big)^2
    =
    \frac{1 - {\rm SFF}(t)}{d+1},
\end{equation}
where ${\rm SFF}(t):=\left|\Tr[{\cal U}(t)]\right|^2/d^2$ is 
the spectral form factor. 
Since $0\le {\rm SFF}(t) \le 1$, the standard deviation $\sigma_{\rm Haar}(t)=\sqrt{{\rm Var}_{\psi \sim {\rm Haar}}[\calL(t)]}$ is bounded as $0 \le \sigma_{\rm Haar}(t) \le 1/\sqrt{d+1}$. 
Moreover, the spectral form factor decays in time 
to zero at $t \sim \sigma_\tagH^{-1}$, where $\sigma_\tagH$ is defined in 
Eq.~(\ref{eq:sigma_H}). 
The decay of ${\rm SFF}(t)$ in the early time $t \lesssim \sigma_\tagH^{-1}$ can be understood by noticing that ${\rm SFF}(t)$ is proportional to the absolute square of the Fourier transform of the density of states, which has an approximate energy extent of $\sigma_\tagH$.
Thus, the standard deviation $\sigma_{\rm Haar}(t)$ can be approximated as $\sigma_{\rm Haar}(t) \approx 1/\sqrt{d+1}$ for $t\gg \sigma_{\tagH}^{-1}$ 
[see Fig.~\ref{fig:sample_m}(b)].

As shown in Fig.~\ref{fig:sample_m}(b), the standard 
deviations of the real and imaginary parts, 
$\sigma_{{\rm Re} {\cal L}(t)}$ and $\sigma_{{\rm Im} {\cal L}(t)}$, 
respectively, align with $\sigma_{\rm Haar}(t)/\sqrt{2}$. 
Consequently, the standard deviation $\sigma_{\calL(t)}$ 
agrees with the corresponding Haar value $\sigma_{\rm Haar}(t)$. 
This agreement between $\sigma_{\calL(t)}$ and $\sigma_{\rm Haar}(t)$ suggests that the ensemble $\Psi$ approximates 
a state 2-design. 

\subsection{Accurate estimate of normalized trace with a single scrambled state}

In the previous subsection, we numerically confirmed that the 
ensemble $\Psi$ in Eq.~\eqref{eq:floquet_ensemble} 
behaves like an approximate state 2-design.
Here, we justify that a single state $(U_\mF)^m\ket{\psi_0}$ 
with a sufficiently large $m$ (specifically, here we consider 
$m\gtrsim N/2$ for one-dimensional systems under periodic 
boundary conditions) suffices to estimate the normalized trace 
$\Tr[\calU(t)]/d$ accurately.
For the sake of numerical study, we randomly sample initial 
product states $|\psi_0\rangle$ with a fixed number $m$ of 
Floquet cycles.  
As described in Sec.~\ref{sec:exp_thermal}, 
we define our Floquet scrambled states as $|\psi_r\rangle = (U_{\mathrm F})^{m} \ket{\psi_0^{(r)}}$ with the initial product 
state 
$\ket{\psi_0^{(r)}}=\prod_{i=1}^{N}{\rm e}^{-\imag\phi_{i,r} Y_i/2}|0^N\rangle$, where $\{\phi_{i,r}\}_{i=1}^{N}$ is the 
$r$th set of independent random angles. We should note here 
that the random initialization in $\ket{\psi_0^{(r)}}$ 
corresponds to the average over a tensor 
product of single-qubit Haar random unitaries, which by itself 
does not form a state 2-design over an $N$-qubit system 
(also see Fig.~\ref{fig:sd_m}).

\begin{figure*}
  \includegraphics[width=0.49\textwidth]{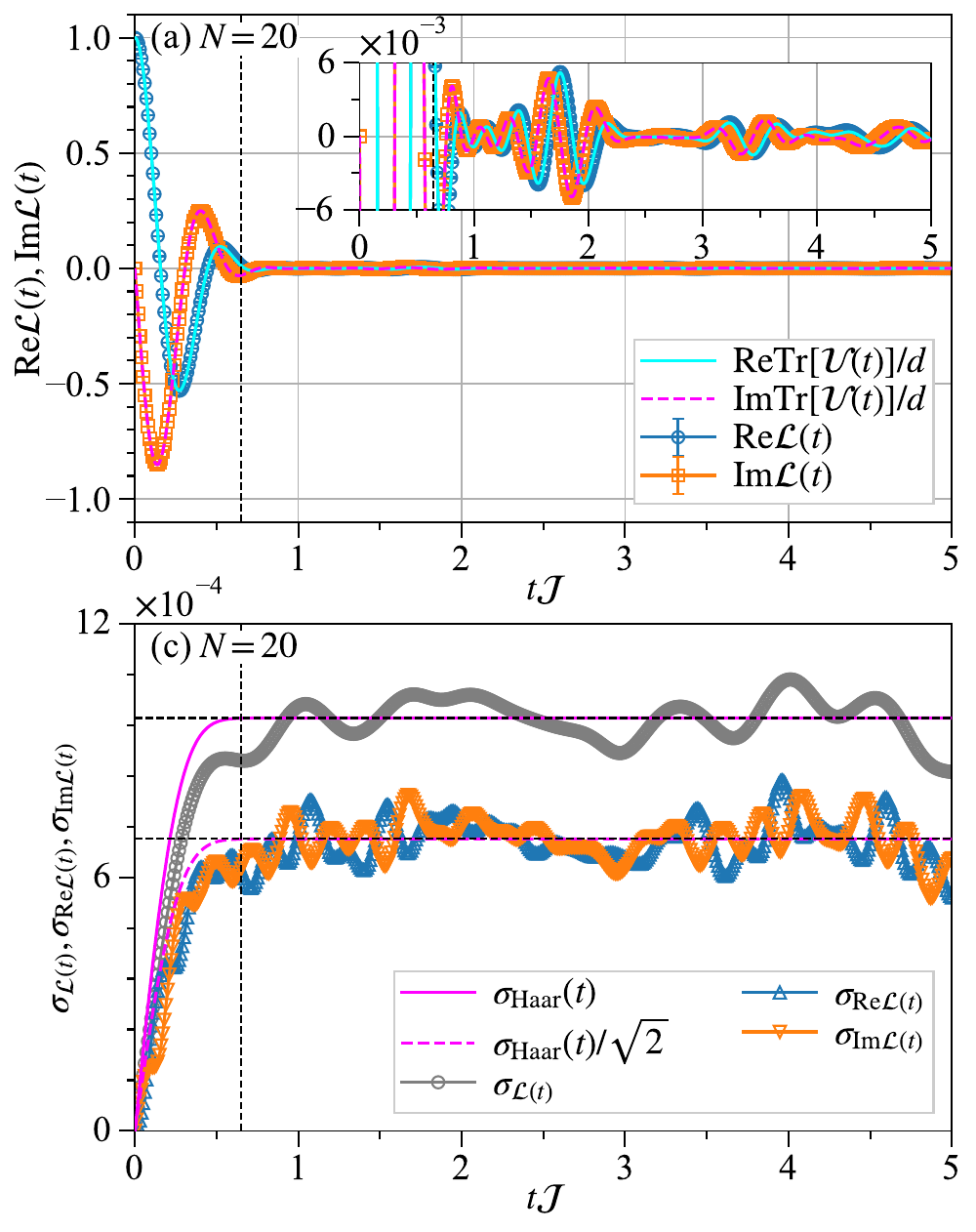}
    \includegraphics[width=0.49\textwidth]{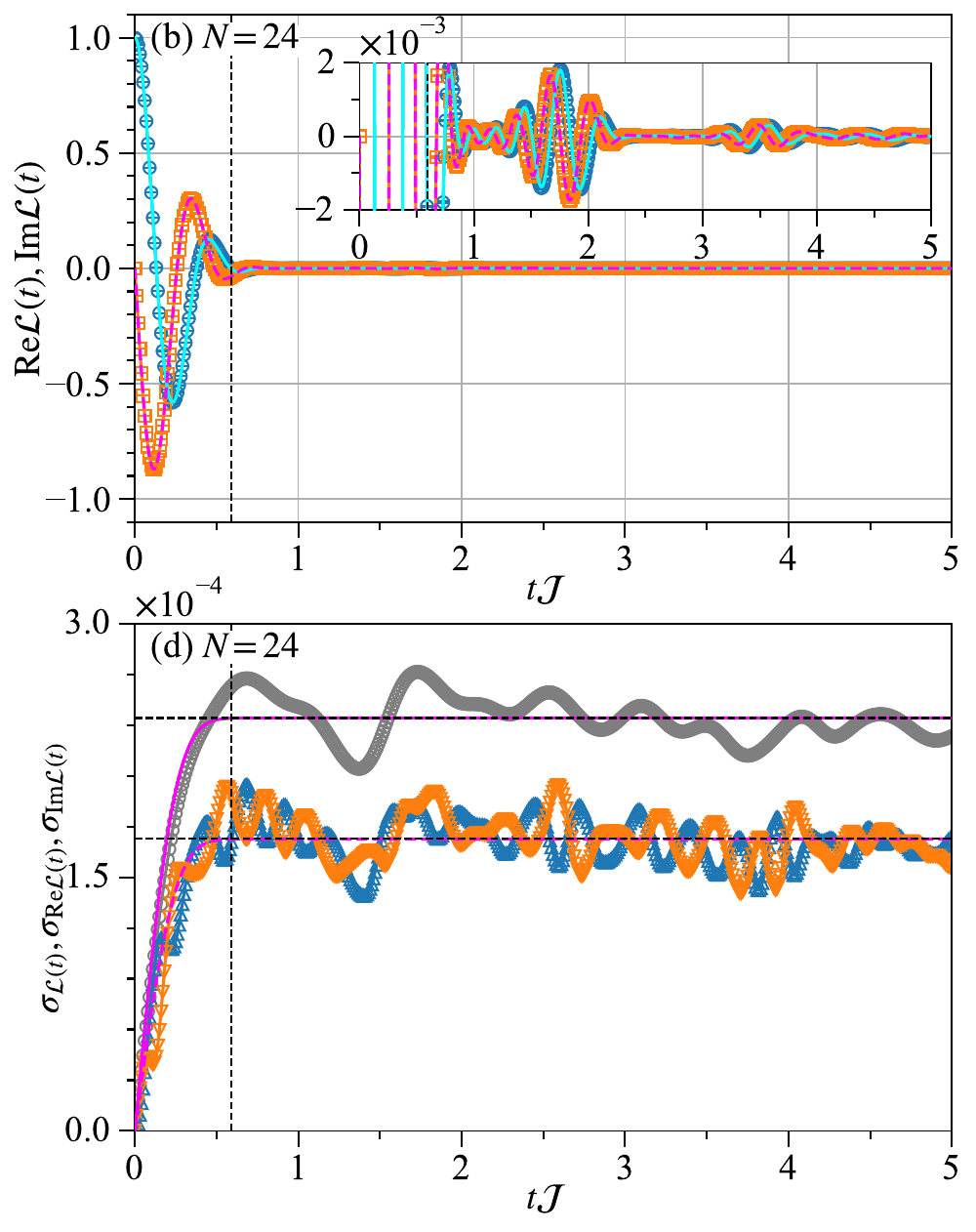}
  \caption{
    \label{fig:Loschmidt}
    (a,b) Sample-averaged real and imaginary parts of the Loschmidt 
    amplitude, $\Esample_{\psi\sim\Psi_m}[\mathrm{Re}\calL(t)]$ and 
    $\Esample_{\psi\sim\Psi_m}[\mathrm{Im}\calL(t)^2]$, denoted by 
    circles and squares, respectively, for the one-dimensional Heisenberg model $\tagH_{XXX}$ with $N=20$ and 24 sites 
    under periodic boundary conditions. 
    The sample average is taken over the ensemble $\Psi_m$ with 
    the number of Floquet cycles $m=N/2$. 
    For comparison, the exact results of 
    ${\rm Re Tr}[{\cal U}(t)]/d$ and 
    ${\rm Im Tr}[{\cal U}(t)]/d$ are also shown by solid 
    and dashed curves, respectively, where $d=2^N$ is the dimension 
    of the Hilbert space.     
    The insets show an enlarged view around zero. 
    (c,d) Sample-averaged standard deviations of the Loschmidt 
    amplitude, 
    $\sigma_{{\cal L}(t)}$, and its real and imaginary parts, 
    $\sigma_{{\rm Re} {\cal L}(t)}$ and 
    $\sigma_{{\rm Im} {\cal L}(t)}$, respectively, corresponding 
    to the results shown in (a,b). 
    The solid and dashed curves indicate the standard deviation 
    $\sigma_{\rm Haar}(t)$ and $\sigma_{\rm Haar}(t)/\sqrt{2}$, 
    respectively, calculated for Haar random states. 
    The two dashed horizontal lines indicate $1/\sqrt{d+1}$ and 
    $1/\sqrt{2(d+1)}$.
    The dashed vertical lines in (a-d) indicate $t=\sigma^{-1}$, 
    where $\sigma$ is defined in Eq.~(\ref{eq:sigma}).
  }
\end{figure*}

\begin{figure}
  \includegraphics[width=.48\textwidth]{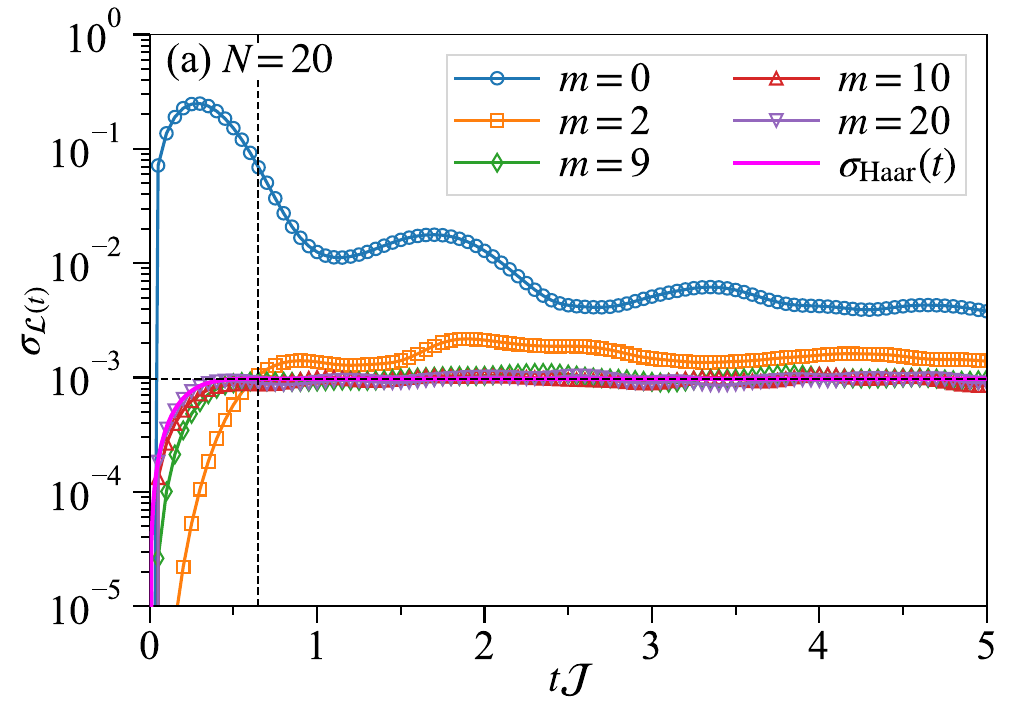}
  \includegraphics[width=.48\textwidth]{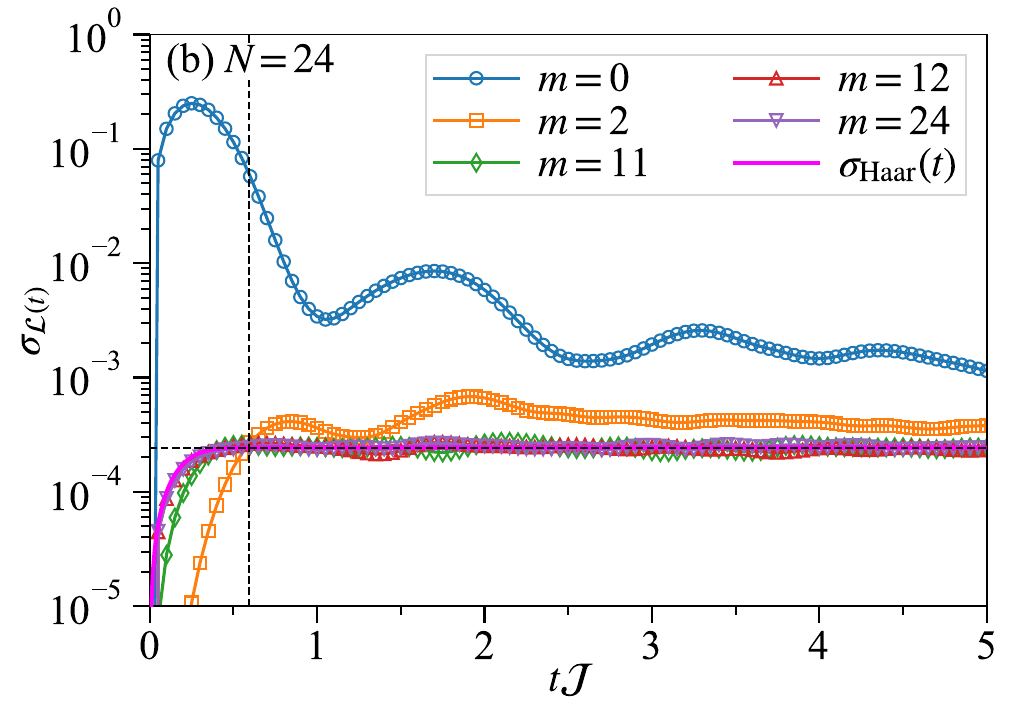}
  \caption{
    \label{fig:sd_m}
    Sample-averaged standard deviations $\sigma_{\calL(t)}$ of the 
    Loschmidt amplitude for the one-dimensional Heisenberg model $\tagH_{XXX}$ with (a) $N=20$ and (b) $N=24$ sites under periodic 
    boundary conditions are evaluated for various values of the Floquet 
    cycles $m$ in the ensemble $\Psi_m$. 
    For comparison, the standard deviations evaluated 
    for Haar random states, $\sigma_{\rm Haar}(t)$, are also plotted 
    by solid curves. 
    The dashed horizontal lines and the dashed vertical lines indicate $\sigma_{\calL(t)}=1/\sqrt{d+1}$ and $t=\sigma^{-1}$, respectively, 
    where $d=2^N$ is the dimension of the Hilbert space and $\sigma$ is 
    defined in Eq.~(\ref{eq:sigma}). Note that the results for $m=N/2$ 
    are the same as those shown in Figs.~\ref{fig:Loschmidt}(c,d). 
  }
\end{figure}

Figures~\ref{fig:Loschmidt}(a,b) show the Loschmidt amplitude 
${\cal L}_r(t) = \bra{\psi_r} {\cal U}(t) \ket{\psi_r}$ averaged 
over $R$ distinct random initializations, i.e.,   
\begin{align}
    \Esample_{\psi\sim\Psi_m}[\calL(t)] 
    :=
    \frac{1}{R} \sum_{r=1}^R {\cal L}_r(t),
\end{align}
for the one-dimensional Heisenberg model $\tagH_{XXX}$ 
with $N=20$ and 24 sites, respectively. 
Here, $\Psi_m = \{(U_{\rm F})^m \prod_{i=1}^{N}{\rm e}^{-\imag\phi_{i,r} Y_i/2}|0^N\rangle \mid r=1,\dots,R \}$ represents 
an ensemble of states with a fixed number $m$ of Floquet cycles. 
We set the number of Floquet cycles to $m=N/2$, while 
$R=100$ is chosen to achieve nearly converged results 
for the standard deviation. 
However, only a small number $R$ of samples is required 
to obtain converged results for the Loschmidt amplitude.
As expected, the statistical average coincides with the normalized 
trace of the time evolution operator within statistical errors, i.e., 
$\Esample_{\psi\sim\Psi_m}[\calL(t)] \approx \Tr[{\cal U}(t)]/d$.

Figures~\ref{fig:Loschmidt}(c,d) show the standard deviations of the 
real and the imaginary parts of the Loschmidt amplitude, denoted 
as $\sigma_{{\rm Re} {\cal L}(t)}:=\sqrt{\Vsample_\psi[{\rm Re} {\cal L}(t)]}$ and $\sigma_{{\rm Im} {\cal L}(t)}:=\sqrt{\Vsample_\psi[{\rm Im} {\cal L}(t)]}$, respectively. They are associated to the variances
\begin{align}
    \Vsample_{\psi\sim\Psi_m}[\mathrm{Re}\calL(t)]
    &= 
        \Esample_{\psi\sim\Psi_m}[\mathrm{Re}\calL(t)^2]
        - (\Esample_{\psi\sim\Psi_m}[\mathrm{Re}\calL(t)])^2
\end{align}
and
\begin{align}
    \Vsample_{\psi\sim\Psi_m}[\mathrm{Im}\calL(t)]
    &= 
        \Esample_{\psi\sim\Psi_m}[\mathrm{Im}\calL(t)^2]
        - (\Esample_{\psi\sim\Psi_m}[\mathrm{Im}\calL(t)])^2. 
\end{align}
Additionally, Figs.~\ref{fig:Loschmidt}(c,d) show the standard 
deviation of the Loschmidt amplitude, denoted as $\sigma_{{\cal L}(t)}:=\sqrt{\Vsample_\psi[{\cal L}(t)]}$, with the variance 
$\Vsample_\psi[{\cal L}(t)]$ calculated as the sum of the variances 
of its real and imaginary parts: 
\begin{equation}
    \Vsample_{\psi\sim\Psi_m}[\calL(t)]
    = 
    \Vsample_{\psi\sim\Psi_m}[\mathrm{Re}\calL(t)] 
    + 
    \Vsample_{\psi\sim\Psi_m}[\mathrm{Im}\calL(t)].
\end{equation}
%
Although mostly invisible because they are smaller than the size of 
symbols, the error bars in Figs.~\ref{fig:Loschmidt}(a,b) represent 
the standard deviations of the mean, calculated as 
$\sqrt{\Vsample_\psi[{\rm Re} {\cal L}(t)]/R}$ and 
$\sqrt{\Vsample_\psi[{\rm Im} {\cal L}(t)]/R}$ for the real and 
imaginary parts of the Loschmidt amplitude, respectively.  
%
As shown in Figs.~\ref{fig:Loschmidt}(c,d), the standard deviation 
of the Loschmidt amplitude $\sigma_{{\cal L}(t)}$ aligns with 
$\sigma_{\rm Haar}(t)$.
Moreover, the standard deviations of the real and imaginary parts, 
$\sigma_{{\rm Re} {\cal L}(t)}$ and $\sigma_{{\rm Im} {\cal L}(t)}$, 
respectively, also closely follow 
$\sigma_{\rm Haar}(t)/\sqrt{2}$.

\begin{figure}
  \includegraphics[width=.4\textwidth]{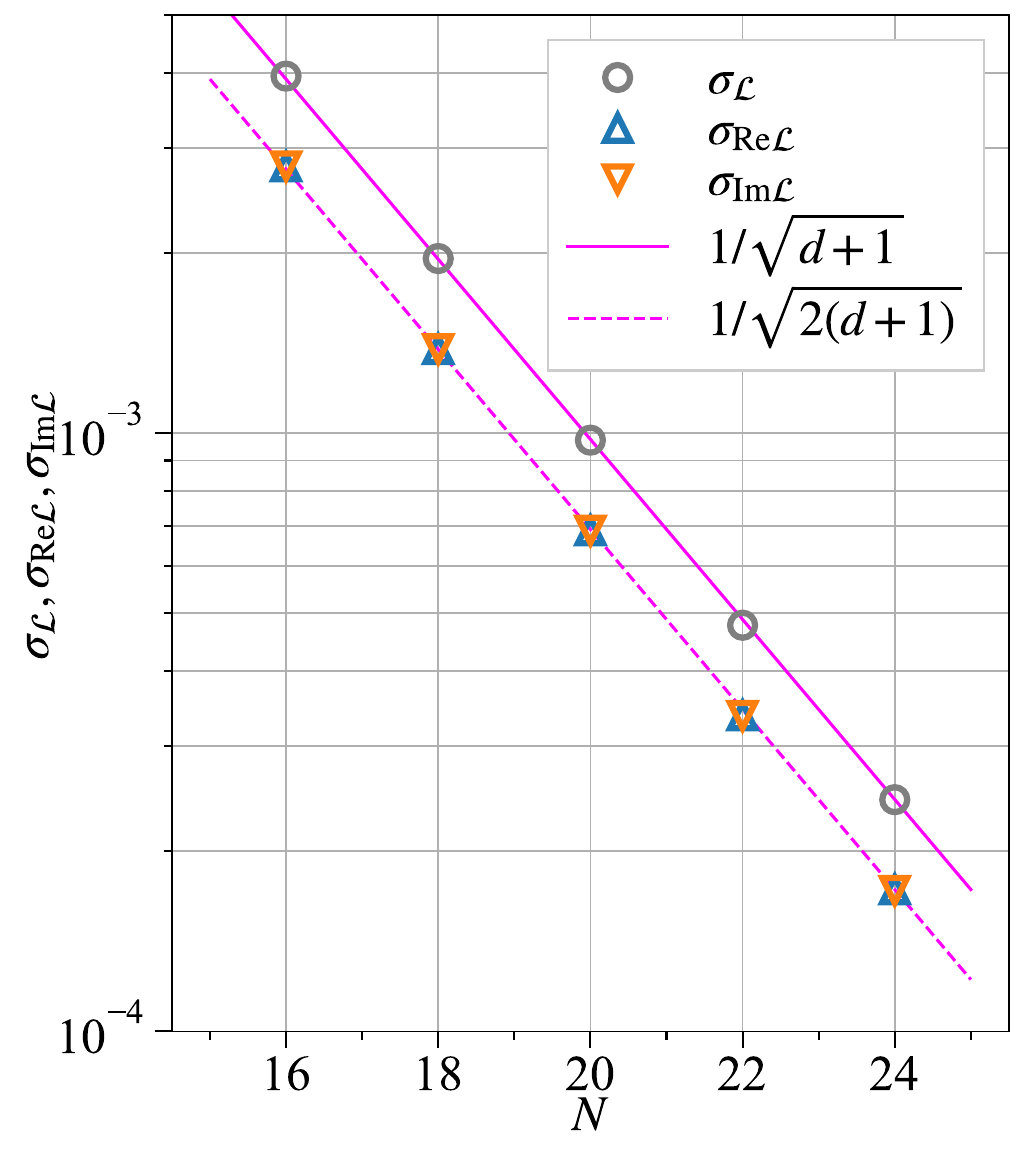}
  \caption{
    \label{fig:sd}
    Standard deviations of the Loschmidt amplitude, 
    and its real and imaginary parts, $\sigma_{{\cal L}}$, 
    $\sigma_{{\rm Re}{\cal L}}$, and $\sigma_{{\rm Im}{\cal L}}$, 
    respectively, averaged over time for 
    $\sigma^{-1}\tagJ < t\tagJ \le 5$, plotted as a function of 
    the system 
    size $N$. Here, $\sigma$ is defined in Eq.~(\ref{eq:sigma}) 
    and the number of Floquet circles in $\Psi_m$ is $m=N/2$. 
    The solid and dashed magenta lines indicate 
    $1\sqrt{d+1}$ and $1/\sqrt{2(d+1)}$, 
    respectively, where $d=2^N$ is the dimension of the Hilbert space. 
  }
\end{figure}

Figure~\ref{fig:sd} shows the system size dependence of 
the standard deviations 
$\sigma_{\cal L}(t)$, $\sigma_{{\rm Re}{\cal L}}(t)$, 
and $\sigma_{{\rm Im}{\cal L}}(t)$ averaged over time 
for $\sigma^{-1}\tagJ < t\tagJ \le 5$, where $\sigma$ is defined 
in Eq.~(\ref{eq:sigma}). Here, these standard 
deviations are evaluated over the ensemble $\Psi_m$ with 
the number of Floquet cycles $m=N/2$, as shown in 
Figs.~\ref{fig:Loschmidt}(c,d) for $N=20$ and 24 sites, respectively. 
The exponential decrease of the standard deviations with respect to 
$N$ is clearly observed in Fig.~\ref{fig:sd}.

Figure~\ref{fig:sd_m} shows the standard deviations $\sigma_{\calL(t)}$ 
of the Loschmidt amplitude with various values of the Floquet 
cycles $m$ in $\Psi_m$ for $N=20$ and $24$ sites. 
The result for $m=0$ corresponds to the standard deviation evaluated 
with respect to a set of the random product states $\{\ket{\psi_0^{(r)}}=\prod_{i=1}^{N}{\rm e}^{-\imag\phi_{i,r} Y_i/2}|0^N\rangle \}_{r=1}^{R}$. 
As shown in Fig.~\ref{fig:sd_m}, with increasing $m$, the standard 
deviation $\sigma_{\calL(t)}$ approaches the Haar value 
$\sigma_{\rm Haar}(t)$, 
and a good agreement is found for $ m\gtrsim N/2$. 
This result, together with the scrambling dynamics of the OTOC shown 
in Fig.~\ref{fig:OTOC_butterfly_pbc}(d), suggests that the ensemble of 
states $\Psi_m$ forms an approximate state 2-design for $m \gtrsim N/2$ 
under periodic boundary conditions.

The above result is consistent with the observations made in Refs.~\cite{Goto2021, Goto2022}, where they found that thermal expectation values can be efficiently evaluated using 
random-phase-product states~\cite{IItaka2020} evolved by a Trotterized 
time evolution operator of a nonintegrable Hamiltonian, with Trotter 
cycles proportional to the system size. 
Their choice of a large Trotter time step implies dynamics resembling 
Floquet dynamics rather than Hamiltonian dynamics. Therefore, 
our result suggests that their observations can be attributed to 
Floquet scrambling of states behaving as an approximate state 2-design.

\subsection{Influence of shot noise on the normalized trace estimation}

In the previous subsections, we assumed that the shot noise was absent, 
which is relevant for classical simulations or quantum simulations in 
the limit of the infinite number of shots. 
Now, we discuss the statistical errors of the Loschmidt amplitude 
in the presence of the shot noise, assuming that the number 
of shots $N_{\rm shots}$ is finite.

\subsubsection{Sampling over shots}

First, let us consider the shot noise when evaluating the Loschmidt 
amplitude, ${\cal L}(t) = \bra{\psi} {\cal U}(t)\ket{\psi}$, 
for a fixed pure state $\ket{\psi}$.  
Using the Hadamard test, the Loschmidt amplitude is estimated as 
$\Esample_\mathrm{shots}[x_\psi] + \imag \Esample_\mathrm{shots}[y_\psi]$, where $\Esample_\mathrm{shots}[x_\psi]$ and 
$\Esample_\mathrm{shots}[y_\psi]$ represent the sample means of 
its real and imaginary parts, respectively, obtained from the 
measurement outcomes on the ancillary qubit, i.e., 
\begin{align}
    \Esample_\mathrm{shots}[x_\psi]
    := 
    \frac{1}{N_{\rm shots}} \sum_{c=1}^{N_{\rm shots}} x_{\psi,c},
    \qquad
    \Esample_\mathrm{shots}[y_{\psi}]
    := 
    \frac{1}{N_{\rm shots}} \sum_{c=1}^{N_{\rm shots}} y_{\psi,c}.
\end{align}
Here, $x_{\psi,c}\in\{+1,-1\}$ denotes the $X$-measurement outcome of 
the ancillary qubit for the $c$-th shot, and similarly  
$y_{\psi,c}\in\{+1,-1\}$ is the $Y$-measurement outcome.
The sample variances of the real and imaginary parts due to the finite 
number of shots are given respectively by 
\begin{align}
    \Vsample_{\rm shots} [x_\psi] 
    = 1 - \big(\Esample_{\rm shots}[{x_{\psi}}]\big)^2, 
    \qquad
    \Vsample_{\rm shots} [y_\psi] 
    = 1 - \big(\Esample_{\rm shots}[{y_{\psi}}]\big)^2. 
\end{align}
The standard deviations of the means are then calculated as  
\begin{align}
    \sigma_{\rm shots}^{\rm mean} [x_\psi] 
    := \sqrt{\frac{\Vsample_{\rm shots}[x_\psi]}{N_{\rm shots}}}, 
    \qquad 
    \sigma_{\rm shots}^{\rm mean} [y_\psi] 
    := \sqrt{\frac{\Vsample_{\rm shots}[y_\psi]}{N_{\rm shots}}}.
\end{align}

\subsubsection{Sampling over states}

The above result is for a single pure state $\ket{\psi}$. Now, we 
extend this analysis to the average over $R$ independent realizations 
of states $\{\ket{\psi_r}\}_{r=1}^{R}$. 
The corresponding estimator of the normalized trace is given by $\Esample_\psi [\Esample_{\rm shots}[x_{\psi}]] + \im\,
\Esample_{\psi} [\Esample_{\rm shots}[y_{\psi}]]$, where 
\begin{align} 
    \Esample_{\psi} [\Esample_{\rm shots}[x_{\psi}]]
    := \frac{1}{R} \sum_{r=1}^{R} \Esample_{\rm shots}[x_{\psi_r}]
    \qquad
    \Esample_{\psi} [\Esample_{\rm shots}[y_{\psi}]]
    := \frac{1}{R} \sum_{r=1}^{R} \Esample_{\rm shots}[y_{\psi_r}].
\end{align}
The sample variances of the real and imaginary parts due to the 
sampling over states $\ket{\psi_r}$ are given respectively by 
\begin{align}
    \Vsample_{\psi} [\Esample_{\rm shots}[x_{\psi}]]
    &:= 
    \Esample_{\psi} [\Esample_{\rm shots}[x_{\psi}]^2] - 
    (\Esample_{\psi} [\Esample_{\rm shots}[x_{\psi}]])^2
\end{align}
and
\begin{align}
    \Vsample_{\psi} [\Esample_{\rm shots}[y_{\psi}]]
    &:= 
    \Esample_{\psi} [\Esample_{\rm shots}[y_{\psi}]^2] - 
    (\Esample_{\psi} [\Esample_{\rm shots}[y_{\psi}]])^2.
\end{align}
The corresponding standard deviations of the mean due to the sampling 
over states are then calculated as  
\begin{align}
    \sigma_{\psi}^{\rm mean} 
    \left[\Esample_{\rm shots}[x_{\psi}]\right]
    := 
    \sqrt{\frac{\Vsample_{\psi} [\Esample_{\rm shots}[x_{\psi}]]}{R}}, 
    \qquad
        \sigma_{\psi}^{\rm mean} \left[\Esample_{\rm shots}[y_{\psi}]\right]
    := 
    \sqrt{\frac{\Vsample_{\psi} [\Esample_{\rm shots}[y_{\psi}]]}{R}}.
\end{align}

Here, we first average over shots and then states, 
to associate the result in the previous subsection with 
the present analysis. 
Alternatively, one could average over shots and states simultaneously: 
$\Vsample[x]:=
\Esample_{\psi} [\Esample_{\rm shots}[x_{\psi}^2]]-
(\Esample_{\psi} [\Esample_{\rm shots}[x_{\psi}]])^2
=1-(\Esample_{\psi} [\Esample_{\rm shots}[x_{\psi}]])^2$ 
with a similar expression for $\Vsample[y]$. 
Then, as discussed in the next subsection, 
one could readily find that the standard deviation 
of the mean 
is given by the right-hand side of Eq.~(\ref{eq:sigma_mean_Nshots}) 
assuming that the states are sampled from a state 2-design.

\subsubsection{Sampling over shots and states}

Recall that each sample mean $\Esample_{\rm shots}[x_{\psi}]$ for a given $\ket{\psi}$ has the standard deviation $\sigma_{\rm shots}^{\rm mean}[x_{\psi}] (\sim 1/\sqrt{N_{\rm shots}})$ around the population mean.
Therefore, the standard deviation of 
$\Esample_{\psi}[\Esample_{\rm shots}[x_{\psi}]]$ should be described 
by both $\sigma_{\psi}^{\rm mean}$ and $\sigma_{\rm shots}^{\rm mean}$ (the same applies to  $\Esample_{\psi}[\Esample_{\rm shots}[y_{\psi}]]$ ). 
From the error-propagation formula, the standard deviations of 
the means  
$\Esample_{\psi}[\Esample_{\rm shots}[x_{\psi}]]$ and 
$\Esample_{\psi}[\Esample_{\rm shots}[y_{\psi}]]$ 
are expressed respectively as 
\begin{align}
    \sigma^{\rm mean}[x] 
    &:= 
    \sqrt{
    \left(\sigma_{\psi}^{\rm mean}[\Esample_{\rm shots}[x_{\psi}]]\right)^2 
    + 
    \frac{1}{R^2}\sum_{r=1}^{R} \left(\sigma_{\rm shots}^{\rm mean}[x_{\psi_r}]\right)^2}
\end{align}
and
\begin{align}
    \sigma^{\rm mean}[y] 
    &:= 
    \sqrt{
    \left(\sigma_{\psi}^{\rm mean}[\Esample_{\rm shots}[y_{\psi}]]\right)^2 + 
    \frac{1}{R^2}\sum_{r=1}^{R} \left(\sigma_{\rm shots}^{\rm mean}[y_{\psi_r}]\right)^2}.
\end{align}

\subsubsection{Standard deviation of the mean in the presence of the 
shot noise assuming a state 2-design}

To proceed further, we consider $\sigma^{\rm mean}[x+\imag y]=\sqrt{(\sigma^{\rm mean}[x])^2 + (\sigma^{\rm mean}[y])^2 }$ for simplicity. 
Additionally, we make the following replacements: 
$\Esample_{\rm shots}[{x}_{\psi}] \to \mathbb{E}_{\rm shots}[x_{\psi}] = {\rm Re} \bra{\psi} U(t) \ket{\psi}$, 
$\Esample_{\rm shots}[{y}_{\psi}] \to \mathbb{E}_{\rm shots}[y_{\psi}] = {\rm Im} \bra{\psi} U(t)\ket{\psi}$, 
and $\Esample_{\psi} [\cdots] \to \bE_{\psi\sim\text{2-design}}[\cdots]$. 
Then, the variance due to the shots averaged over states is replaced with
\begin{equation}
    \frac{1}{R^2}\sum_{r=1}^{R}
    \left(
    {\sigma}_{\rm shots}^{\rm mean}[x_{\psi_r} + \imag y_{\psi_r}] 
    \right)^2
    \to 
    \frac{1}{R N_{\rm shots}} 
    \left(2 - \frac{1}{d+1} - \frac{d {\rm SFF}(t)}{d+1}\right),
\end{equation}
and the variance due to the sampling of states is replaced with
\begin{equation}
    \left({\sigma}_{\psi_r}^{\rm mean}[\Esample_{\rm shots}[x_{\psi_r} + \imag y_{\psi_r}]] \right)^2\to \frac{1-{\rm SFF}(t)}{R(d+1)}.
\end{equation}
The standard deviation of the mean is then replaced with
\begin{equation}
\sigma^{\rm mean}[x+\imag y] \to
    \sqrt{
    \frac{1-{\rm SFF}(t)}{R(d+1)}
    +
    \frac{1}{R N_{\rm shots}}
    \left(2 - \frac{1}{d+1} - \frac{d {\rm SFF}(t)}{d+1}\right)
    }.
    \label{eq:sigma_mean}
\end{equation}

Let us first consider a situation relevant to classical state-vector simulations where the output of the Loschmidt amplitude is given by a double-precision (single-precision) floating-point number with a relative error due to round-off of at most $2^{-53}$ ($2^{-24}$).
Assuming $R=O(1)$ and requiring the shot noise to be no larger than this round-off error, the corresponding number of shots would be around 
$N_{\rm shots}\sim 2^{106}$ ($2^{48}$).  
On the other hand, the classically tractable dimension of the Hilbert space is at most $d \sim 2^{48}$~\cite{DeReadt2019}. 
Therefore, in most cases, classical state-vector simulations with double-precision arithmetic satisfy the relation $R \ll d \ll N_{\rm shots}$. 
By taking the limit of $N_{\rm shots}\to \infty$ with finite $d$ and $R$, we obtain the result consistent with Eq.~(\ref{eq:var_Haar}) in the previous subsection: 
\begin{equation}
    \sigma^{\rm mean}[x+\imag y] 
    \overset{N_{\rm shots}\to \infty}{\to}  \sqrt{\frac{1-{\rm SFF}(t)}{R(d+1)}}.
\end{equation}

Next, we consider a situation relevant to quantum computation, where 
the relation $R \ll N_{\rm shots} \ll d$ holds. In this case, 
by taking the limit of $d\to \infty$ with finite $N_{\rm shots}$ 
and $R$, 
the standard deviation of the mean simplifies to 
\begin{equation}
\label{eq:sigma_mean_Nshots}
    \sigma^{\rm mean}[x+\imag y] 
    \overset{d \to \infty}{\to }
    \sqrt{\frac{2-{\rm SFF}(t)}{R N_{\rm shots}}}.
\end{equation}
Thus, unless $N_{\rm shots}$ is as large as $O(d)$, the shot noise 
dominates $\sigma^{\rm mean}$ .

\section{Error in microcanonical expectation value due to a state 2-design}
\label{app:TPQ}

In this Appendix, we shall prove the inequality stated 
in Eq.~\eqref{eq:success_prob}.
We introduce a filter operator defined as 
\begin{align}
\label{eq:filter_op}
    G_\sigma(E) := \e^{-(E-\tagH)^2/2\sigma^2},
\end{align}
and examine a thermal state $\rho_\sigma=G_\sigma/\Tr[G_\sigma]$ representing the microcanonical ensemble parametrized by the internal energy $E$ and width $\sigma$.
The thermal entropy is expressed as 
\begin{align}
\label{eq:entropy}
    S_\sigma(E) = \log\Tr[G_\sigma(E)],
\end{align}
while the microcanonical expectation value of a Hermitian operator $O$ is calculated by
\begin{align}
\label{eq:exp_O}
    \langle O\rangle = \Tr[O \rho_\sigma].
\end{align}

Here, we estimate the expectation value $\langle O\rangle$ using a state 2-design.
Letting $\ket{\psi}$ be a state sampled from a state 2-design, we compute
\begin{align}
\label{eq:app_O}
    \langle O\rangle^\text{(est)}
    =
    \frac{\bra{\psi}OG_\sigma\ket{\psi}}
    {\bra{\psi}G_\sigma\ket{\psi}}.
\end{align}
We first prove that this gives an accurate estimator of $\langle O\rangle$ in Eq.~\eqref{eq:exp_O} with high probability.
To this end, we show the mean-square error 
\begin{align}
\label{eq:mse}
    \bE_{\psi}\big[|\langle O\rangle^\text{(est)} - \langle O\rangle|^2\big]
    =
    \big|\bE_{\psi}[\langle O\rangle^\text{(est)}] - \langle O\rangle\big|^2
    +
    \Var_{\psi}[\langle O\rangle^\text{(est)}],
\end{align}
is exponentially small in the system size $N$ 
and then we employ the Chebyshev inequality, as shown in Eq.~\eqref{eq:TPQ_Chebyshev}.
Here, The first and second terms on the right-hand side in Eq.~\eqref{eq:mse} represent the squared bias and the statistical uncertainty, respectively.

Following Refs.~\cite{Jin2021,Coopmans2023,Watts2023}, we introduce the quantities 
\begin{align}
    f:= \bra{\psi}OG_\sigma(E)\ket{\psi},
    \qquad
    g:= \bra{\psi}G_\sigma(E)\ket{\psi}.
\end{align}
Upon averaging $\ket{\psi}$ over the state 2-design, we obtain
\begin{align}
    \bE_{\psi} [f]
    = \frac{\Tr[O G_\sigma]}{d},
    \qquad
    \bE_{\psi} [g]
    = \frac{\Tr[G_\sigma]}{d},
\end{align}
along with the variances
\begin{align}
    \label{eq:var_numer}
    \mathrm{Var}_{\psi} [f]
    &= 
    \bE_{\psi} [|\bra{\psi}OG_\sigma\ket{\psi}|^2]
    - \big|\bE_{\psi} [\bra{\psi}OG_\sigma\ket{\psi}]\big|^2
    \nonumber\\
    &= 
    \frac{\Tr[O^\dagger O G_\sigma^2] + \big|\Tr[O G_\sigma]\big|^2}{d(d+1)}
    - \frac{\big|\Tr[O G_\sigma]\big|^2}{d^2}
    \le
    \| O\|^2(\Tr[G_\sigma])^2\frac{\Tr[\rho_\sigma^2]}{d^2}
\end{align}
and
\begin{align}
    \label{eq:var_denom}
    \mathrm{Var}_{\psi} [g]
    &= 
    \bE_{\psi} [\bra{\psi}G_\sigma\ket{\psi}^2]
    - \left( \bE_{\psi} [\bra{\psi}G_\sigma\ket{\psi}] \right)^2
    \le
    (\Tr[G_\sigma])^2\frac{\Tr[\rho_\sigma^2]}{d^2},
\end{align}
respectively. Here,  
$\| O\|:=\sqrt{\max_{\ket{v}}\frac{\bra{v}O^\dagger O\ket{v}}{\braket{v}{v}}}$ is the spectral norm with $\ket{v}$ being the eigenstate of $O^\dagger O$.

Our objective now is to assess $\bE_\psi[\langle O\rangle^{\mathrm{(est)}}]=\bE_\psi[f/g]$ and its variance.
One might initially consider the expansion
\begin{align}
    \bE_\psi\left[\frac{f}{g}\right]
    =
    \bE_\psi\left[\left(\frac{\bE_\psi[f]}{\bE_\psi[g]}+\Delta f\right)\frac{1}{1+\Delta g}\right]
    =
    \frac{\bE_\psi[f]}{\bE_\psi[g]}\bE_\psi\left[\left(\frac{\bE_\psi[f]}{\bE_\psi[g]}+\Delta f\right)(1+\Delta g+\calO(\Delta g^2))\right],
\end{align}
assuming that $\Delta f:= (f-\bE_\psi[f])/\bE_\psi[g]$ and $\Delta g:= (g-\bE_\psi[g])/\bE_\psi[g]$ are small.
However, this expansion is not convergent because $|\Delta g|$ can be larger than 1.
To circumvent the issue, we partition the ensemble of $\ket{\psi}$ (2-design) into a subset $\Delta$ with a small positive number $\delta$, where $|\Delta g|$ is small, and its complement $\bar{\Delta}$:  
\begin{align}
\label{eq:subsets}
    \Delta := \left\{
        \ket{\psi}\big|\,
        |\Delta f| \le \delta 
        \text{ and }
        |\Delta g| \le \delta
    \right\},
    \qquad
    \bar{\Delta} := \left\{
        \ket{\psi}\big|\,
        |\Delta f| > \delta 
        \text{ or }
        |\Delta g| > \delta
    \right\}.
\end{align}
Two important observations are that one can safely use expansions in terms of $\Delta g$ in the subset $\Delta$, and that the probability of $\ket{\psi}$ falling into $\bar{\Delta}$ is very small. The latter observation is supported by the Chebyshev inequalities
\begin{align}
    \Pr\big[ | \Delta f | > \delta \big]
    <
    \frac{\Var_\psi[f]}{\delta^2(\bE_\psi[g])^2}
    \le
    \| O\|^2\frac{\Tr[\rho_\sigma^2]}{\delta^2},
    \qquad
    \Pr\big[ | \Delta g | > \delta\big]
    <
    \frac{\Var_\psi[g]}{\delta^2(\bE_\psi[g])^2}
    \le
    \frac{\Tr[\rho_\sigma^2]}{\delta^2},
\end{align}
leading to 
\begin{align}
\label{eq:prob_subsets}
    \Pr[\Delta]
    \ge
    1- (1+\| O\|^2)\frac{\Tr[\rho_\sigma^2]}{\delta^2},
    \qquad
    \Pr[\bar{\Delta}]
    <
    (1+\| O\|^2)\frac{\Tr[\rho_\sigma^2]}{\delta^2},
\end{align} 
where $\Pr[\Delta]+\Pr[\bar{\Delta}]=1$.
We ascertain that $\Pr[\bar{\Delta}]$ is indeed small because the purity $\Tr[\rho_\sigma^2]$ diminishes exponentially with the system size $N$, 
as elaborated later in Eq.~\eqref{eq:purity}.
The inequalities in Eq.~\eqref{eq:prob_subsets} will be 
used repeatedly to bound the bias and variance in the subsequent analysis.

\subsection*{Bounding bias}

Noticing that $\langle O\rangle=\bE_\psi[f]/\bE_\psi[g]$, we first establish the upper bound for the bias: 
\begin{align}
\label{eq:bias}
\begin{split}
    \big|\bE_\psi\big[\langle O\rangle^{\mathrm{(est)}}\big] - \langle O\rangle\big|
    &=
    \Pr[\Delta] \left|\bE_{\psi_\Delta}\left[
        \frac{f}{g} - \frac{\bE_\psi[f]}{\bE_\psi[g]}
    \right]\right|
    +
    \Pr[\bar{\Delta}] \left|\bE_{\psi_{\bar{\Delta}}}\left[
        \frac{f}{g} - \frac{\bE_\psi[f]}{\bE_\psi[g]}
    \right]\right|
    \\
    &\le
    \left|\bE_{\psi_\Delta}\left[
        \frac{f}{g} - \frac{\bE_\psi[f]}{\bE_\psi[g]}
    \right]\right|
    +
    \frac{(1+\|O\|^2)\Tr[\rho_\sigma^2]}{\delta^2}2\|O\|,
\end{split}
\end{align}
where $\bE_{\psi_{\Delta}}[\cdot]$ represents the average of $\ket{\psi}$ over the subset of a state 2-design in Eq.~\eqref{eq:subsets} and $\bE_{\psi_{\bar{\Delta}}}[\cdot]$ is defined analogously. 
In deriving the inequality, we used the bound $\big|f/g - \bE_\psi[f]/\bE_\psi[g]\big|\le 2\|O\|$.
The first term is further bounded as
\begin{align}
\label{eq:bias_1st_term}
    \left|\bE_{\psi_\Delta}\left[
        \frac{f}{g} - \frac{\bE_\psi[f]}{\bE_\psi[g]}
    \right]\right|
    =
    \left|\bE_{\psi_\Delta}\left[
        \left(\frac{\bE_\psi[f]}{\bE_\psi[g]} + \Delta f\right) 
        \left(1 - \frac{\Delta g}{1+\Delta g}\right)
        - 
        \frac{\bE_\psi[f]}{\bE_\psi[g]}
    \right]\right| 
    \le
    \delta\left(1 +\frac{\|O\|+\delta}{1-\delta}\right),
\end{align}
where the following was applied: 
\begin{align}
    \left|\frac{\Delta g}{1+\Delta g}\right|\le \frac{\delta}{1-\delta}
\end{align}    
for $|\Delta g|\le\delta\le1$.
By substituting the bound from Eq.~\eqref{eq:bias_1st_term} into the bias given in Eq.~\eqref{eq:bias}, we derive 
\begin{align}
\begin{split}
\big|\bE_\psi\big[\langle O\rangle^{\mathrm{(est)}}\big] - \langle O\rangle\big|=
    \left|\bE_\psi\left[\frac{f}{g}\right] - \frac{\bE_\psi[f]}{\bE_\psi[g]}\right|
    \le
    \delta\left(1 +\frac{\|O\|+\delta}{1-\delta}\right)
    +
    \frac{2\|O\|(1+\|O\|^2)\Tr[\rho_\sigma^2]}{\delta^2}.
\end{split}
\end{align}
Setting $\delta=(\Tr[\rho_\sigma^2])^{1/3}/2$, we finally establish the following bound on the bias:
\begin{align}
\label{eq:bias_bound}
\begin{split}
    \big|\bE_\psi\big[\langle O\rangle^{\mathrm{(est)}}\big] - \langle O\rangle\big|
    \le
    (\Tr[\rho_\sigma^2])^{1/3}
    \big(
        1 + 9\|O\| + 8\|O\|^3
    \big),
\end{split}
\end{align}
where we used the condition that $0<\delta\le1/2$.
As we will show later in Eq.~\eqref{eq:purity}, the purity $\Tr[\rho_\sigma^2]$ is exponentially small in the system size $N$.

\subsection*{Bounding variance}

Using the same strategy, we bound the variance, which is the second term on the right-hand side of Eq.~\eqref{eq:mse}, as follows: 
\begin{align}
\label{eq:var_psi}
\begin{split}
    \Var_\psi\big[\langle O\rangle^{\mathrm{(est)}}\big]
    &=
    \bE_\psi\left[\left(
        \frac{f}{g} - \bE_\psi\left[\frac{f}{g}\right]
    \right)^2\right]
    \\
    &=
    \Pr[\Delta] \bE_{\psi_\Delta}\left[\frac{f^2}{g^2}\right]
    +
    \Pr[\bar{\Delta}] \bE_{\psi_{\bar{\Delta}}}\left[\frac{f^2}{g^2}\right] 
    - \left(
        \Pr[\Delta] \bE_{\psi_\Delta}\left[\frac{f}{g}\right]
        +
        \Pr[\bar{\Delta}] \bE_{\psi_{\bar{\Delta}}}\left[\frac{f}{g}\right] 
    \right)^2
    \\
    &\le
    \Pr[\Delta] \bE_{\psi_\Delta}\left[\frac{f^2}{g^2}\right]
    - \left(
        \Pr[\Delta] \bE_{\psi_\Delta}\left[\frac{f}{g}\right]
    \right)^2
    +
    \Pr[\bar{\Delta}]\cdot 3\|O\|^2 
    \\
    &\le
    \left(
        \bE_{\psi_\Delta}\left[\frac{f^2}{g^2}\right]
        -
        \left(\bE_{\psi_\Delta}\left[\frac{f}{g}\right]\right)^2
    \right)
    +
    \Pr[\bar{\Delta}]\cdot 4\|O\|^2.
\end{split}
\end{align}
The first term in the last equation is expanded as
\begin{align}
\begin{split}
    &\bE_{\psi_\Delta}\left[\frac{f^2}{g^2}\right]
    -
    \left(\bE_{\psi_\Delta}\left[\frac{f}{g}\right]\right)^2
    \\
    &=
    \bE_{\psi_\Delta}
    \left[\left(
        \frac{\bE_\psi[f]}{\bE_\psi[g]} 
        + \Delta f
        - \frac{\bE_\psi[f]}{\bE_\psi[g]} \frac{\Delta g}{1+\Delta g}
        - \frac{\Delta f\Delta g}{1+\Delta g}
    \right)^2\right]
    -
    \left(\bE_{\psi_\Delta}
    \left[
        \frac{\bE_\psi[f]}{\bE_\psi[g]} 
        + \Delta f
        - \frac{\bE_\psi[f]}{\bE_\psi[g]} \frac{\Delta g}{1+\Delta g}
        - \frac{\Delta f\Delta g}{1+\Delta g}
    \right]\right)^2
    \\
    &=
    \bE_{\psi_\Delta}
    \left[
        \Delta f^2 
        + \frac{\bE_\psi[f]^2}{\bE_\psi[g]^2}\Big(\frac{\Delta g}{1+\Delta g}\Big)^2
        - 2 \frac{\bE_\psi[f]}{\bE_\psi[g]}\frac{\Delta f\Delta g}{1+\Delta g}
    \right]
    \\
    &\quad -
    \left(
        \bE_{\psi_\Delta}[\Delta f]^2
        + \frac{\bE_\psi[f]^2}{\bE_\psi[g]^2}
        \bE_{\psi_\Delta}\left[\frac{\Delta g}{1+\Delta g}\right]^2
        - 2 \frac{\bE_\psi[f]}{\bE_\psi[g]}
        \bE_{\psi_\Delta}[\Delta f]
        \bE_{\psi_\Delta}\left[\frac{\Delta g}{1+\Delta g}\right]
    \right)
    +
    \calO(\delta^3)
    \\
    &\le
    \delta^2 
    + \|O\|^2\frac{\delta^2}{(1-\delta)^2}
    + 4 \|O\|\frac{\delta^2}{1-\delta}
    + \calO(\delta^3).
\end{split}
\end{align}
Thus, the upper bound of the variance in Eq.~\eqref{eq:var_psi} is obtained as 
\begin{align}
    \Var_\psi\big[\langle O\rangle^{\mathrm{(est)}}\big]
    \le
    \delta^2\left(
        1 + \frac{\|O\|^2}{(1-\delta)^2} + \frac{4\|O\|}{1-\delta}
    \right)
    +
    \frac{4(1+\|O^2\|)\|O\|^2\Tr[\rho_\sigma^2]}{\delta^2}
    +
    \calO(\delta^3).
\end{align}
Setting $\delta=(\Tr[\rho_\sigma^2])^{1/4}/2$, we find the bound 
\begin{align}
\label{eq:var_psi_bound}
    \Var_\psi\big[\langle O\rangle^{\mathrm{(est)}}\big]
    \le
    (\Tr[\rho_\sigma^2])^{1/2}
    \left(
        \frac{1}{4} + 2\|O\| + 17\|O\|^2 + 4\|O\|^4
    \right)
    +
    \calO\big((\Tr[\rho_\sigma^2])^{3/4}\big).
\end{align}

\subsection*{Scaling of purity}

Let us evaluate the scaling of the purity $\Tr[\rho_\sigma^2]$ as follows:  
\begin{align}
\label{eq:purity}
    \Tr[\rho_\sigma^2]
    =
    \frac{\Tr[G_\sigma(E)^2]}{(\Tr[G_\sigma(E)])^2}
    =
    \frac{\Tr[G_{\sigma/\sqrt{2}}(E)]}{(\Tr[G_\sigma(E)])^2}
    =
    \e^{S_{\sigma/\sqrt{2}}-2S_{\sigma}}
    =
    \calO(\e^{-\alpha N}), 
\end{align}
where $\alpha$ is a positive constant independent of $N$. Here, we used that the entropy $S_\sigma$ defined in Eq.~\eqref{eq:entropy} is an extensive quantity, $S_\sigma=\Theta(N)$, and monotonically increases with $\sigma$, i.e., 
\begin{align}
    \frac{\p S_\sigma}{\p\sigma}
    =
    \frac{1}{\sigma^3}\frac{\Tr[(E-\tagH)^2 G_\sigma(E)]}{\Tr[G_\sigma(E)]}
    \ge0, 
\end{align}
as shown in Ref.~\cite{Seki2022}.

\subsection*{Success probability}

Substituting the bounds from Eqs.~\eqref{eq:bias_bound} and \eqref{eq:var_psi_bound} into the mean-square error in Eq.~\eqref{eq:mse}, we obtain
\begin{align}
\label{eq:mse_bound}
    \bE_{\psi}\big[|\langle O\rangle^\text{(est)} - \langle O\rangle|^2\big]
    \le
    (\Tr[\rho_\sigma^2])^{1/2}
    \left(
        \frac{1}{4} + 2\|O\| + 17\|O\|^2 + 4\|O\|^4
    \right)
    +
    \calO\big((\Tr[\rho_\sigma^2])^{2/3}\big)
\end{align}
Note that the bound on the squared bias in Eq.~\eqref{eq:bias_bound} is subleading relative to that on the variance in Eq.~\eqref{eq:var_psi_bound}.

Applying the Chebyshev inequality, we can bound the success probability as follows: 
\begin{align}
\label{eq:TPQ_Chebyshev}
    \mathrm{Pr}_{\ket{\psi}}\left[
        \left|
            \langle O\rangle^{\mathrm{(est)}}
            -
            \langle O\rangle 
        \right| 
        \le \epsilon
    \right]
    \ge
    1 - \frac{\bE_{\psi}\big[|\langle O\rangle^\text{(est)} - \langle O\rangle|^2\big]}{\epsilon^2}
    \ge
    1-\calO\left(\frac{\e^{-\alpha N/2}}{\epsilon^2}\right), 
\end{align}
as stated in Eq.~(\ref{eq:success_prob}). 
Thus, we conclude that $\langle O\rangle^\text{(est)}$ provides an $\epsilon$-precise estimator of the microcanonical expectation value of $O$ with high probability.

\section{Resources for imaginary time evolution}
\label{app:imaginary}

We approximate the integral form of $G_\sigma(E)$~\cite{Chowdhury2016,Seki2022} 
\begin{align}
    G_\sigma(E)
    =
    \frac{\sigma}{\sqrt{2\pi}}\int_{-\infty}^{\infty} \diff  t\,
    \e^{-\frac{\sigma^2 t^2}{2}}\e^{-\im (E-\tagH) t}
\end{align}
by truncating and discretizing the integral as 
\begin{align}
    \tilde{G}'_{\sigma}(E)
    =
    \frac{\sigma}{\sqrt{2\pi}}\sum_{s=-S}^{S} \Delta  t\,
    \e^{-\frac{\sigma^2(s\Delta t)^2}{2}}\e^{-\im (E-\tagH) s\Delta t}.
\end{align}

We restate two useful results from Ref~\cite{Chowdhury2016} for our analysis.
Lemma~1 of Ref.~\cite{Chowdhury2016} states that there exist parameters
\begin{align}
\label{eq:discretization_params}
    S=\Theta\left(\frac{\|E-\tagH\|}{\sigma} \log\frac{1}{\epsilon'}\right),
    \qquad
    \Delta t = \Theta\left(\frac{1}{\|E-\tagH\|\sqrt{\log\frac{1}{\epsilon'}}}\right),
\end{align}
such that
\begin{align}
    \| \tilde{G}'_{\sigma}(E) - G_{\sigma}(E)\|
    \le \frac{\epsilon'}{2}.
\end{align}
The filter $\tilde{G}'_{\sigma}(E)$ is an approximation to $G_{\sigma}(E)$ provided that $\e^{-\im H s\Delta  t}$ is treated exactly.
Furthermore, Corollary~1 of Ref.~\cite{Chowdhury2016} states that under an approximate implementation $U_s(\tagH)$ of $\e^{-\im H s\Delta  t}$,
\begin{align}
    \| U_s(\tagH) - \e^{-\im \tagH s\Delta  t}\|
    \le 
    \frac{\epsilon'}{4}
\end{align}
for all $s\in[-S,S]$. Consequently, the approximate filter operator
\begin{align}
\label{eq:G_tilde}
    \tilde{G}_{\sigma}(E)
    =
    \frac{\sigma}{\sqrt{2\pi}}\sum_{s=-S}^{S} \Delta  t\,
    \e^{-\frac{\sigma^2(s\Delta t)^2}{2}}U_s(\tagH)
\end{align}
has an error upper-bounded by
\begin{align}
\label{eq:error_G}
    \| \tilde{G}_{\sigma}(E) - G_{\sigma}(E)\|
    \le 
    \epsilon'.
\end{align}

From Eq.~\eqref{eq:error_G}, one can show 
\begin{align}
    |\bra{\psi_0}O \tilde{G}_{\sigma}\ket{\psi_0}
        - \bra{\psi_0} O G_{\sigma}\ket{\psi_0}|
    =
    |\bra{\psi_0}O (\tilde{G}_{\sigma} - G_{\sigma})\ket{\psi_0}| 
    \le
    \|O\|\, \|\tilde{G}_{\sigma} - G_{\sigma}\|
    \le 
    \epsilon'\|O\|,
\end{align}
where $\ket{\psi_0}$ is an arbitrary quantum state.
Recall that for positive real values $a, \tilde{a}, b, \tilde{b}$ satisfying $|\tilde{a}-a|\le\epsilon'\|O\|$ and $|\tilde{b}-b|\le \epsilon'$, the following inequality holds:
\begin{align}
    \left|
        \frac{\tilde{a}}{\tilde{b}}
        - \frac{a}{b}
    \right|
    \le
    \frac{|a-\tilde{a}|\tilde{b}}{\tilde{b}b}
    +
    \frac{\tilde{a}|b-\tilde{b}|}{\tilde{b}b}
    \le
    \frac{(\tilde{a}+\|O\|\tilde{b})\epsilon}{\tilde{b}b}.
\end{align}
Setting
\begin{align}
    a =
    \bra{\psi_0} O G_{\sigma}\ket{\psi_0},
    \qquad
    \tilde{a} =
    \bra{\psi_0} O \tilde{G}_{\sigma}\ket{\psi_0},
    \qquad
    b =
    \bra{\psi_0}G_{\sigma}\ket{\psi_0},
    \qquad
    \tilde{b} =
    \bra{\psi_0}\tilde{G}_{\sigma}\ket{\psi_0},
\end{align}
we find
\begin{align}
\label{eq:error_exp}
\begin{split}
    &\left|
        \frac{\bra{\psi_0}O \tilde{G}_{\sigma}\ket{\psi_0}}
    {\bra{\psi_0}\tilde{G}_{\sigma}\ket{\psi_0}}
        -
        \frac{\bra{\psi_0}O G_{\sigma}\ket{\psi_0}}
    {\bra{\psi_0}G_{\sigma}\ket{\psi_0}}
    \right|
    \le
    \frac{\epsilon'\|O\|
    (\bra{\psi_0}\tilde{G}_\sigma\ket{\psi_0}+\bra{\psi_0}\tilde{G}_\sigma\ket{\psi_0})}
    {\bra{\psi_0}G_\sigma\ket{\psi_0}\bra{\psi_0}\tilde{G}_\sigma\ket{\psi_0}}
    \underset{\text{w.h.p}}{\le}
    \frac{\epsilon'\|O\|}
    {\Tr[G_\sigma]/2^N}
    =:
    \epsilon,
\end{split}
\end{align}
where the second inequality holds with high probability. Here, we have defined $\epsilon=\epsilon'2^N\|O\|/\Tr[G_\sigma]=\epsilon'2^N\|O\|/\e^{S_\sigma}$ using Eq.~\eqref{eq:entropy}.



Finally, we estimate the number of shots required to estimate $\bra{\psi_0}O\tilde{G}_\sigma\ket{\psi_0}$ with the statistical uncertainty $\epsilon'$.
The variance of $\bra{\psi_0} O U_s\ket{\psi_0}$ is given by 
\begin{align}
\begin{split}
    \mathrm{Var_{shots}}\left[\bra{\psi_0} O U_s\ket{\psi_0}\right]
    &:=
    \bra{\psi_0} (O U_s)^\dag (O U_s)\ket{\psi_0}
    - |\bra{\psi_0} O U_s\ket{\psi_0}|^2
    \\
    &\le
    \bra{\psi_0}U_s^\dag O^\dag O U_s\ket{\psi_0}
    \le
    \|O\|^2.
\end{split}
\end{align}
Then, the variance of $\bra{\psi_0}O\tilde{G}_\sigma\ket{\psi_0}$, where 
$\tilde{G}_\sigma$ is given in Eq.~\eqref{eq:G_tilde}, is calculated as 
\begin{align}
    &\frac{\sigma^2}{2\pi}\sum_{s=-S}^{S} (\Delta  t)^2\,
    \e^{-\sigma^2(s\Delta t)^2}
    \mathrm{Var_{shots}}\left[\bra{\psi_0} O U_s\ket{\psi_0}\right]
    \le
    \calO\big(\sigma\Delta t\|O\|^2\big).
\end{align}
Refer to Eq.~\eqref{eq:discretization_params} for $\Delta t$.
To achieve a standard deviation of $\calO(\epsilon') = \calO(\epsilon\e^{S_\sigma}/(2^{N}\|O\|))$, the number of shots required is 
\begin{align}
    N_\mathrm{shots}
    =
    \calO\left(
        \frac{\sigma\Delta t\|O\|^2}{\epsilon'^2}
    \right)
    \approx
    \calO\left(
        \frac{\|O\|^4 2^{2N}}{\epsilon^2 \e^{2S_\sigma}}
    \right), 
\end{align}
where polynomial dependence on $N$ is ignored in the approximate equality.

\section{Additional experimental data for thermal expectation values}
\label{app:LZZ}

In this appendix, we present additional experimental results of 
the amplitudes $\calL(t)$ and $\calL_{Z_1 Z_2}(t)$, which 
complement those shown in Fig.~\ref{fig:H12_Z1Z2}, and are also 
obtained using the Quantinuum H1-2 system. 
The experimental setup is described in Sec.~\ref{sec:exp_thermal}.

First, we briefly outline an error mitigation scheme based 
on the depolarizing noise. 
We evaluate the mitigated value ${\cal L}_O(t)^{\rm (mit)}$ by 
renormalizing the experimentally obtained amplitude 
${\cal L}_O(t)^{\rm (noisy)}$ as follows: 
\begin{equation}\label{eq:LOmit}
    {\cal L}_O(t)^{\rm (mit)} = \frac{{\cal L}_O(t)^{\rm (noisy)}}
    {[1-p^\mathrm{(ent)}_\mathrm{2Q}(\tagJ t/2)]^{N_\mathrm{2Q}}}.  
\end{equation}
Here, $O$ represents either the identity or the operator $Z_1 Z_2$, 
and $p^\mathrm{(ent)}_\mathrm{2Q}(\theta)=\frac{5}{4}p_\mathrm{2Q}(\theta)$ is the entanglement infidelity~\cite{Horodecki1998,Nielsen2002,Proctor2020, Baldwin2022} with $N_\mathrm{2Q}$ denoting the number of two-qubit gates involved 
in the controlled $O{\cal U}(t)$ gate (see Fig.~\ref{circ:TPQ}). 
The average two-qubit gate error rate $p_\mathrm{2Q}(\tagJ t/2)$ 
is given in Eq.~\eqref{eq:pt}. 
Specifically, in our experiments, $N_\mathrm{2Q}=112$ for the 
controlled ${\cal U}(t)$ gate and 
$N_\mathrm{2Q}=114$ for the controlled $Z_1 Z_2 {\cal U}(t)$ gate. 

\begin{figure*}[th]
  \includegraphics[width=1.0\textwidth]
  {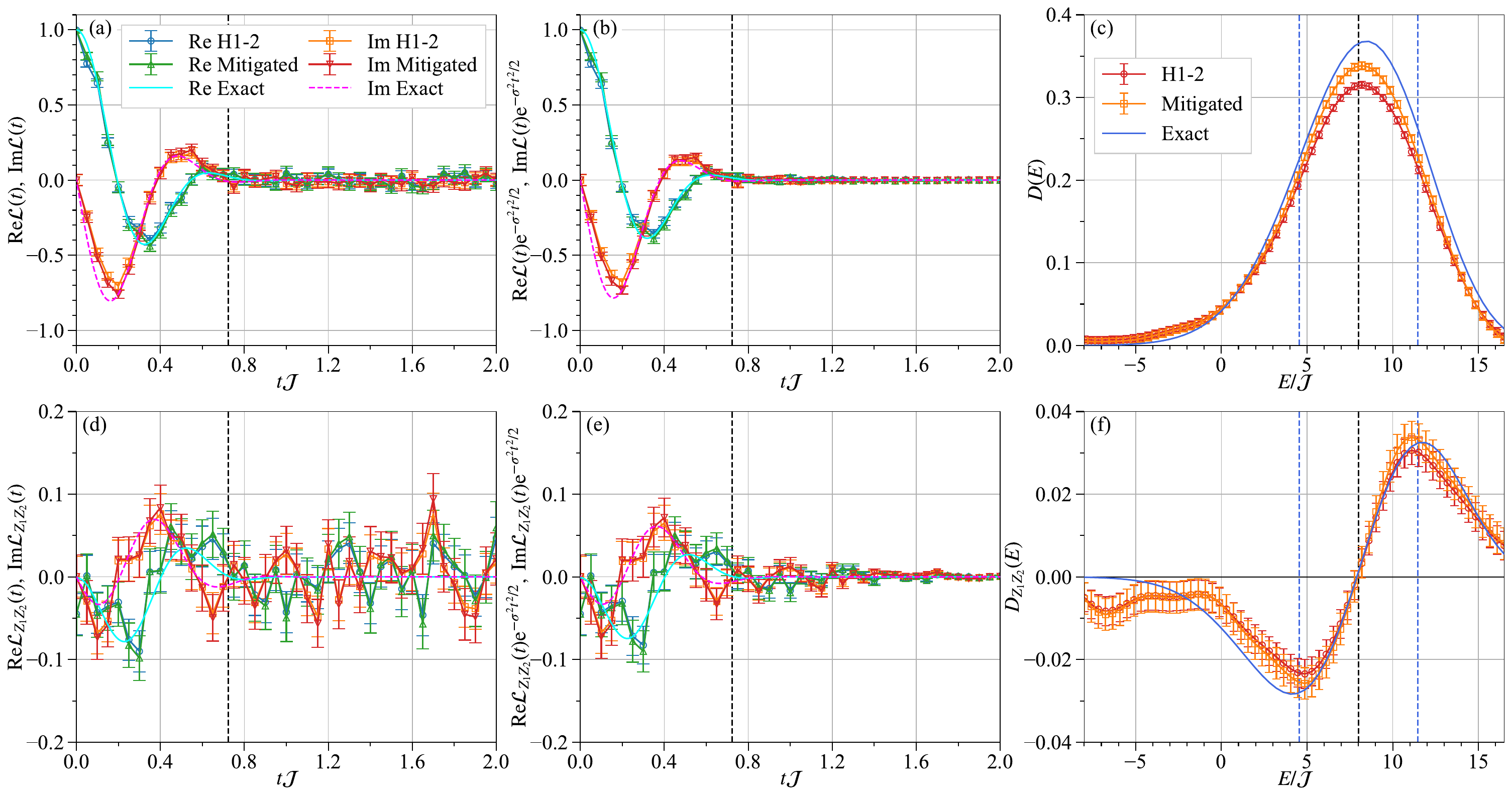}
  \caption{
    \label{fig:Loschmidt_H12}
    (a) Real and imaginary parts of the Loschmidt amplitude 
    $\calL(t)$ for the one-dimensional Heisenberg model $\tagH_{XXX}$ with $N=16$ sites under periodic boundary conditions. 
    The experimental results ${\cal L}_O(t)^{\rm (noisy)}$ 
    are obtained for the Floquet scrambled 
    states $|\psi(m)\rangle$ with $m=8$ 
    and $16$ using the Quantinuum H1-2 system. 
    The error-mitigated results ${\cal L}_O(t)^{\rm (mit)}$ are 
    obtained by renormalizing the experimental values as in 
    Eq.~(\ref{eq:LOmit}). 
    The vertical line indicates $t=\sigma^{-1}$, where $\sigma$ is 
    defined in Eq.~(\ref{eq:sigma}). For comparison, the ideal 
    values obtained from classical calculations are also shown 
    by solid and dashed curves for the real and imaginary parts, 
    respectively. 
    (b) Same as (a) but with the Loschmidt amplitude multiplied by the Gaussian weight $\e^{-\sigma^2t^2/2}$. 
    (c) The associated density of states $D(E)$ obtained via 
    Eq.~(\ref{eq:D(E)}) as a Fourier transform of the 
    Gaussian-weighted Loschmidt amplitude 
    $\calL(t) \e^{-\sigma^2t^2/2}$ shown in (b).  
    The black dashed vertical line indicates $E=E_{\infty}=8\tagJ$, 
    and the blue dashed vertical lines indicate 
    $E=E_{\infty}\pm \sigma_\tagH = 8\tagJ \pm 2\sqrt{3}\tagJ$, 
    where $\sigma_\tagH$ is defined in Eq.~(\ref{eq:sigma_H}). 
    The ideal values obtained from classical 
    calculations are also shown by a blue solid curve.    
    (d-f) Same as (a-c) but for $\calL_{Z_1 Z_2}(t)$ and 
    $D_{Z_1 Z_2}(E)$. 
    Note that the unmitigated results in (a) and (c) are the same 
    as those shown in Figs.~\ref{fig:H12_Z1Z2}(a) and 
    \ref{fig:H12_Z1Z2}(b), respectively.     
  }
\end{figure*}

Figures~\ref{fig:Loschmidt_H12}(a,d) show the amplitudes 
${\cal L}(t)$ and ${\cal L}_{Z_1 Z_2}(t)$ with and without 
the error mitigation, along with the corresponding ideal values 
obtained from classical calculations. 
While the difference is subtle, we observe that the 
error-mitigated ${\cal L}(t)^{\rm (mit)}$ tends 
towards the ideal value, particularly at early times 
$t\lesssim \sigma^{-1}$, where $\sigma$ is defined 
in Eq.~(\ref{eq:sigma}). 
In contrast, the error-mitigated $\calL_{Z_1 Z_2}(t)^{\rm (mit)}$ 
is practically indistinguishable from the unmitigated counterpart  
$\calL_{Z_1 Z_2}(t)^{\rm (noisy)}$ within the statistical errors 
due to the shot noise. 
This is because, 
at early times $t\lesssim \sigma^{-1}$, the signal of 
$\calL_{Z_1 Z_2}(t)$ is nearly ten times smaller than that of 
$\calL(t)$, implying that 100 times more shots are required to 
observe the difference, which is impractical. 

Figures~\ref{fig:Loschmidt_H12}(b,e) show the amplitudes multiplied 
by the Gaussian weight $\e^{-\sigma^2 t^2/2}$, i.e., 
$\calL(t) \e^{-\sigma^2t^2/2}$ and $\calL_{Z_1 Z_2}(t) \e^{-\sigma^2t^2/2}$, respectively. 
The Gaussian factor $\e^{-\sigma^2 t^2/2}$ 
in Eq.~\eqref{eq:amplitude} attenuates these amplitudes and 
their statistical fluctuations for $t\gtrsim \sigma$. 
This illustrates that only the amplitudes at early times 
$t \lesssim \sigma$ significantly contribute to the associated 
density of states 
in Eqs.~(\ref{eq:D_O(E)}) and (\ref{eq:D(E)}).

Figures~\ref{fig:Loschmidt_H12}(c,f) show $D(E)$ and $D_{Z_1 Z_2}(E)$, 
respectively. 
Although the difference in the the Loschmidt amplitude 
$\calL(t)$ with and without the error mitigation, as shown in 
Fig.~\ref{fig:Loschmidt_H12}(a), is small, 
the error mitigated $D(E)$ shows better agreement with 
the ideal results compared to the unmitigated one. 
However, as expected, 
no significant improvement beyond the statistical errors is observed 
in $D_{Z_1 Z_2}(E)$ due to the small signals of $\calL_{Z_1 Z_2}(t)$ 
relative to the statistical errors.
Despite the small signals of $\calL_{Z_1 Z_2}(t)$, as discussed for 
$\langle Z_1 Z_2 \rangle^{(\rm est)}$ in Sec.~\ref{sec:exp_thermal}, we observe excellent agreement with the ideal results 
within the statistical uncertainty 
in the energy range $E_\infty-\sigma_\tagH\lesssim E\lesssim E_\infty+\sigma_\tagH$ due to the concentrated energy spectrum 
around $E_\infty$. Here, $\sigma_\tagH$ is defined in 
Eq.~(\ref{eq:sigma_H}). 
The experimental data of $D_{Z_1 Z_2}(E)$ indeed 
changes its sign such that 
$D_{Z_1 Z_2}(E)<0$ for $E\lesssim E_{\infty}$ and 
$D_{Z_1 Z_2}(E)>0$ for $E\gtrsim  E_{\infty}$, as theoretically 
expected. 


Importantly, the effect of depolarizing noise mostly cancels out 
between the numerator and denominator in the thermal expectation 
value $\langle Z_1 Z_2 \rangle^{(\rm est)}=D_{Z_1 Z_2}(E)/D(E)$, 
as generally given in Eq.~(\ref{eq:estimator_circuit}), 
similarly to the normalized OTOC in Eq.~\eqref{eq:normalized_OTOC}.
We expect that this error mitigation scheme is effective 
even for more complex observables $O$ where $N_{2Q}$ in 
$\calL_O(t)$ is significantly larger than that of $\calL(t)$, 
including higher-weight Pauli strings and dynamical correlation 
functions.

\end{document}